\documentclass[a4paper,fleqn,usenatbib,onecolumn]{mnras} 

\usepackage[T1]{fontenc}
\usepackage{ae,aecompl}

\usepackage{comment}
\usepackage{graphicx}	
\usepackage{amsmath}	
\usepackage{amssymb}	
\usepackage{lmodern}
\usepackage{bm}
\usepackage{cases}
\usepackage{float}
\usepackage[caption = false]{subfig}
\usepackage{colortbl}
\usepackage{multirow}
\usepackage{booktabs}
\usepackage{tabularx}
\usepackage{makecell}
\usepackage{hyperref}
\usepackage[textsize=small,textwidth=1in]{todonotes}

\newcommand{\semibold}[1]{{\fontseries{b}\selectfont{#1}}}
\newcommand{\para}[1]{\par\vspace{2mm}\noindent\semibold{{#1.}---}\ignorespaces}

\renewcommand{\leq}{\leqslant}
\renewcommand{\geq}{\geqslant}

\definecolor{ERCorange}{HTML}{F26623}
\definecolor{SussexCobaltBlue}{HTML}{1E428A}
\definecolor{SussexDeepAquamarine}{HTML}{487A7B}
\definecolor{SussexMidGrey}{HTML}{94A596}
\definecolor{SussexFlint}{HTML}{013035}

\hypersetup{colorlinks=true,citecolor=SussexDeepAquamarine,linkcolor=SussexCobaltBlue,urlcolor=SussexCobaltBlue}

\newcommand{\be}{\begin{equation}}
\newcommand{\ee}{\end{equation}}

\newcommand{\Mpc}{\text{Mpc}}
\newcommand{\Gpc}{\text{Gpc}}

\newcommand{\DiracD}{\delta_{\text{D}}}
\newcommand{\Kronecker}{\bm{1}}
\DeclareMathOperator{\BigO}{\mathcal{O}}
\newcommand{\e}[1]{\mathrm{e}^{#1}}
\newcommand{\im}{\mathrm{i}}

\newcommand{\B}[1]{\ensuremath{\bm{#1}}}
\newcommand{\D}[1]{\ensuremath{\text{d}#1}}

\newcommand{\CovMatrix}{\bm{\mathsf{C}}}
\newcommand{\FisherMatrix}{\bm{\mathsf{F}}}
\newcommand{\CorrMatrix}{\bm{\mathsf{r}}}

\DeclareMathOperator{\Tr}{\mathrm{tr}}

\DeclareMathOperator{\CovGauss}{\mathrm{Cov}_{\mathrm{G}}}

\DeclareMathOperator{\sinc}{\mathrm{sinc}}

\newcommand*{\llangle}{\langle\kern-2\nulldelimiterspace \langle}
\newcommand*{\rrangle}{\rangle\kern-2\nulldelimiterspace \rangle}

\newcommand{\bk}{\bm{\mathrm{k}}}
\newcommand{\bq}{\bm{\mathrm{q}}}
\newcommand{\bx}{\bm{\mathrm{x}}}
\newcommand{\br}{\bm{\mathrm{r}}}
\newcommand{\bZero}{\bm{\mathrm{0}}}

\newcommand{\kmax}{k_{\text{max}}}
\newcommand{\kmin}{k_{\text{min}}}

\newcommand{\kNy}{k_{\text{Ny}}}
\newcommand{\kf}{k_{\mathrm{f}}}

\newcommand{\nmax}{n_{\text{max}}}
\newcommand{\Ntriangles}{N_{\text{triangles}}}
\newcommand{\Ngrid}{N_{\text{grid}}}
\newcommand{\Nbin}{N_{\text{bin}}}
\newcommand{\Nreal}{N_{\text{real}}}

\newcommand{\sigmamax}{\sigma_{\text{max}}}

\newcommand{\WCIC}{W_{\text{\textsc{cic}}}}

\newcommand{\LambdaCDM}{$\Lambda$CDM}

\newcommand{\data}{\text{data}}
\newcommand{\theory}{\text{theory}}
\newcommand{\tree}{\text{tree}}
\newcommand{\oneloop}{\text{1-loop}}
\newcommand{\halo}{\text{halo}}
\newcommand{\phase}{\epsilon}
\newcommand{\modal}{\text{modal}}

\newcommand{\Ptheory}{P^{\theory}}
\newcommand{\Ptree}{P^{\tree}}
\newcommand{\Ploop}{P^{\oneloop}}
\newcommand{\Phalo}{P^{\halo}}

\newcommand{\Pgaltheory}{\Ptheory_{\text{gal}}}
\newcommand{\Pgaltree}{\Ptree_{\text{gal}}}
\newcommand{\Pgalloop}{\Ploop_{\text{gal}}}

\newcommand{\Bmodal}{B_{\modal}}
\newcommand{\Bphase}{B_{\phase}}
\newcommand{\Btheory}{B^{\theory}}
\newcommand{\Bdata}{B^{\data}}
\newcommand{\Btree}{B^{\tree}}
\newcommand{\Bloop}{B^{\oneloop}}
\newcommand{\Bhalo}{B^{\halo}}

\newcommand{\Bgaltheory}{\Btheory_{\text{gal}}}
\newcommand{\Bgaltree}{\Btree_{\text{gal}}}
\newcommand{\Bgalloop}{\Bloop_{\text{gal}}}

\newcommand{\Bphasetheory}{\Btheory_{\phase}}

\newcommand{\iB}{iB}
\newcommand{\ib}{ib}

\newcommand{\iBtheory}{\iB^{\theory}}

\newcommand{\LCF}{\ell}

\newcommand{\LCFtheory}{\LCF^{\theory}}
\newcommand{\LCFtree}{\LCF^{\tree}}

\newcommand{\BispectrumDegeneracy}{\mathsf{N}}
\newcommand{\TriangleRegion}{\mathcal{V}}
\newcommand{\SignalToNoise}{\frac{\mathcal{S}}{\mathcal{N}}}

\newcommand{\Nbody}{N-body}
\newcommand{\NbodyUpper}{N-BODY}    

\newcommand{\Pestimator}{\mathcal{P}}
\newcommand{\Bestimator}{\mathcal{B}}
\newcommand{\Likelihood}{\mathcal{L}}

\newcommand{\transpose}{\mathsf{T}}

\newcommand{\Dfactor}[1]{\mathcal{D}_{#1}}
\newcommand{\Mfactor}[1]{\mathcal{M}_{#1}}

\newcommand{\softwarefont}{\sffamily\fontseries{sbc}\selectfont}
\newcommand{\Vegas}{\href{http://www.feynarts.de/cuba}{\softwarefont Vegas}}
\newcommand{\PTHalos}{{\softwarefont PTHalos}}
\newcommand{\HMcode}{\href{https://github.com/alexander-mead/HMcode}{\softwarefont HMcode}}
\newcommand{\CUBA}{\href{http://www.feynarts.de/cuba}{\softwarefont CUBA}}
\newcommand{\CMBFAST}{\href{https://lambda.gsfc.nasa.gov/toolbox/tb_cmbfast_ov.cfm}{\softwarefont CMBFAST}}
\newcommand{\ZBOX}{\href{http://www.ics.uzh.ch/\%7Estadel/doku.php?id=zbox:zbox4}{\softwarefont ZBOX}}
\newcommand{\Gadget}{\href{http://wwwmpa.mpa-garching.mpg.de/gadget/}{\softwarefont Gadget-2}}

\title[Towards optimal cosmological parameter recovery from compressed bispectrum statistics]
{Towards optimal cosmological parameter recovery\\ from compressed bispectrum statistics}
\author[Byun, Eggemeier, Regan, Seery, Smith]
{
  \parbox{\textwidth}{Joyce Byun\thanks{joyce.byun@sussex.ac.uk}, Alexander Eggemeier\thanks{a.eggemeier@sussex.ac.uk},
  Donough Regan\thanks{d.regan@sussex.ac.uk}, \\
  David Seery\thanks{d.seery@sussex.ac.uk} \& Robert E. Smith\thanks{r.e.smith@sussex.ac.uk}\\\mbox{}} \\
  Astronomy Centre, School of Mathematical and Physical Sciences, 
  University of Sussex, Brighton BN1 9QH, United Kingdom
}

\date{Accepted XXX. Received YYY; in original form ZZZ}

\pubyear{2017}

\begin{document}
\label{firstpage}
\pagerange{\pageref{firstpage}--\pageref{lastpage}}
\maketitle

\begin{abstract}
Over the next decade, improvements in cosmological parameter constraints
will be driven by surveys of large-scale
structure in the Universe.
The information they contain
is encoded in a hierarchy of correlation functions, and
tools to utilize the two-point function are
already well-developed.
But the inherent non-linearity of large-scale structure
suggests that further information will be
embedded in higher correlations, of which the bispectrum
is currently the most accessible.
Extracting this information is extremely challenging:
it requires accurate theoretical modelling
and significant computational resources to estimate
the covariance matrix describing correlations between
different configurations of Fourier modes.
We investigate whether it is possible to reduce the
covariance matrix without significant loss of
information by using a proxy that aggregates the bispectrum
over a subset of Fourier configurations.
Specifically, we study the constraints on
$\Lambda$CDM
parameters from combining the power spectrum
with
(\emph{a}) the modal decomposition of the bispectrum,
(\emph{b}) the line correlation function
and
(\emph{c}) the integrated bispectrum.
We forecast the error bars achievable on
$\Lambda$CDM parameters in a future galaxy survey
that measures one of these proxies and compare them to those
obtained from measurements of the Fourier bispectrum,
including simple estimates of their degradation in the presence of
shot noise.
Our results
demonstrate that the modal bispectrum
performs as well as the Fourier bispectrum,
even with considerably fewer modes than Fourier configurations.
The line correlation function has good performance
but does not match the modal bispectrum.
The integrated bispectrum is
comparatively insensitive to changes in the background cosmology.
We find that the addition of bispectrum data
can improve constraints on bias parameters and the normalization
$\sigma_8$ by a factor between 3 and 5 compared to power spectrum 
measurements alone.
For other parameters, improvements of up to $\sim$ 20\%
are possible.
Finally, we use a range of theoretical models to explore
how the sophistication required for realistic predictions
varies with each proxy.
\end{abstract}

\begin{keywords}
Cosmology: theory, Large-scale structure of the Universe
\end{keywords}
\par\vspace{6mm}\mbox{}\hfill
\raisebox{-0.5\height}{\includegraphics[scale=0.2]{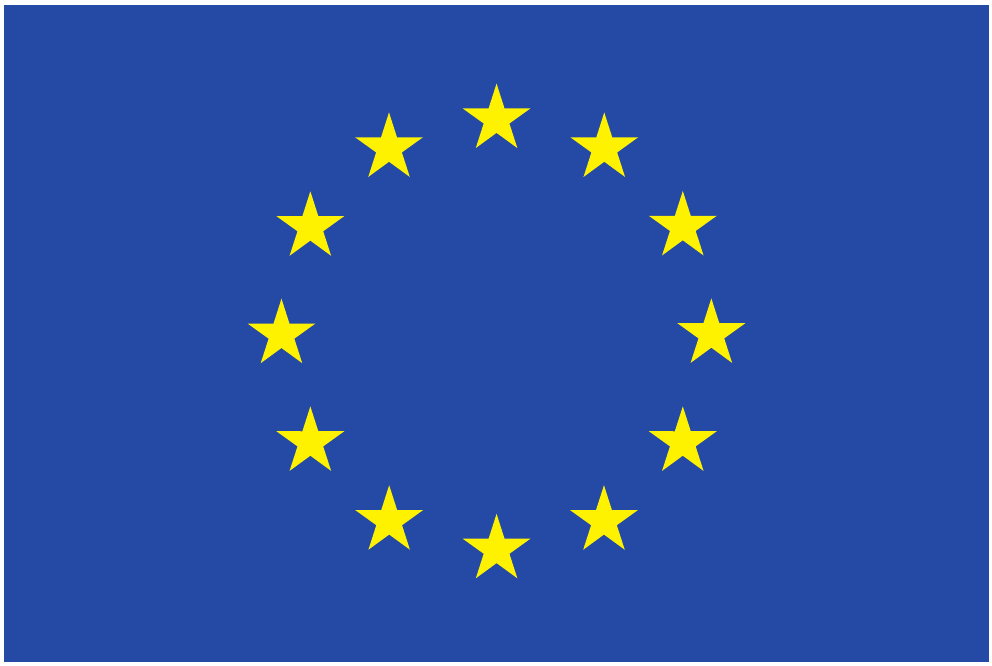}}
\;\textcolor{ERCorange}{\vrule width 1pt}\;
\raisebox{-0.5\height}{\includegraphics[scale=0.1]{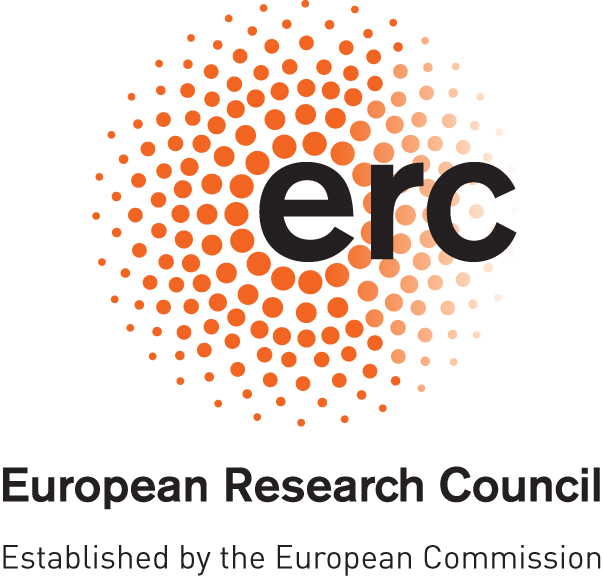}}
\quad
\raisebox{-0.5\height}{\includegraphics[scale=0.5]{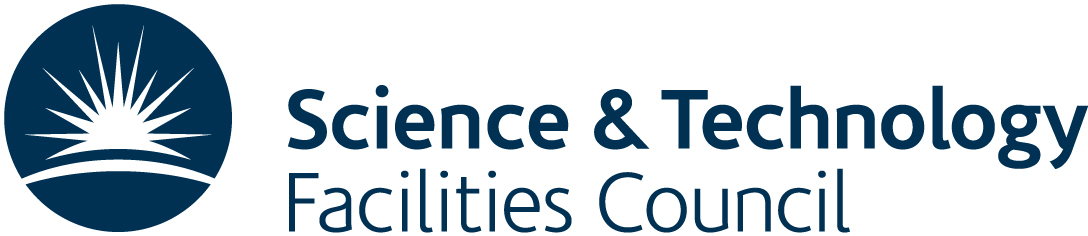}}

\newpage
\section{Introduction}
\label{sec:intro}
Constraints on cosmological
parameters have improved significantly over the last two decades,
driven by high-precision data from the cosmic microwave
background (`CMB') temperature and polarization
anisotropies \citep{Bennett:2003ca,Ade:2013ydc}.
But the capacity of CMB observations to sustain this
rate of progress is now nearly exhausted.
Measurements of the temperature anisotropy have become
limited by cosmic variance down to very small scales,
and therefore
future large-scale measurements will furnish little
new information.
Meanwhile,
on small scales, cosmological information begins to be erased
by astrophysical processes.
Modest improvements may still come from better polarization data,
perhaps shrinking current uncertainties by a factor of a few,
but
eventually these measurements
will also approach the limit of cosmic variance.
Further progress will be possible only with new sources of information.
In the decade 2020--2030 we expect such a source to be provided by
surveys of cosmological large-scale structure---but only if the
information these surveys
contain can be extracted and understood~\citep{Silk:2016srn}.

\para{The bispectrum: challenges}
The statistical information contained in a galaxy survey is
carried by its hierarchy of correlation functions, of which typically only a
few lowest-order functions can be measured accurately.
Tools to extract information from the two-point function
were developed early and are now mature.
The development of tools to extract information from higher-order correlation
functions has proceeded more slowly
\citep{Fry:1983cj,Goroff:1986ep,Scoccimarro:2000sn,Sefusatti:2006pa},
but because structure formation is non-linear it is likely
that these carry an important fraction of the information content.
To make good use of our investment in costly observational programmes it will be
necessary to find a means of using information from at least the
three-point function.

What are the challenges?
A first difficulty arises from combinatorics.
We write the matter overdensity at time $t$ as
$\delta(\bx,t) = \delta\rho(\bx, t) / \bar{\rho}(t)$,
where $\delta\rho(\bx, t) = \rho(\bx,t) - \bar{\rho}(t)$ is the
density perturbation and $\rho(t)$ is the uniform background.
Allowing angle brackets $\langle \cdots \rangle$ to denote an
ensemble average,
statistical homogeneity
makes the two- and three-point functions
$\langle \delta(\bx) \delta(\bx + \br) \rangle$
and $\langle \delta(\bx) \delta(\bx + \br_1) \delta(\bx + \br_2) \rangle$
independent of the origin $\bx$.
After translation to Fourier space this enforces conservation of momentum
for the wavenumbers that participate in the expectation value,
\begin{subequations}
\begin{align}
    \label{eq:defP}
    \langle \delta(\bk_1) \delta(\bk_2) \rangle
    & = (2\pi)^3 \DiracD(\bk_1 + \bk_2) P(k) , \\
    \label{eq:defB}
    \langle \delta(\bk_1) \delta(\bk_2) \delta(\bk_3) \rangle
    & = (2\pi)^3 \DiracD(\bk_1 + \bk_2 + \bk_3) B(k_1, k_2, k_3) ,
\end{align}
\end{subequations}
where $k = |\bk_1| = |\bk_2|$ is the common magnitude of the
wavenumbers appearing in the two-point function.
In Equations~\eqref{eq:defP}--\eqref{eq:defB}
and the remainder of this paper we suppress the time $t$
labelling the hypersurface of evaluation.
Isotropy makes the power spectrum $P$ a function only of $k$,
while the bispectrum $B$ is a function of the three wavenumbers $k_1$, $k_2$, $k_3$
subject to the closure condition $\bk_1 + \bk_2 + \bk_3 = 0$.
Therefore a fixed volume of space yields many more distinct configurations
of the bispectrum than of the spectrum.
If we choose to measure all of them then we must provide an estimate
for their covariance,
and
beyond the Gaussian approximation this
typically requires {\Nbody} simulations.
Since we require at least as many simulations as the
number of independent covariances,
the number of simulations to be performed grows
at least linearly in the number of configurations.
This makes it very expensive to use more than a fraction of the
available bispectrum measurements.

Second, we must estimate 
typical values for $B(k_1, k_2, k_3)$
in a particular cosmological model.
While such estimates
are already necessary for the power spectrum $P(k)$, accurate
estimates
for the bispectrum are substantially more challenging.
There are two key reasons.
No matter what methods we use, the algebraic complexity associated with
high-order correlation functions is usually worse than at lower order.
Also, many of our standard tools have a reduced range of validity as we move
up the correlation hierarchy.
We must therefore work harder to obtain trustworthy predictions from our models,
and in some cases we can do so only by giving up analytic methods altogether.

These problems have hampered the development of a toolkit
that would make use of bispectrum measurements routine.
Nevertheless, they are difficulties of practice and not obstructions of principle---%
if necessary, we could determine both covariances and typical values of $P$ or $B$
from {\Nbody} simulations, at least over a certain range of scales.
But such determinations would require a very large number of realizations.
The sheer computational resource entailed by this strategy makes it unattractive
on timescales of interest for surveys such Euclid,
DESI, or LSST.

\para{Alternative strategies}
To build a practical methodology we must cut the size of the covariance matrices
and avoid simulations where possible.
Simulations are not needed when analytic methods suffice to predict $P$ or $B$,
or when a Gaussian approximation to the covariance is acceptable.
Meanwhile,
an obvious way to reduce the number of configurations is simply not to measure
them all. Depending how aggressively we choose to cut, this may mean
accepting a significant loss of information.
A more nuanced option is to aggregate
groups of configurations into weighted averages,
effectively \emph{compressing} the data carried by the
bispectrum rather than discarding it.
Such averages could be computed directly.
But there are also observables whose
statistics can naturally be expressed
as weighted averages of this kind.
Measuring these will often be
simpler than measuring amplitudes of the Fourier
bispectrum---simultaneously reducing the effort
required to estimate and invert their covariance matrices.
We describe these observables as `proxies'
or `proxy statistics' for the full Fourier
bispectrum.

Each proxy represents a
compromise between (\emph{a}) information loss due to compression,
(\emph{b}) the type of Fourier configurations over which it
aggregates, and therefore the physics to which
it is sensitive, and (\emph{c}) its accessibility to analytical modelling,
either for covariances or to estimate typical measurements.
In this paper we select three proxies that have already been
described in the literature and
characterize their performance in each of these categories.
Our aim is not to find an optimal proxy for any particular measurement,
but rather to demonstrate that their use represents a feasible
strategy for upcoming surveys without unacceptable degradation
in information recovery.

\para{Summary}
Our principal results are forecasts for the parameter error bars achievable
from combinations of the galaxy power spectrum and bispectrum, or its proxies.
The parameter set we study comprises the background quantities
of a $\Lambda$CDM model with evolving dark energy, supplemented by
two parameters describing the bias model~\citep{McDonald:2009dh}.
We study how these forecasts change
when they are estimated using the
complete non-Gaussian covariance matrix
or its Gaussian approximation.
We characterize their dependence
on the method used to predict typical values
for $P(k)$ and $B(k_1, k_2, k_3)$
by sampling the results using
tree-level
and one-loop standard perturbation theory (`SPT'),
and an implementation of the halo model.
We compare these estimates with values measured
directly from simulations.
These results can be used to determine, for each observable,
the degree of modelling sophistication that is required
to obtain accurate forecasts.

Our analysis does not
include the effect of survey geometry or incompleteness, or redshift-space
effects, and should be regarded as a determination of the performance of each proxy
under idealized conditions. We include a simple analysis that indicates how our
results would change in the presence of shot noise.

Fisher forecasts including Fourier bispectrum measurements
have previously been reported by
\cite{Sefusatti:2006pa},
assuming $1{,}015$ bispectrum
configurations and measuring covariances
from a suite of $6{,}000$ mock catalogues generated
by the {\PTHalos} algorithm~\citep{Scoccimarro:2001cj}
and
second-order Lagrangian perturbation theory
(`2LPT').
Their results suggested that the bispectrum contains
significant cosmological information.
For comparison,
in our analysis we use $95$ bispectrum configurations
in order to keep the size of the covariance matrix within
plausible bounds,
and measure it directly from a suite of full {\Nbody} simulations.

More recently, \cite{Chan:2016ehg}
estimated the extra constraining power of Fourier bispectrum measurements
by computing their contribution to the signal-to-noise, but did not
make forecasts for error bars on cosmological parameters.
They found that the bispectrum contributed up to a $\sim 30\%$ increase in  
signal-to-noise above the power spectrum
and concluded that the information gain would be modest,
perhaps being principally useful to break degeneracies.
One of our aims is to clarify the relationship between this conclusion
and the more nuanced outcomes found by~\cite{Sefusatti:2006pa}.
We find that estimates based on signal-to-noise alone
generally give only a rough indication 
compared to the full Fisher calculation
because they do not account for variations in the sensitivity to
background cosmology between observables.

\para{Organization}
Our presentation is organized as follows.
In Section~\ref{sec:estimators}
we introduce the three bispectrum proxies to be studied in the
remainder of the paper.
These are:
(\emph{a}) the \emph{modal bispectrum}, which can be regarded
as an alternative to the Fourier bispectrum obtained by
exchanging the Fourier modes
$\e{\im \bk \cdot \bx}$ for an alternative basis
\citep{Fergusson:2010ia,Reganetal2012};
(\emph{b}) the
\emph{line correlation function},
which samples three-point statistics of the phase
of the density fluctuation~\citep{Obreschkow:2012yb,Wolstenhulme:2014cla},
and (\emph{c})
the \emph{integrated bispectrum}~\citep{Chiang:2014oga},
which measures variation of the power spectrum in subsampled
regions.
Each of these measures can be expressed as a weighted average over
particular configurations of the Fourier bispectrum.

In Sections~\ref{sec:predict-ib}--\ref{sec:predict-modal}
we explain how each proxy can be
predicted using the halo model or
a flavour of SPT.
In Section~\ref{sec:galaxy-bias} we explain our prescription
to obtain the biased galaxy density field from the underlying
matter density field,
which is the quantity predicted by these
analytic models.
In Section~\ref{sec:estimation} we describe our procedure
to recover estimates for each proxy statistic
from {\Nbody} simulations,
and
in Section~\ref{sec:comparison} we compare these
estimates
(and estimates for their deriatives with respect
to the cosmological parameters)
with theoretical predictions.
Readers familiar with the measures of 3-point correlations
described in Section~\ref{sec:estimators}
and the modelling technologies of Section~\ref{sec:modelling}
may choose to begin reading at this point.
In Section~\ref{sec:covariance} we present signal-to-noise estimates
for the information content of each proxy.
Our Fisher forecasts appear in Section~\ref{sec:paramEstim}.
In Section~\ref{sec:discussion} we collect a number of
topics for discussion,
including the compression efficiency of each proxy statistic
and the impact of shot noise on our forecasts.
We conclude in Section~\ref{sec:conclusions}.

\para{Notation}
Our Fourier convention is
$f(\bx) = \int \D^3 k \, (2\pi)^{-3} f(\bk) \e{\im \bk \cdot \bx}$.
To avoid confusion we distinguish the Dirac $\delta$-function
$\DiracD(\bx)$
or $\DiracD(\bk)$
and the Kronecker symbol $\Kronecker_{ij}$
from the matter overdensity
$\delta \equiv \delta\rho / \rho$.

\section{The Fourier bispectrum and its proxies}
\label{sec:estimators}

In this section we introduce the proxy statistics
to which we compare the Fourier bispectrum.
This has already been defined---together with the power spectrum---%
in Equations~\eqref{eq:defP}--\eqref{eq:defB}.
We describe the integrated bispectrum
in Section~\ref{ssec:iB},
the line correlation function
in Section~\ref{ssec:lcf}
and the modal decomposition of the bispectrum in Section~\ref{ssec:modal}.
Each of these represents a
possible compression of the Fourier bispectrum,
in the sense described in Section~\ref{sec:intro}.

\subsection{Integrated bispectrum}
\label{ssec:iB}

The integrated bispectrum (or `position-dependent power spectrum')
was developed by \citet{Chiang:2014oga}
as a tool to search for primordial non-Gaussianity in large-scale structure.
It has several convenient features: it is easily estimated using standard power-spectrum
codes
and it has a clear physical interpretation.
As we shall see in Section~\ref{sec:predict-ib},
it represents a weighted average of the Fourier bispectrum
dominated by `squeezed' configurations---%
that is, wavenumbers
$(\bk_1, \bk_2, \bk_3)$ where one $k_i$ is much smaller than the other
two.
If we assume $k_3 \ll k_1, k_2$
then
the bispectrum $\langle \delta(\bk_1) \delta(\bk_2) \delta(\bk_3) \rangle$
expresses correlations between a single long-wavelength mode
$\delta(\bk_3)$ and the two-point function
$\langle \delta(\bk_1) \delta(\bk_2) \rangle$.
This makes it sensitive to `local-type' non-Gaussianity produced
by inflationary models with
more than one active field.
However, because gravitational collapse correlates
modes with comparable wavenumbers, the
bispectrum produced during mass assembly
is typically concentrated away from squeezed configurations.
For this reason it is not clear how sensitive
the integrated bispectrum
might be to the cosmological parameters that influence this assembly
process.

To define the integrated bispectrum
divide the total survey volume into $N_s$ cubic subvolumes,
each of volume $V_s \equiv L_s^3$ and centred at positions $\br_L$.
Compute the power spectrum and average overdensity for each subvolume,
which we denote $P(\bk,\br_L)$ and $\bar{\delta}(\br_L)$, respectively.
(The power spectrum $P(\bk, \br_L)$ may depend on the orientation of
$\bk$ if the subvolumes are not isotropic.)
Finally, the integrated
bispectrum is defined to be the expectation of
$P(\bk, \br_L) \bar{\delta}(\br_L)$,
averaged over the orientation of $\bk$,
\be 
\iB(k) \equiv \int \frac{\D^2\hat{k}}{4\pi} \; \langle P(\bk,\br_L) \bar{\delta}(\br_L) \rangle_{N_s} \,.
\label{eq:iB}
\ee
The notation $\langle \cdots \rangle_{N_s}$ indicates that the expectation
is to be taken over all subvolumes.

To compute this expectation we Taylor expand $P(\bk, \br_L)$
in powers of $\bar{\delta}(\br_L)$~\citep{Chiang:2014oga}.
The leading contribution is
\be \langle P(\bk,\br_L)\bar{\delta}(\br_L)\rangle_{N_s} =
\bigg\langle
    \bigg[
        P(\bk)\big|_{\bar{\delta}=0}
        + \frac{\D P(\bk)}{\D \bar{\delta}}\bigg|_{\bar{\delta}=0}
          \bar{\delta}(\br_L)
        + \cdots
    \bigg]
    \bar{\delta}(\br_L)
\bigg\rangle_{N_s}
\approx
\frac{\D \ln P(\bk)}{\D \bar{\delta}}\bigg|_{\bar{\delta}=0} P(\bk)
\sigma^2_{L}\ ,
\label{eq:iB-lowest}
\ee
where $\sigma^2_{L}\equiv \langle\bar{\delta}^2(\br_L)\rangle_{N_s}$ is the
variance in mean overdensity over the subvolumes. 
Therefore, at lowest order, the integrated bispectrum
describes variation of the power spectrum in
response to changes in the large-scale
overdensity.%
    \footnote{In field theory this is the `operator product expansion'.}
We conclude that measurements of $\iB$
contain both the power spectrum and its variance.
Since these can be measured directly, any new information contained
in the integrated bispectrum
must reside in its normalized component~\citep{Chiang:2014oga},
\be \ib(k) \equiv \frac{\iB(k)}{P(k)\sigma_{L}^2} \approx
\left. \frac{\D \ln P(k)}{\D
  \bar{\delta}}\right|_{\bar{\delta}=0} ,
  \label{eq:ib}
\ee
where the second approximate equality
applies when only the lowest-order contribution from the Taylor
expansion need be retained.
This is the linear response approximation.
The quantity
$\D\ln P(k)/\D\bar{\delta}$ is the \emph{linear response function}
and provides a good approximation to $\ib$ for large $k$.

\subsection{Line Correlation Function}
\label{ssec:lcf}

Equation~\eqref{eq:defP} shows that the power spectrum is
sensitive only to information carried by the amplitude of each
Fourier mode.
In contrast, higher-order statistics generally
encode information carried by both amplitudes and phases.
Phase correlations are
an exclusive signature of non-Gaussian density fields. For
instance, they may arise through processes in the primordial
Universe or from mode coupling in the non-linear regime of gravitational
collapse. Therefore, unlike
the amplitudes, phases directly probe cosmological information
that is absent from the two-point function.

With this motivation, \citet{Obreschkow:2012yb} proposed the
line correlation function (often abbreviated as `LCF').
It measures
a subset of
three-point phase correlations of the density field---%
specifically,
correlations between collinear points, each separated by a distance
$r$. \citet{Obreschkow:2012yb} demonstrated that the
LCF is a robust tracer of
filamentary structures, and showed that
it could be used as a phenomenological tool to distinguish between cold
and warm dark matter scenarios. Subsequent work established its
connection to conventional higher-order statistics
\citep{Wolstenhulme:2014cla,Eggemeier:2015ifa,Eggemeier:2016asq}.

The line correlation function
can be understood as follows: for a given density field
$\delta(\bx)$ in some
volume $V$, its real-space phase field $\epsilon_r(\bx)$ smoothed on a
scale $r$ satisfies
\begin{equation}
\label{eq:estimators.epsilon}
    \epsilon_r(\bx)
    =
    \int \frac{\D{^3 k}}{(2\pi)^3} \epsilon(\bk) \e{\im\bk\cdot\bx} W(k | r) 
    \equiv
    \int \frac{\D{^3 k}}{(2\pi)^3}\frac{\delta(\bk)}{|\delta(\bk)|}
    \e{\im\bk\cdot\bx} W(k | r)
    ,
\end{equation}
where $W(k | r)$ is the Fourier transform of the smoothing window
function. We take this to be a spherical top-hat in $k$-space,
$W(k | r) \equiv \Theta(1-k\,r/2\pi)$, where $\Theta(x)$ denotes the
Heaviside step function.
The phase at $\bk=0$ is defined so that $\epsilon(\bZero) \equiv 0$.
Following \citet{Obreschkow:2012yb} the
LCF is defined by
\be \LCF(r) \equiv
\frac{V^3}{(2\pi)^9}\left(\frac{r^3}{V}\right)^{3/2}
\int \frac{\D^2\hat{r}}{4\pi}
\big\langle
    \epsilon_{r}(\bx)
    \epsilon_{r}(\bx+\br)
    \epsilon_r(\bx-\br)
\big\rangle
\label{eq:l1} ,
\ee
where the factor
$V^3/(2\pi)^9$
represents a volume regularization.
After taking Fourier transforms
we require the three-point function of the
$\epsilon_r(\bk)$ in order to evaluate
this integral.
\citet{Wolstenhulme:2014cla} and
\citet{Eggemeier:2016asq} demonstrated
that, at lowest order in the expansion of the
probability density function for Fourier phases, this
three-point function is directly
related to the Fourier bispectrum. Therefore the
LCF must contain some fraction of
the information in $B$,
but because $\LCF(r)$
is an average over specific collinear configurations
it represents a compression.
Specifically, the
number of LCF bins will vary linearly with changes in the
effective cut-off on Fourier modes.

\subsection{Modal bispectrum}
\label{ssec:modal}

Our final proxy is a `modal' expansion of the
three-point function.
This is very similar to the Fourier bispectrum,
except that we exchange the Fourier basis $\e{\im\bx\cdot\bk}$
for a set of alternative modes that are better
adapted to the structure of $B$.
The exchange is helpful if
we can represent the bispectrum to the same accuracy
using fewer modes than required by the
Fourier representation.
This approach
was originally developed by
\citet{FergussonShellard2009} and \citet{Reganetal2010} to analyse
microwave background data, and subsequently applied
to large-scale structure by \citet*{Fergusson:2010ia} and
\citet{Reganetal2012}.

In the alternative basis we represent the Fourier bispectrum
in the form
\be B(k_1,k_2,k_3)\approx \Bmodal(k_1,k_2,k_3)
    \equiv
    \frac{1}{w(k_1,k_2,k_3)} \sum_{n=0}^{\nmax-1} \beta_n^{Q}
    Q_n(k_1,k_2,k_3)
    ,
    \label{eq:Bmodal}
\ee 
where the $Q_n$ are basis functions that span the space of configurations
compatible with a triangle condition
on $(k_1, k_2, k_3)$,
but can otherwise be chosen freely provided they are linearly independent.
The $\beta_n^Q$ are numbers that we describe as `modal coefficients'.
They can be regarded as averages of the Fourier bispectrum
over a set of configurations picked out by the corresponding $Q_n$.
The function $w(k_1,k_2,k_3)$ is an arbitrary weight that will be
chosen in Section~\ref{sec:predict-modal}.

If the $Q_n$ form a complete basis
we expect $B$ and $\Bmodal$
to become equivalent in the limit $\nmax \rightarrow \infty$.
In this limit the modal expansion is merely a reorganization
of the Fourier representation.
But if we select the lowest $Q_n$ to average over
the most relevant Fourier configurations
then it may be possible to represent a typical
$B$ using only a small number of modes.%
	\footnote{Here, `most relevant' is defined by the features
	of the bispectrum for which we wish to search.
	For example, inspection of the formulae appearing
	in Sections~\ref{sec:predict-ib}--\ref{sec:predict-lcf} below
	shows that both the integrated bispectrum and
	line correlation function can be regarded
	as instances of~\eqref{eq:Bmodal},
	with $Q_n$ adjusted to prioritize specific
	groups of Fourier configurations.
	For these cases, however, the resulting
	$Q$-basis is not complete.
	In this paper we distinguish the modal decomposition,
	for which the $Q$-basis is intended to be complete,
	from proxies such as $\ib$ and $\LCF$ which are
	intended to be projections.}
Taking $\nmax$ to be of order this number,
the outcome yields useful compression whenever
$\nmax \ll \Ntriangles$, where
$\Ntriangles$ is the number
of Fourier configurations
contained in the volume under discussion.
At least for reasonably smooth bispectra,
\citet*{Schmittfull:2012hq}
found that this could be done
with no more than modest
loss of signal.

\para{Orthonormal basis}
Given a choice of $Q_n$ we may redefine the basis
by taking arbitrary linear combinations.
For example, we will use this freedom
in Section~\ref{sec:predict-modal}
to obtain a basis for which the $\beta$-coefficients
are uncorrelated. The covariance matrix in this redefined
basis is especially simple.

Such a redefinition can be performed
using an invertible matrix $\lambda_{mn}$.
We define $R_n \equiv \sum_m \lambda_{nm}^{-1} Q_m$.
The $\beta$-coefficients in the $R$-basis
now satisfy
$\beta_n^R \equiv \sum_m \lambda_{mn} \beta_m^Q$.
Since the $Q$- and $R$-bases are reorganizations
of each other, the modal bispectrum defined using
either basis is equivalent,
\begin{equation}
	B(k_1, k_2, k_3)
	\approx
	\frac{1}{w(k_1, k_2, k_3)}
	\sum_{n=0}^{\nmax-1} \beta_n^Q Q_n(k_1, k_2, k_3)
	=
	\frac{1}{w(k_1, k_2, k_3)}
	\sum_{n=0}^{\nmax-1} \beta_n^R R_n(k_1, k_2, k_3) .
	\label{eq:Brecdef}
\end{equation}

\section{Predicting typical values and covariances for the proxies}
\label{sec:modelling}

In this section we explain how to obtain predictions for the typical values
and covariances of $\ib(k)$, $\LCF(r)$ and $\beta^R_m$
in a given cosmological model.
This can be done with different degrees of sophistication, corresponding---%
for example---%
to truncations at different levels in
the loop expansion
of standard perturbation theory~\citep{Bernardeau:2001qr},
or by using fitting functions calibrated to match 
the output of {\Nbody} simulations~\citep{Mead:2015yca}.
Since each proxy aggregates a different group of Fourier configurations,
and these configurations vary in their response to
features of the background cosmology,
the sophistication needed to adequately capture
the behaviour of the proxies may vary.

This is both a challenge and an opportunity.
Proxies that require delicate modelling to obtain accurate predictions
are harder to use, and may be expensive to deploy in a parameter-estimation
Monte Carlo.
In favourable cases, however,
the payoff will be sensitive discrimination between nearby cosmological
models.
On the other hand, proxies that can be modelled robustly using
simple methods are easy to use and cheap to deploy,
but may offer correspondingly coarse discrimination.
We study these trade-offs by
contrasting predictions
made using tree-level and one-loop SPT,
and the halo model.
For the halo-model power spectrum we choose the
{\HMcode} implementation~\citep{Mead:2015yca}.
For the halo-model bispectrum we use the standard formulae
given by~\citet{CooraySheth2002} with a Sheth--Tormen mass
function~\citep{Sheth:1999mn} and
Navarro--Frenk--White halo profile~\citep{Navarro:1995iw}.
In Section~\ref{sec:comparison}
we study the performance of each method
compared to numerical estimates
extracted directly from {\Nbody} simulations,
which enables us to characterize the minimum adequate sophistication
for each proxy.
For simplicity our analysis is framed in terms of the
underlying dark matter density field, although
in Section~\ref{sec:galaxy-bias}
we explain how this can be extended to predict
galaxy clustering.

\para{Covariance}
To compute a likelihood for a given proxy, either for the purposes of parameter estimation
or to make forecasts,
we require an estimate for the covariance between different configurations.
Therefore the minimum sophistication needed to adequately predict this
covariance matrix will play an additional role in
determining the relative expense of each proxy.
In practice the covariance
matrix is typically estimated
by taking measurements
from a large suite of {\Nbody} simulations or 2LPT catalogues,
or, if this is cannot be done, by falling
back to a Gaussian approximation.
{\Nbody} simulations give accurate results,
but are expensive enough that
assembling sufficient
independent realizations 
to determine the inverse covariance
is often not feasible.
In comparison, catalogues based on 2LPT
are significantly cheaper but become inaccurate in the non-linear regime,
while the Gaussian prediction breaks down even earlier
and may miss cross-correlations that significantly
affect the outcome.

The relative importance of these cross-correlations
varies between proxies.
In Sections~\ref{sec:covariance}--\ref{sec:paramEstim}
we estimate their significance by comparing
results from {\Nbody} and Gaussian covariances.
We describe our procedure to estimate covariance
matrices from the simulations in Section~\ref{sec:comparison},
but collect formulae for the Gaussian approximation
here.

For comparison, the Gaussian covariance for the
power spectrum and Fourier bispectrum,
measured on a grid of spacing $\Delta k$ with fundamental frequency $\kf=2\pi/V^{1/3}$,
can be written
\begin{equation}
    \label{eq:CovGauss-P}
    \CovGauss[P(k_i),P(k_j)] \approx \Kronecker_{ij} \frac{2\kf^3}{4\pi k_i^2 \Delta k} P^2(k_i) ,
\end{equation}
where $\Kronecker_{ij}$ is the Kronecker symbol,
and
\begin{equation}
    \label{eq:CovGauss-B}
    \CovGauss[B(\bk_1,\bk_2,\bk_3),B(\bq_1,\bq_2,\bq_3)] \approx
        \Kronecker_{\bk,\bq}
        \frac{\BispectrumDegeneracy \pi \kf^3}{k_1 k_2 k_3 (\Delta k)^3}
        P(k_1)P(k_2)P(k_3) .
\end{equation}
The Kronecker symbol $\Kronecker_{\bk,\bq}$ should be
interpreted to equal unity if the triangles
defined by $\{ \bk_1, \bk_2, \bk_3 \}$ and $\{ \bq_1, \bq_2, \bq_3 \}$ are equal,
and zero otherwise.
The degeneracy factor $\BispectrumDegeneracy$ equals unity for a scalene triangle,
two for an isosceles triangle and
six for an equilateral triangle.

\subsection{Integrated bispectrum}
\label{sec:predict-ib}

To evaluate the expression~\eqref{eq:ib}
we first establish its relation to the underlying 3-point function.
The overdensity within the subvolume labelled by $\br_L$
can be written
\begin{equation}
  \delta(\bk,\br_L)=\int \frac{\D^3 q}{(2\pi)^3} \delta(\bk - \bq) W_L(\bq) \e{-\im \bq \cdot \br_L }
    ,
\end{equation}
where $W_L(\bq)=V_s \prod_{i=1}^3 \sinc(q_i L_s/2)$ is the Fourier
transform of the cubic window function with side length $L_s$,
and $\sinc x \equiv (\sin x) / x$.
The power spectrum in this subvolume
is $P(\bk, \br_L) \equiv \langle |\delta(\bk, \br_L)|^2 \rangle / V_s$
and the mean overdensity is
$\bar{\delta}(\br_L) \equiv \delta(\bZero,\br_L)/V_s$. 
Combining these with equation~\eqref{eq:iB} yields~\citep{Chiang:2014oga}
\begin{equation}
\iBtheory(k)
  = \frac{1}{V_s^2}\int \frac{\D^2\hat{k}}{4\pi} \int \frac{\D^3 q_1}{(2\pi)^3} \int
  \frac{\D^3 q_2}{(2\pi)^3} \Btheory(\bk-\bq_1,-\bk+\bq_1+\bq_2,-\bq_2)
  W_L(\bq_1)W_L(-\bq_1-\bq_2)W_L(\bq_2)\,. \label{eq:iBtheory}
\end{equation}
Because $\sinc x$ is strongly peaked for $|x| \lesssim \pi$ the window functions
$W_L$ effectively constrain the $q_i$ integrals to
$q_i \lesssim 1/L_s$.
Since $k \gtrsim 1/L$ within each subvolume,
the integral receives
significant contributions only from
squeezed configurations of the Fourier bispectrum
that are of order the subvolume size or larger,
because in the limit $q_1, q_2 \ll k$
we have
$\Btheory(\bk-\bq_1,-\bk+\bq_1+\bq_2,-\bq_2) \approx \Btheory(\bk,-\bk,-\bq_2)$.

\citet{Chiang:2014oga} computed the linear response function
using~\eqref{eq:iBtheory}
and tree-level SPT,
and verified that it reproduces
equation~\eqref{eq:ib}
to within $2\%$ for $k \gtrsim 0.2\,h^{-1}\,\Mpc$.
For our purposes we require accurate estimates at
smaller $k$,
and therefore we perform
a numerical integration using~\eqref{eq:iBtheory} directly.
The integral is 8-dimensional and its evaluation is challenging;
we implement it
using the {\Vegas} algorithm
provided by the {\CUBA} package \citep{Hahn:2016ktb}.
To make the integration time feasible
we densely sample $\Btheory$ on a 3-dimensional cubic mesh
in coordinates
$(k_1, k_2, \mu_{12})$, where
$\mu_{12} \equiv ( k_1^2 + k_2^2 - k_3^2 ) / (2 k_1 k_2)$ 
is the cosine of the angle
between $\bk_1$ and $\bk_2$ and can be used in place
of the third wavenumber $k_3$.
We construct a 3-dimensional cubic spline that
interpolates between lattice points
and use this spline to evaluate the integrand.
To validate this procedure
we have verified that our numerical
results match the
analytic prediction from the linear response function
at large $k$.

Although we have not written subvolume labels explicitly,
$\sigma_{L}^2$ and all power spectra 
in~\eqref{eq:ib}
refer to subsampled quantities,
and therefore should be computed by appropriate convolution with the
subvolume window function $W_L(\bq)$.

\para{Halo model}
This procedure yields good results for tree-level and one-loop SPT,
but does not perform well when applied to the halo model.
In this case
we we do not recover equivalence between
our evaluation of~\eqref{eq:iBtheory}
and the linear response function,
which we compute by numerical differentiation of the
{\HMcode} power spectrum.
We interpret this disagreement as an indication that
the standard halo model
makes inconsistent predictions for
the modulation of the power spectrum with $\bar{\delta}$,
or the squeezed limit of the bispectrum, or both.
Moreover,
comparison of the halo-model $\ib$ computed
using~\eqref{eq:iBtheory}
to our {\Nbody} simulations shows poor agreement,
suggesting that estimates based on~\eqref{eq:iBtheory}
will be inaccurate.
Therefore, for the halo model only, we estimate $\ib$ by assuming
the linear response approximation~\eqref{eq:ib}
and computing $\D\ln P / \D\bar{\delta}$.
We calculate the derivative using the simulation-calibrated
formula proposed by~\citet{Chiang:2014oga},
\begin{equation}\label{eq:ibsqlim_halo}
  \frac{\D\ln \Phalo(k)}{\D\bar{\delta}} =
  \frac{13}{21}\frac{\D\ln \Phalo(k)}{\D\ln \sigma_8} + 2 -
  \frac{1}{3}\frac{\D\ln k^3 \Phalo(k)}{\D\ln k} \, ,
\end{equation}
which gives reasonable agreement with our simulations.

\para{Covariance}
In the absence of shot noise,
the Gaussian covariance
for estimates of
$\ib$ constructed from data
can be written
\begin{equation}
    \CovGauss\big[\ib(k_i),\ib(k_j) \big]
    =
    \frac{V_s}{V N_{ks}}
    \frac{1}{\sigma_L^2}
    \Kronecker_{ij} .
\label{eq:ib_covg}
\end{equation}
In this expression
$V_s$ is the volume of a subsampled region
and
$V$ denotes the total survey volume.
The quantity
$N_{ks}=2\pi k^2 \Delta k V_s$ is the number of Fourier modes in a
subvolume
$k$-bin.

\subsection{Line correlation function}\label{sec:predict-lcf}

\citet{Wolstenhulme:2014cla}
used tree-level SPT to predict the line correlation function.
Their result was generalized to an arbitrary bispectrum
by~\citet{Eggemeier:2016asq}, who gave the formula
\begin{equation}
\label{eq:models.lcf-perturbative}
    \LCFtheory(r)
    \simeq \Big(\frac{r}{4\pi}\Big)^{9/2}
    \iint\displaylimits_{\substack{|\bk_1|,|\bk_2|,\\|\bk_1+\bk_2|\leq 2\pi/r}}
    \D{^3 k_1}\,\D{^3 k_2} \;
    \Bphasetheory(k_1,k_2,k_3)
    j_0\big(\left|\bk_1-\bk_2\right| r\big)\,,
\end{equation}
where $j_0(x)=\sin(x)/x$ is the spherical Bessel function
of order zero
and the integrals over $\bk_1$ and
$\bk_2$ are cut off at the scale $k_i = 2\pi/r$.
The quantity $\Bphase$
is defined by
\be
    \Bphase(k_1,k_2,k_3)
    \equiv
    \frac{B(k_1,k_2,k_3)}{\sqrt{P(k_1)P(k_2)P(k_3)}}
    \label{eq:phaseB}
\ee
and gives the dominant contribution to the bispectrum of the phase field
$\epsilon(\bk) = \delta(\bk) / |\delta(\bk)|$ in the limit of large volume $V$.
For smaller volumes there are corrections
scaling as powers of $V^{-1/2}$ compared to the dominant term~\citep{Eggemeier:2016asq}.

\para{Evaluation}
To evaluate~\eqref{eq:models.lcf-perturbative} we must perform a 
6-dimensional integral.
We use a strategy similar to that described in
Section~\ref{sec:predict-ib},
by sampling the bispectrum over a cubic lattice and
interpolating between lattice sites.
The integration is again performed using~{\Vegas}.

In the special case of tree-level SPT,
\citet{Wolstenhulme:2014cla}
showed that~\eqref{eq:models.lcf-perturbative}
could be reduced to a 3-dimensional integral,
\begin{equation}
\begin{split}    
\label{eq:models.lcf-tree2}
\LCFtree(r) = \mbox{} & 16\pi^2\Big(\frac{r}{4\pi}\Big)^{9/2}
\int_0^{\frac{2\pi}{r}}\D{k_1}\,k_1^2
\int_0^{\frac{2\pi}{r}}\D{k_2}\,k_2^2
\int_{-1}^{\mu_{\mathrm{cut}}}\D{\mu}\,F_2^{(s)}(k_1,\,k_2,\,\mu)\,
\sqrt{\frac{\Ptree(k_1) \Ptree(k_2)}{\Ptree(|\bk_1+\bk_2|)}}\, \\
& \mbox{} \times \left[j_0{\big(\left|\bk_2-\bk_1\right|r\big)}+2j_0{\big(\left|\bk_1+2\bk_2\right|r\big)}\right]\,,
\end{split}
\end{equation}
where $\Ptree$ is the tree-level power spectrum, and the upper limit of the
$\mu$-integral is chosen to guarantee
$|\bk_1 + \bk_2| \leq 2 \pi / r$.
That requires
\begin{equation}
    \mu_{\text{cut}} =
    \min\left\{
        1,
        \max\left\{
            -1,
            \frac{(2\pi/r)^2-k_1^2-k_2^2}{2 k_1 k_2}
        \right\}
    \right\} .  
\end{equation}
Equation~\eqref{eq:models.lcf-tree2}
is useful because it provides a means to test the accuracy of our
6-dimensional {\Vegas} integrations, and the 3-dimensional
interpolations they entail.
We have compared estimates
for the tree-level line correlation function
using both~\eqref{eq:models.lcf-perturbative}
and~\eqref{eq:models.lcf-tree2} and find good agreement.

\para{Covariance}
To determine the Gaussian covariance we require
the two-point function of the phase field,
\begin{equation}
    \langle \epsilon(\bk_1) \epsilon(\bk_2)\rangle
    = \frac{(2\pi)^3}{V}\,\DiracD(\bk_1+\bk_2) .
\end{equation}
It follows that,
in the absence of shot noise, the covariance between
estimators for the
the line correlation function on scales $r_i$ and $r_j$ can be
written~\citep{Eggemeier:2016asq}
\begin{equation}
\label{eq:models.lcf-cov}
    \CovGauss\big[ \LCF(r_i),\LCF(r_j) \big] = \frac{(r_i r_j)^{9/2}}{V^3}
    \iint\displaylimits_{\substack{|\bk_1|,|\bk_2|,\\|\bk_1+\bk_2| \leq 2\pi/r}}
    \frac{\D^3 k_1}{\kf^3}\,\frac{\D^3 k_2}{\kf^3}
    \Big(
        j_0(|2\bk_1+\bk_2|\,r_i)
        \big[
            2j_0(|\bk_1-\bk_2|\,r_j)+j_0(|2\bk_1+\bk_2|\,r_j)
        \big]
        +
        r_i \leftrightarrow r_j
    \Big) ,
\end{equation}
where  $\kf = 2\pi/V^{1/3}$ denotes the fundamental frequency
(defined above equation~\eqref{eq:CovGauss-P}),
and $r = \max\{r_i,\,r_j\}$.
Note that~\eqref{eq:models.lcf-cov}
is not diagonal;
the integral that defines the line correlation
function depends on a range of Fourier modes for
any scale $r_i$,
and any Fourier modes that are common between
$\LCF(r_i)$ and $\LCF(r_j)$ will contribute
a nonzero covariance.
Moreover, equation~\eqref{eq:models.lcf-cov}
shows that the Gaussian covariance is independent
of redshift and all cosmological parameters.

\subsection{Modal bispectrum}\label{sec:predict-modal}
It was explained in Section~\ref{ssec:modal}
that the modal decomposition is defined by choice of
a basis $Q_n$ that samples groups of
relevant Fourier configurations.
The structure and ordering of the $Q_n$ determine those
configurations we wish to prioritize.
But unless we carefully
adjust the $Q_n$ they will
be correlated, and these correlations will be inherited
by the $\beta_n^Q$.
The outcome is that the covariance
matrix for estimators of the $\beta_n^Q$
is rather complex.

\para{Construction of $R$-basis}
To avoid this we redefine the basis,
as in equation~\eqref{eq:Brecdef},
to simplify the covariance matrix
for estimators of the
corresponding $\beta_n^R$.
The construction proceeds in stages.
First, consider the expected signal-to-noise
with which it is possible to measure
a single mode $Q_n/w$ from~\eqref{eq:Bmodal}.
Using a Gaussian approximation for the noise this can be
written
\begin{equation}
	6\left(\SignalToNoise	\right)^2_{Q_n}
	=
	\int
	\frac{\D^3 k_1}{(2\pi)^3}
	\frac{\D^3 k_2}{(2\pi)^3}
	\frac{\D^3 k_3}{(2\pi)^3}
	(2\pi)^3 
	\frac{\DiracD(\bk_1 + \bk_2 + \bk_3)}{w(k_1, k_2, k_3)^2}
	\frac{Q_n(k_1, k_2, k_3)^2}{P(k_1) P(k_2) P(k_3)} .
	\label{eq:Qn-signal-to-noise}
\end{equation}
We are free to choose the weight $w$ to simplify
this integral.
We define
\begin{equation}
	w(k_1, k_2, k_3) = \sqrt{\frac{k_1 k_2 k_3}{P(k_1) P(k_2) P(k_3)}} ,
	\label{eq:weight-function}
\end{equation}
after which the computation of the expected signal-to-noise reduces to
\begin{equation}
	6\left(\SignalToNoise	\right)^2_{Q_n}
    =
    \llangle Q_n | Q_n \rrangle .
\end{equation}
To write this and similar expressions economically we have introduced the
notation
\begin{equation}
    \llangle f | g \rrangle
    \equiv
    \int
    \frac{\D{^3 k_1}}{(2\pi)^3}
    \frac{\D{^3 k_2}}{(2\pi)^3}
    \frac{\D{^3 k_3}}{(2\pi)^3}
    (2\pi)^3 \DiracD(\bk_1 + \bk_2 + \bk_3)
    \frac{f(\bk_1, \bk_2, \bk_3) g(\bk_1, \bk_2, \bk_3)}{k_1 k_2 k_3}
    \label{eq:inner-product}
\end{equation}
for any $f$ and $g$.
In the special case that these depend only on the wavenumbers $k_i$
and not their orientations $\hat{\bk}_i$
some of the angular integrations are trivial
and we obtain the simpler expression
\begin{equation}
	\llangle f | g \rrangle
	\equiv
    \frac{1}{8\pi^4} \int_{\TriangleRegion} \D{k_1} \, \D{k_2} \, \D{k_3} \;
    f(k_1, k_2, k_3) g(k_1, k_2, k_3) .
    \label{eq:inner-product-simple}
\end{equation}
Here, $\TriangleRegion$ represents the set
of points $(k_1, k_2, k_3)$ where lines
of length $k_1$, $k_2$ and $k_3$ can be arranged to form a triangle,
ie. $2 \max \{ k_i \} \leq \sum_i k_i$;
for details, see~\citet{Fergusson:2009nv}.
In principle the integral can be carried over all $k_i$,
but
in practice it will be cut off at upper and lower limits
$\kmax$ and $\kmin$.
The expressions~\eqref{eq:inner-product}
and~\eqref{eq:inner-product-simple}
can be regarded as an inner product on the $Q_n$
that weights each contributing Fourier configuration
according to its individual signal-to-noise.

Second,
the $R$-basis is chosen to be diagonal with respect
to this inner product.
As we will see below, because the resulting $R_n$ modes
are orthogonal when weighted by signal-to-noise,
the covariance matrix for estimators
of the coefficients $\beta_n^R$ becomes diagonal under the same
approximation of Gaussian noise used to determine the weighting
in~\eqref{eq:Qn-signal-to-noise}.
Specifically, we define
\begin{equation}
	\llangle Q_m | Q_n \rrangle
	\equiv
	\gamma_{mn}
	\equiv
	\frac{(\kmax - \kmin)^3}{8\pi^4} \bar{\gamma}_{mn} .
	\label{eq:gamma-matrix-def}
\end{equation}
It is sometimes preferable to express results in terms
of $\bar{\gamma}_{mn}$, which is independent of $\kmin$
and $\kmax$.
For any suitable $Q$-basis
both
$\gamma_{mn}$ and $\bar{\gamma}_{mn}$
will be symmetric and positive-definite
and may be factored into the product
of a matrix and its transpose.
Therefore there exists a matrix $\lambda_{mn}$
such that
$\bar{\gamma}_{mn} = \sum_r \lambda_{mr} \lambda_{nr}$.
Application of~\eqref{eq:Brecdef} with $\lambda_{mn}$ as the
transformation matrix yields
$R_n = \sum_{n'} \lambda_{nn'}^{-1} Q_{n'}$,
and these modes are orthogonal in the sense
\begin{equation}
    \llangle R_m | R_n \rrangle
    =
    \frac{(\kmax - \kmin)^3}{8\pi^4} \Kronecker_{mn} .
    \label{eq:R-basis-inner-product}
\end{equation}

\para{Determination of modal coefficients}
Whether we work with the $Q$- or $R$-basis,
we must predict the corresponding $\beta$-coefficients
for each model of interest.
In practice the extra matrix operations needed to obtain
the $R$-basis mean that it is simplest to perform calculations
in the $Q$-basis, before translating to the $R$-basis to
interpret the results. We adopt this procedure whenever
concrete calculations using the modal decomposition are required.
We use the $Q$-basis constructed by~\citet{Fergusson:2009nv}.
(The details are summarized in Appendix~\ref{app:poly}.)
It is not intended to prioritize any single class
of Fourier configurations,
but rather attempts to provide a good description of reasonably
smooth bispectra over a range of shapes and scales.

To extract the $\beta_n^Q$ we use~\eqref{eq:inner-product}.
Assuming~\eqref{eq:Bmodal}
can be interpreted as an equality, we conclude
that for an arbitrary bispectrum $\Btheory(k_1, k_2, k_3)$
\begin{equation}
    \llangle w \Btheory | Q_m \rrangle
	=
	\sum_{n=0}^{\nmax-1} \beta_n^{Q,\theory} \gamma_{nm} .
	\label{eq:Qbasis-project}
\end{equation}
Finally, the individual $\beta_n^Q$ should be extracted by contraction with
the inverse matrix $\gamma_{mn}^{-1}$.
If the bispectrum has no angular dependence then the inner product
can be computed using the simplified expression~\eqref{eq:inner-product-simple},
which yields
\begin{equation}
	\beta_n^{Q,\theory}
	=
	\frac{1}{8\pi^4}
	\sum_m \gamma^{-1}_{nm}
	\int_{\TriangleRegion} \D{k_1} \, \D{k_2} \, \D{k_3} \;
	\sqrt{k_1 k_2 k_3}
	\Bphasetheory(k_1, k_2, k_3) Q_m(k_1, k_2, k_3) ,
    \label{eq:modalBT}
\end{equation}
where we have used the quantity
$\Bphase$ defined in~\eqref{eq:phaseB}.
The $\beta_n^{R,\theory}$ may be obtained
by the transformation $\beta_n^R = \sum_m \lambda_{mn} \beta^Q_m$.
The appearance of the phase bispectrum in~\eqref{eq:modalBT}
is a consequence of our choice of weight $w$.

Equation~\eqref{eq:Qbasis-project} would continue to apply
were we to change the definition of the `inner product'
$\llangle \cdot | \cdot \rrangle$,
and an analogue of~\eqref{eq:modalBT}
would continue to give the individual $\beta_n^{Q,\theory}$.
Our choice of signal-to-noise weighting
in $\llangle \cdot | \cdot \rrangle$
is
important only
for construction of the $R$-modes
and the covariance inherited by the $\beta_n^{R,\theory}$.

\para{Numerical evaluation}
In practice, equation~\eqref{eq:modalBT}
requires evaluation of a 3-dimensional integral over the region
$\TriangleRegion$.
To implement it we compute $wB$ on a $200^3$ cubic lattice
in $(k_1, k_2, k_3)$ and estimate the integral by volume-weighted
cubature over this lattice. Some work is
required to account for irregular boundary orientations;
we give these details in Appendix~\ref{app:voxel}.

\para{Covariance}
Finally we compute the covariance of
estimators for the $\beta_n^R$ coefficients
under the assumption of Gaussian covariance for the
bispectrum estimator
$\delta(\bk_1) \delta(\bk_2) \delta(\bk_3)
V^{-1} \Kronecker_{\bk_1 + \bk_2 + \bk_3, \bZero}$.
Using equation~\eqref{eq:inner-product},
and~\eqref{eq:Qbasis-project} with $R$ exchanged for 
$Q$,
we obtain
\begin{equation}
\label{eq:modalcov-pre}
\begin{split}
    \langle \beta_m^R \beta_n^R \rangle
    & =
    (2\pi)^3 \delta(\bZero)
    \frac{6}{V^2}
    \frac{(8\pi^4)^2}{(\kmax - \kmin)^6}
    \int \frac{\D{^3 k_1} \, \D{^3 k_2} \, \D{^3 k_3}}{(2\pi)^9}
    (2\pi)^3 \delta(\bk_1 + \bk_2 + \bk_3)
    \frac{R_m(k_1, k_2, k_3) R_n(k_1, k_2, k_3)}{k_1 k_2 k_3} ,
    \\
    & =
    \frac{6}{V}
    \frac{(8\pi^4)^2}{(\kmax - \kmin)^6}
    \llangle R_m | R_n \rrangle .
\end{split}
\end{equation}
The weighting for each Fourier configuration matches the signal-to-noise,
making
this correlator diagonal as a consequence of our construction of the $R$-basis.
Therefore we conclude
\begin{equation}
    \label{eq:modalcov}
    \CovGauss(\beta_m^R, \beta_n^R) =
    \frac{6}{V}\frac{8 \pi^4}{(\kmax-\kmin)^3}\Kronecker_{mn} .
\end{equation}
As for the line correlation function, it is independent of redshift
and cosmological parameters.
If we were to abandon the approximation of Gaussian covariance
then~\eqref{eq:modalcov-pre}
would no longer be proportional to exactly
$\llangle R_m | R_n \rrangle$.
In this case the amplitude of the diagonal
elements would be modified, and non-diagonal
components would appear.

\subsection{Galaxy bias}\label{sec:galaxy-bias}

The discussion in
Sections~\ref{sec:predict-ib}--\ref{sec:predict-modal}
was framed in terms of the dark matter overdensity $\delta$,
but this is not what is measured by
surveys of large-scale structure.
Instead, they record the abundance of
galaxies or some other population of tracers
whose density
responds to the dark matter density but need not match it.

On large scales the relation between the galaxy ($\delta_g$)
and dark matter ($\delta$)
density fields is well-described by the linear model
$\delta_g = b_1 \delta$~\citep{Kaiser1984,Fry:1992vr}.
The \emph{linear bias parameter} $b_1$
may be redshift-dependent,
and varies between different populations of galaxies.
On small scales the overdensities are larger,
and both non-linear and non-local corrections become
important.
To obtain a satisfactory description we must
typically include terms at least
quadratic (or higher) in $\delta$~\citep{Fry:1992vr,Smith:2006ne},
together with terms involving the tidal
gravitational field~\citep{Catelan:2000vn,McDonald:2009dh,Chan:2012jj,Baldauf:2012hs}.

In what follows we assume the local Lagrangian bias model,
in which the galaxy overdensity at early times is taken to be
a local function of the dark matter overdensity.
At later times the bias is determined by propagating
this relationship along
the dark matter flow.
\citet{McDonald:2009dh}
demonstrated that this implies the Eulerian
galaxy overdensity
at the time of observation can be written
\begin{equation}
	\label{eq:models.bias}
	\delta_g(\bx)
	=
	b_1 \delta(\bx)
	+ \frac{1}{2}b_2 \big[
		\delta^2(\bx)
		- \langle\delta^2(\bx)\rangle
	\big]
	+ \frac{1}{2}b_{s^2} \big[
		s^2(\bx)
		-\langle s^2(\bx)\rangle
	\big]
	+ \cdots ,
\end{equation}
where `$\cdots$' denotes terms of third order and higher that
we have not written explicitly. The field $s^2(\bx) =
s^{ij}(\bx)\,s_{ji}(\bx)$ is a contraction of the tidal tensor,
defined by $s_{ij}(\bx) \equiv
\left[\partial_i\partial_j\nabla^{-2}-\frac{1}{3}\Kronecker_{ij}\right] \delta(\bx)$.
Therefore, up to second order in $\delta$, we require two additional
redshift- and population-dependent bias
parameters: the \emph{quadratic bias} $b_2$, as well as the
\emph{non-local bias}
$b_{s^{2}}$.
In the local Lagrangian model
the non-local bias satisfies
$b_{s^2} = - 4 (b_1 - 1)/7$~\citep{Chan:2012jj,Baldauf:2012hs},
although in more general biasing prescriptions
it could be allowed to vary independently.

\para{Power spectrum}
After translating to Fourier space
it follows that the tree-level galaxy power spectrum
can be written
\begin{equation}
	\Pgaltree(k)
	=
	 b_1^2 \Ptree(k)
	 .
	 \label{eq:Ptree}
\end{equation}
To obtain a consistent result at one-loop we should include
the unwritten
third-order contributions in~\eqref{eq:models.bias},
which generate multiplicative renormalizations
of the linear power spectrum in the same way as the
`13' terms of one-loop SPT.
\citet{McDonald:2009dh} showed that these
could be collected into a single new
parameter
which we denote $b_{3\text{nl}}$ to match~\citet{Gil-Marin:2014sta}.
Therefore
\begin{equation}
\begin{split}
	\Pgalloop(k)
	= \mbox{} &
	b_1^2 \Ploop(k)
	+ 2 b_1 b_2 P_{b2}(k) 
	+ 2 b_1 b_{s^2} P_{bs2}(k)
	+ b_2^2 P_{b22}(k)
	+ 2 b_2b_{s^2} P_{b2,bs2}(k)
	+ b_{s^2}^2 P_{bs22}(k)
	\\
	& \mbox{}
	+ 2 b_1 b_{3\text{nl}} \sigma_3^2(k) \Ptree(k)
	.
	\label{eq:models.Pg-1loop}
\end{split}
\end{equation}
\citet{Saito:2014qha}
showed that
in the local Lagrangian model
$b_{3\text{nl}}$ satisfies
$b_{3\text{nl}} =
32 (b_1-1) / 315$.
Explicit expressions for all terms
appearing in~\eqref{eq:models.Pg-1loop}
were given by
\citet{McDonald:2009dh}.
Note that contributions from the
non-linear bias appear only in the one-loop
power spectrum.

\para{Bispectrum}
In contrast to the power spectrum, the bispectrum receives corrections
from non-linear bias terms even at tree-level.
Specifically,
\begin{equation}
	\label{eq:models.Bg-tree}
	\Bgaltree(\bk_1,\,\bk_2,\,\bk_3) 
	=
	b_1^3 \Btree(\bk_1, \bk_2, \bk_3)
	+ b_1^2 \Ptree(k_1) \Ptree(k_2) \left[
		b_2
		+ b_{s^2} S_2(\bk_1,\bk_2)
	\right]
	+ \text{cyclic}
	,
\end{equation}
where $S_2(\bk_1,\bk_2) \equiv (\bk_1 \cdot \bk_2)^2 / (k_1 k_2)^2 - 1/3$
is the kernel appearing in the Fourier transform of the
contracted tidal field,
$s^2(\bk) = (2\pi)^{-3} \int \D{^3 q} \, S_2(\bq, \bk - \bq) \delta(\bq) \delta(\bk - \bq)$.

To obtain the galaxy bispectrum consistently at one
loop one should compute the dark matter overdensity
to fourth order in perturbation theory
and develop the bias expansion to the same order.
This procedure has been adumbrated in the
literature~\citep{Assassi:2014fva}
but not developed completely.
Therefore to obtain an estimate of the one-loop
bispectrum we make the approximation
\begin{equation}
	\label{eq:models.Bg-1loop}
	\Bgalloop(\bk_1,\,\bk_2,\,\bk_3)
	=
	b_1^3 \Bloop(\bk_1,\,\bk_2,\,\bk_3)
	+ b_1^2 \Ploop(k_1) \Ploop(k_2) \left[
		b_2
		+ b_{s^2} S_2(\bk_1,\bk_2)
	\right]
	+
	\text{cyclic}
	.
\end{equation}
This is consistent with the prescriptions
used by~\citet{Gil-Marin:2014sta}
and~\cite{Baldauf:2016sjb}.

\para{Application to bispectrum proxies}
The outcome of this discussion is that,
to predict the integrated
bispectrum, line correlation function, or modal bispectrum
for the galaxy density field,
we should make the
replacements $\Ptheory(k)\rightarrow \Pgaltheory(k)$
and $\Btheory(k_1,k_2,k_3) \rightarrow \Bgaltheory(k_1,k_2,k_3)$
where necessary
in equations~\eqref{eq:iBtheory},
\eqref{eq:models.lcf-perturbative} and~\eqref{eq:modalBT}.

To obtain theory predictions at tree-level
we use
equations~\eqref{eq:Ptree} and~\eqref{eq:models.Bg-tree},
whereas to obtain perdictions at one-loop
we use equations~\eqref{eq:models.Pg-1loop} and~\eqref{eq:models.Bg-1loop}.
Finally, to evaluate predictions using the halo model
we apply
equations~\eqref{eq:models.Pg-1loop} and~\eqref{eq:models.Bg-tree},
but with $\Ploop\rightarrow \Phalo$ and
$\Btree\rightarrow \Bhalo$ for the dark matter correlations.

\section{Estimating bispectrum proxies from {\NbodyUpper} simulations}
\label{sec:estimation}

In this section we briefly describe our {\Nbody} simulations and
explain how they are used to estimate
the Fourier bispectrum
and its proxies
$\ib$, $\LCF$ and $\beta_n^Q$.

\subsection{Simulations}
\label{ssec:simulations}

Our measurements are based on two sets of simulations: (1) $200$
{\Nbody} simulations containing
dark matter only, with a fixed choice of fiducial
cosmological parameters; (2) a total of
$60$ simulations constructed by varying one cosmological
parameter at a time, with four realizations per model including
the fiducial set. These
simulations were performed on the {\ZBOX} supercomputer at the
University of Zurich and were described in \citet{Smith:2008ut} and
\citet{Smith:2012uz}. Each set uses a comoving boxsize
of $L = 1500\,h^{-1}\,\Mpc$ and contains $N = 750^3$
particles. Initial conditions for the particles were set at
redshift $z=49$
using second-order Lagrangian perturbation theory
acting on a realization of a Gaussian random field~\citep{Crocce:2006ve}
with transfer functions from {\CMBFAST}~\citep{Seljak:1996is}.
The particles are evolved to $z=0$ under
the influence of gravity using the {\Gadget} code
\citep{Springel:2005mi}, modified to allow a
time-evolving equation of state for dark energy.

\begin{table}
\centering
\caption{Fiducial values of the cosmological parameters, together with
the stepsize $\Delta\theta$ used to vary each parameter in the simulations.
We perform one simulation with offset $+\Delta\theta$ and one with increment
$-\Delta\theta$,
giving two offset simulations per parameter. With seven parameters and four
realizations per model this gives $4 + 2 \times 7 \times 4 = 60$ simulations
in the suite.
The bias parameters are assumed to be $b_1 = 1$ and $b_2 = 0$.}
  \begin{tabular}{cccccccc}
    \toprule
    Parameter $\theta$ & $\Omega_m$ & $\Omega_b$ & $w_0$ & $w_a$ & $\sigma_8$ & $n_s$ & $h$
    \\ \midrule
    Fiducial value & $0.25$ & $0.040$ & $-1.0$ & $0.0$ & $0.8$ & $1.00$ & $0.70$ \\
    $\Delta\theta$ & $\pm\,0.05$ & $\pm\,0.005$ & $\pm\,0.2$ & $\pm\,0.1$ & $\pm\,0.1$ & $\pm\,0.05$ & $\pm\,0.05$ \\
    \bottomrule
  \end{tabular}
  \label{tab:parameters}
\end{table}

The fiducial cosmological parameters correspond to a
flat $\Lambda$CDM model and are summarized in Table~\ref{tab:parameters}.
Specifically,
$\Omega_m$ and $\Omega_b$ are the matter and baryon density parameters;
$w_0$ and $w_a$ parametrize the equation of
state for dark energy, viz. $w(a) \equiv w_0 + (1-a)\,w_a$; $\sigma_8$ is
the amplitude of density fluctuations smoothed on a scale
$8\,h^{-1}\,\Mpc$; $n_s$ is the spectral index of the
primordial power spectrum; and $h$ is the dimensionless Hubble
parameter.
We collectively write these as
a vector $\theta_\alpha$
with index $\alpha$ labelling the
different parameters.
To construct set (2)
each parameter is offset by
$+\Delta\theta_\alpha$ and $-\Delta\theta_\alpha$,
with all other parameters held fixed.
The stepsizes $\Delta \theta_\alpha$
are listed in Table~\ref{tab:parameters}.
To reduce noise when estimating parameter derivatives,
we construct initial conditions for each of
the four realizations using the same
Gaussian random field as its fiducial partner.
Since we vary over seven cosmological parameters
this gives a total of $4 + 2 \times 7 \times 4 = 60$ simulations
in the suite.

\subsection{Density field}

To compute the overdensity field in each simulation
we
use the cloud-in-cell assignment scheme to distribute
particles over a regular Cartesian grid.
We apply a fast Fourier transform
and
extract the discrete
real-space density field
by deconvolving the cloud-in-cell
window function.
The result is
\begin{equation}
	\delta^{\text{disc}}(\bk) = \frac{\delta^\text{grid}(\bk)}{\WCIC(\bk)}
	,
	\quad \text{where} \quad
	\WCIC(\bk)
	=
	\prod_{i=1}^3 \left[\frac{\sin{\left(\pi k_i/ 2\kNy\right)}}{\pi k_i/ 2\kNy}\right]^2
	.
\end{equation}
The labels `disc' and `grid' label Fourier-space fields
in the full volume $V$ and on the cloud-in-cell grid, respectively. 
The Nyquist frequency $\kNy = \pi \Ngrid / L$
is determined by the number of grid cells per dimension.
For our numerical results we use $\Ngrid = 512$. 

\subsection{Estimating the power spectrum}\label{sec:stdPEst}

Given a realization of the $\delta$-field within
a simulation volume $V = L^3 = (2\pi)^3 \DiracD(\bZero)$,
a simple estimator for the power at wavevector $\bk_1$
can be written
$\hat{\Pestimator}(\bk_1,\bk_2) = \delta(\bk_1)\delta(\bk_2)\Kronecker_{\bk_1,-\bk_2}/V$.%
	\footnote{In the remainder of this paper
	we assume it is understood that we are dealing with the
	discrete density field whenever we refer to measured quantities,
	and drop the label `disc'.}
Unfortunately this procedure is very noisy.
An improved estimate can be obtained
by summing over a set of modes
satisfying the closure criterion $\sum_{i}\bk_i=\bZero$
within a thin $\bk$-shell.
Since we are working in finite volume
the available modes are discretized in units of
the fundamental frequency
$\kf = 2\pi/L$,
and therefore the thin-shell average should be written
\begin{equation}\label{eq:estimPSE}
  \hat{P}(k)
  = \frac{1}{V_P(k)}
  \int \D{^3 q_1} \, \D{^3 q_2} \; \DiracD(\bq_1 + \bq_2)
  \hat{\Pestimator}(\bq_1,\bq_2)
  \tilde\Pi_k(\bq_1)\tilde\Pi_k(\bq_2)
  ,
\end{equation}
where $\Delta k \geq \kf$ represents a bin width,
and
we have introduced the binning function $\tilde\Pi_k(\bq)$
which is defined to be unity
if $|\bq|\in [k-\Delta k/2, k+\Delta k/2]$ and zero otherwise.
Finally, the quantity $V_P$ represents the volume of the spherical
shell accounting for discretization,
\begin{equation}
	V_P(k)
	\equiv \int \D{^3 q_1} \, \D{^3 q_2} \; \DiracD(\bq_1 + \bq_2)
	\tilde\Pi_{k}(\bq_1)\tilde\Pi_k(\bq_2)
	=
	\int \D{^3 q} \; \tilde\Pi^2_k(\bq)
	=
	\int \D{^3 q} \; \tilde\Pi_k(\bq)
	=
	4 \pi k^2 \Delta k
	\bigg[
		1
		+ \frac{1}{12}
		\Big(
			\frac{\Delta k}{k}
		\Big)^2
	\bigg]
	.
\end{equation}

\subsection{Estimating the bispectrum}\label{sec:stdEst}

In analogy with the power spectrum, an estimator for
a single configuration of the
Fourier bispectrum
can be written
$\hat{\Bestimator}(\bk_1,\bk_2,\bk_3) =  \delta(\bk_1)\delta(\bk_2)\delta(\bk_3)\Kronecker_{\bk_1+\bk_2+\bk_3,\bZero}/V$.
[This expression was already used in Section~\ref{sec:predict-modal}
to obtain the Gaussian covariance for estimators of the $\beta_n^R$.]
To obtain an acceptable signal-to-noise we should again
average over a set of configurations
whose wavenumbers lie within suitable
discretized $\bk$-shells.
After doing so we obtain
the estimator
\begin{equation}
  \hat{B}(k_1,k_2,k_3)
  =
  \frac{1}{V_B(k_1,k_2,k_3)}
  \int \D{^3 q_1} \, \D{^3 q_2} \, \D{^3 q_3} \;
  \DiracD(\bq_1 + \bq_2 + \bq_3)
  \hat{\Bestimator}(\bq_1,\bq_2,\bq_3)
  \tilde\Pi_{k_1}(\bq_1)\tilde\Pi_{k_2}(\bq_2)
  \tilde\Pi_{k_3}(\bq_3)
  ,
  \label{eq:BSE}
\end{equation}
where the normalization $V_B$ should now be evaluated
using~\citep{Sefusatti:2006pa,Joachimietal2009}
\begin{equation}
	V_B(k_1,k_2,k_3)
	\equiv
	\int \D{^3 q_1} \, \D{^3 q_2} \, \D{^3 q_3} \;
	\DiracD(\bq_1 + \bq_2+ \bq_3)
	\tilde\Pi_{k_1}(\bq_1)\tilde\Pi_{k_2}(\bq_2)\tilde\Pi_{k_3}(\bq_3)
	\approx
	8 \pi^2 k_1 k_2 k_3 (\Delta k)^3
	.
\end{equation}
Dividing by the square of the fundamental cell volume
shows that the number of configurations scales as
$\Ntriangles(k_1,k_2,k_3)=V_B(k_1,k_2,k_3)/\kf^6\propto
N_1N_2N_3$, where $N_i\equiv k_i/\kf$ is the length of the side $k_i$
in units of the fundamental mode.
Hence, if we scale the configuration by
$k_i \rightarrow \lambda k_i$
then the number of available configurations
scales as $\lambda^3$.

\citet{Sefusatti2005},
\citet*{Fergusson:2010ia}
and~\citet{Scoccimarro:2015bla}
observed that~\eqref{eq:BSE} could be implemented
efficiently by rewriting the Dirac $\delta$-function
using its Fourier representation,
$(2\pi)^3\DiracD(\bq)=\int \D{^3 x} \, \e{\im \bq\cdot\bx}$, and
factorizing the dependence on the $\bq_i$.
This yields
\begin{equation}
  \hat{B}(k_1,k_2,k_3)
  =
  \frac{\kf^3}{(2\pi)^6 V_B(k_1,k_2,k_3)}
  \int \D{^3 x} \; \Dfactor{k_1}(\bx) \Dfactor{k_2}(\bx) \Dfactor{k_3}(\bx) ,
  \quad \text{where} \quad
  \Dfactor{k}(\bx)
  \equiv
  \int \D{^3 q} \; \e{\im \bx\cdot\bq} \delta(\bq)\tilde\Pi_{k}(\bq)
  .
  \label{eq:BSE-factorized}
\end{equation}
Similarly,
\begin{equation}
  V_B(k_1,k_2,k_3) =  \int \frac{\D{^3 x}}{(2\pi)^3} \Pi_{k_1}(\bx) \Pi_{k_2}(\bx) \Pi_{k_3}(\bx)
  ,
\end{equation}
where $\Pi_k(\bx)$ is the inverse Fourier transform of $\tilde\Pi_k(\bq)$.

Equation~\eqref{eq:BSE-factorized}
is numerically more efficient
than a direct implementation of~\eqref{eq:BSE}, because
it requires
only three Fourier transforms
to compute $\Dfactor{k}$
for each wavenumber in the
triplet $\{ k_1, k_2, k_3 \}$.
Moreover, once each
$\Dfactor{k}$
has been obtained it can be re-used
for any configuration that shares the same wavenumber.
In spite of this improvement, however,
it remains a formidable computational challenge
to estimate all bispectrum configurations contained
within a large volume $V$.
Different strategies have been employed
to make the calculation feasible.
One option is to coarsely bin configurations
with binning width equal to several times
the fundamental mode.
This drastically reduces the number of
configurations to be measured.
An alternative is to search only among
a limited subset of configurations.
This may be helpful if we wish to search for
specific physical effects, but risks
overlooking important signals if we
are searching blindly.
In either case the analysis is unlikely to be optimal
because information is lost.

\subsection{Estimating the integrated bispectrum}

Our procedure to estimate the integrated bispectrum
is based directly on its definition.
We separate
the total volume into $N_s$ subvolumes,
enumerated by the labels $i = 1, \ldots, N_s$.
We compute the mean overdensity
$\hat{\bar{\delta_i}}$ and power spectrum
$\hat{P}(k)_i$
within each subvolume.
Finally, we average the product
$\hat{P}(k)_i \hat{\bar{\delta}}_i$
over all subvolumes.
Therefore,
\begin{equation}
	\widehat{\iB}(k) = \frac{1}{N_s} \sum_{i=1}^{N_s}\hat{P}(k)_i \hat{\bar{\delta}}_i .
\end{equation}
The normalized integrated bispectrum can be obtained by rescaling,
\begin{equation}
	\widehat{ib}(k) = \frac{\widehat{\iB}(k)}{\hat{P}(k)\hat{\sigma}_L^2} ,
\end{equation}
where here $\hat{P}(k) = \sum_{i=1}^{N_s} \hat{P}(k)_i /N_s$ is the average
subvolume power spectrum and 
$\hat{\sigma}_L^2 = \sum_{i=1}^{N_s} \hat{\bar{\delta}}_i^2 /N_s$ is the average
variance of the mean overdensity.

\subsection{Estimating the line correlation function}

A procedure to estimate the line correlation function was
outlined by~\citet{Eggemeier:2016asq}.
We evaluate
\begin{equation}
	\hat{\ell}(r)
	=
	\Big(
		\frac{r^3}{V}
	\Big)^{3/2}
	\hspace{-1.5em}
	\sum_{\substack{|\bk_1|,|\bk_2|,\\|\bk_1+\bk_2| \leq 2\pi/r}}
	\hspace{-1.5em}
	\overline{j_0}(|\bk_1-\bk_2|r)\,
	\epsilon(\bk_1)\,\epsilon(\bk_2)\,\epsilon(-\bk_1-\bk_2)
	,
	\label{eq:est1}
\end{equation}
where $\overline{j_0}(|\bk|r)$ denotes an average of $j_0(kr)$ taken over the volume
of a fundamental $k$-space cell centred at $\bk$. The sum scales as
$\sim (2L/r)^6$, making its evaluation fast on large scales but challenging
on small ones, where the sum includes the majority of Fourier modes.
On scales below $\sim 105\,h^{-1}\,\Mpc$ we find that the real space estimator
described by~\citet{Eggemeier:2016asq} becomes more efficient and
therefore we use it within that regime. For scales accessible to both
schemes we verified that both estimators yield the same result.

\subsection{Estimating the modal bispectrum}\label{sec:modalEst}

Equation~\eqref{eq:modalBT}
shows that
an estimate of the modal coefficient $\beta^{Q}_m$
requires evaluation of
$\llangle w \hat{\Bestimator}|Q_n\rrangle$, where
$\hat{\Bestimator}$ is the bispectrum estimator defined in Section~\ref{sec:stdEst}.
Using equation~\eqref{eq:inner-product},
writing the $\delta$-function
using its Fourier representation,
and factorizing the integral as described in Section~\ref{sec:stdEst},
we find
\begin{equation}
	\label{eq:wBhat_Qn}
	\llangle w \hat{\Bestimator} | Q_n \rrangle
	=
	\frac{1}{V}
	\int \D{^3 x} \;
	\Mfactor{n_1}(\bx) \Mfactor{n_2}(\bx) \Mfactor{n_3}(\bx)
	,
	\quad \text{where} \quad
	\Mfactor{n}(\bx)
	\equiv
	\int \frac{\D{^3 k}}{(2\pi)^3}
	\e{\im\bk\cdot\bx}
	\frac{q_{n}(k)}{\sqrt{k \hat{P}(k)}}\delta(\bk)
	.
\end{equation}
Here, $q_n(k)$ is a polynomial used in the construction of
the modes
$Q_n$; see Appendix~\ref{app:poly}.
Equation~\eqref{eq:wBhat_Qn}
shows that the computation can be reduced to a single 3-dimensional
integral over the $\Mfactor{n}(\bx)$, which
are themselves weighted Fourier transforms of $\delta$.
Finally,
$\beta_m^Q$
can be estimated by contracting with the
inverse inner product matrix
$\gamma_{mn}^{-1}$
defined in~\eqref{eq:gamma-matrix-def},
\begin{equation}
	\hat{\beta}_m^{Q} 
	= \sum_{n=0}^{\nmax-1}
	\llangle w \hat{\Bestimator} | Q_n\rrangle\gamma^{-1}_{nm}
	.
	\label{eq:betaEst}
\end{equation}
To obtain the corresponding $R$-basis coefficients
requires a further linear transformation
\begin{equation}
	\hat{\beta}_n^R=\sum_{m}\lambda_{m n}\hat{\beta}_m^{Q} , 
\end{equation}
where $\lambda_{mn}$ is the matrix defined
above~\eqref{eq:R-basis-inner-product}.
As explained in Section~\ref{sec:predict-modal},
we generally perform numerical calculations in the $Q$-basis
in order to preserve the simplicity of~\eqref{eq:wBhat_Qn},
but present results in the $R$-basis
because their covariance properties make these
coefficients simpler to interpret.
In either basis, the measured coefficients can be used
to reconstruct the bispectrum for any
required Fourier configuration
using equation~\eqref{eq:Brecdef}.

Note that, because the matrix $\gamma_{nm}$ can be tabulated,
measuring a single modal coefficient has the same computational
complexity as measuring a single configuration of the
Fourier bispectrum.

\subsection{Choice of bins}

In Table~\ref{tab:binning} we summarize the
parameters used in implementing estimators
for each of these
statistical quantities.
The power spectrum and Fourier bispectrum are binned
by averaging over shells of width $\Delta k$
as explained in Sections~\ref{sec:stdPEst}--\ref{sec:stdEst}.
For the same reasons we also average the subvolume power spectra
used to construct the integrated bispectrum.
The line correlation function and modal coefficients do not
involve averaging over shells,
but instead are evaluated using equations~\eqref{eq:est1}
and~\eqref{eq:wBhat_Qn}
which are themselves aggregates over groups of
configurations.
For each statistic we report
the minimum and maximum $k$-modes that contribute,
and the total number of measurements
or bins.
Note that the bispectrum bin width corresponds
to $\Delta k = 8\,\kf$.

In what 
follows we will label the Fourier
configurations for the bispectrum
using the scheme of
\citet{Gil-Marin:2016wya}. 
We assign the label (or `index') zero
to the equilateral configuration with $k_1 = k_2 = k_3 =
\kmin$.
The remaining configurations
are ordered so that $k_1 \leq k_2 \leq k_3$
and $k_3 \leq k_1 + k_2$.
Their labels are assigned by sequentially
increasing $k_3$, $k_2$ and $k_1$
(in this order)
and incrementing the index
for each valid triangle.

In our measurements of the integrated bispectrum
we split the simulation
box into $125$ subcubes, corresponding
to a side of
$300\,h^{-1}\,\Mpc$.
This increases $\kmin$ by a factor
of five compared to the full box.
Finally, for the line correlation function we use a
non-regular $r$-spacing, spanning the range from $10$ to
$200\,h^{-1}\,\Mpc$. The first seven bins are separated by
$2.5\,h^{-1}\,\Mpc$, which doubles to $5\,h^{-1}\,\Mpc$
for the next eleven and to $10\,h^{-1}\,\Mpc$ for the remaining
twelve bins.

\begin{table}
  \centering
  \caption{Shell widths $\Delta k$
  used to average
  estimators for the
  power spectrum and bispectrum
  (where used),
  together with minimum and maximum modes $\kmin$, $\kmax$
  and the total number of bins or measurements
  $\Nbin$.}
  \begin{tabular}{ccccc}
    \toprule
    & $\Delta k$ [$h\,\Mpc^{-1}$] & $\kmin$ [$h\,\Mpc^{-1}$] & $\kmax$
    [$h\,\Mpc^{-1}$] & $\Nbin$ \\ \midrule
    $P$ & $0.010$ & $0.004$ & $0.300$ & $30$ \\
    $B$ & $0.034$ & $0.004$ & $0.302$ & $95$ \\
    $\beta$ & $-$ & $0.004$ & $0.302$ & $50$ \\
    $\ib$ & $0.010$ & $0.021$ & $0.306$ & $29$ \\
    $\LCF$ & $-$ & $0.016$ & $0.314$ & $30$ \\
    \bottomrule
  \end{tabular}
  \label{tab:binning}
\end{table}

\section{Comparison of theoretical predictions and simulations}
\label{sec:comparison}

In this section we present estimates of the typical
values for each bispectrum
proxy introduced in Section~\ref{sec:estimators},
and
implemented using the formulae of Section~\ref{sec:estimation}.
We
derive
these
from the 200 simulations of our fiducial cosmology in set (1)---%
see Section~\ref{ssec:simulations}---%
at redshifts $z = 0$, $z = 0.52$ and $z=1$.
Also,
using the simulation set (2)
we determine how each proxy responds to changes in
the cosmological parameters (Section~\ref{sec:derivs}).
These measurements enable us to characterize
the accuracy of the theoretical predictions
for these typical values
discussed in Section~\ref{sec:modelling}.
Finally, in Section~\ref{sec:ngcovariance}
we discuss measurements of the
covariances and cross-covariances for each pair of
proxies.

\subsection{Mean values in the fiducial cosmology}
\label{sec:means}

\subsubsection{Comparison of measurements and theoretical predictions}
\label{sec:means.comparison-measurement-theory}

\begin{figure}
    \centering
    \includegraphics{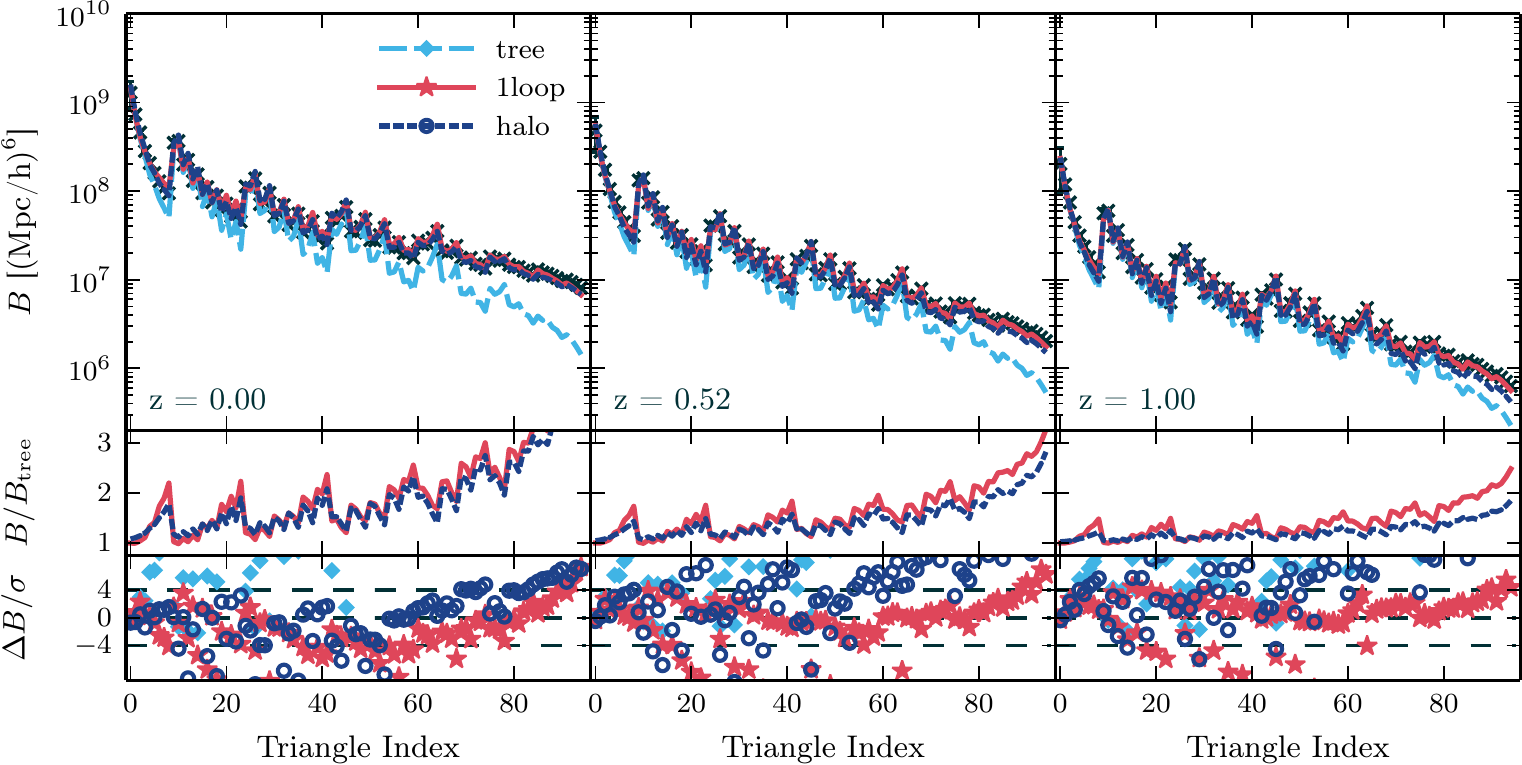}
    \caption{\semibold{Top row}:
    Measurements of the bispectrum as a function of
    configuration index (see text), estimated from $200$ {\Nbody} simulations
    at redshifts $z=0$, $0.52$ and $1$.
    We compare these measurements to the theoretical estimates
    of Section~\ref{sec:modelling}:
    the tree-level predictions are shown as dashed light-blue lines,
    the one-loop predictions are shown as solid red lines, and the halo
    model predictions are shown as short-dashed dark-blue lines.
    Black crosses mark the measured values.
    \semibold{Middle row}:
    One-loop and halo model predictions relative to the
    tree-level prediction.
    \semibold{Bottom row}:
    Differences between {\Nbody}
    measurements and theoretical
    predictions
    (ie., $\Delta B = \Bdata-\Btheory$),
    normalized to the corresponding $1\sigma$ standard deviation
    in the {\Nbody} value.}
\label{fig:bspec}
\end{figure}

\begin{figure}
    \centering
    \includegraphics{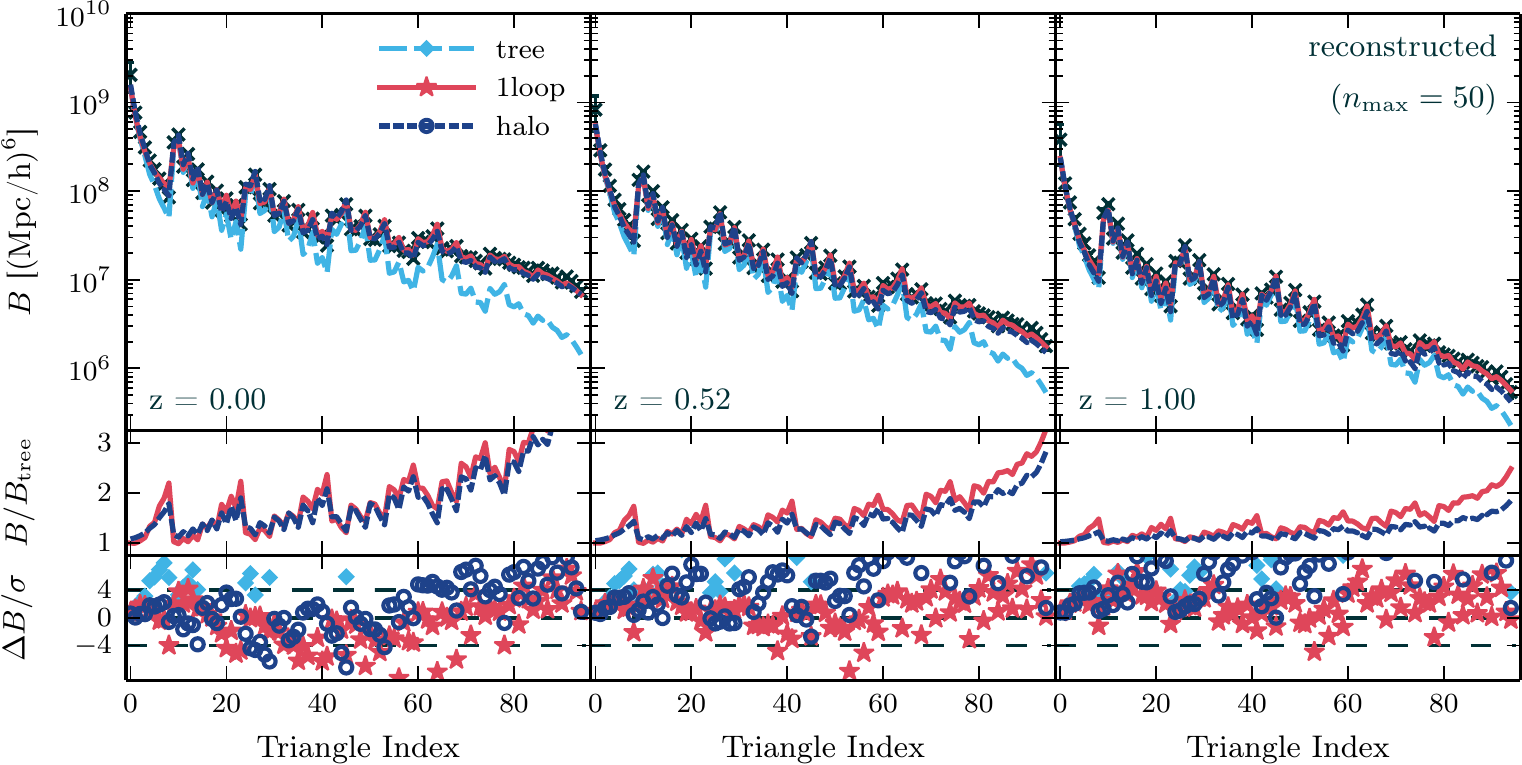}
    \caption{Same configuration as Fig.~\ref{fig:bspec},
    showing values for the Fourier bispectrum reconstructed
    from the modal coefficients $\beta_n^Q$ using equation~\eqref{eq:Bmodal}.
    In the bottom row we plot
    differences computed using $\Delta B = \Bmodal - \Btheory$.}
    \label{fig:bspec-modes}
\end{figure}

\begin{figure}
\centering
\includegraphics{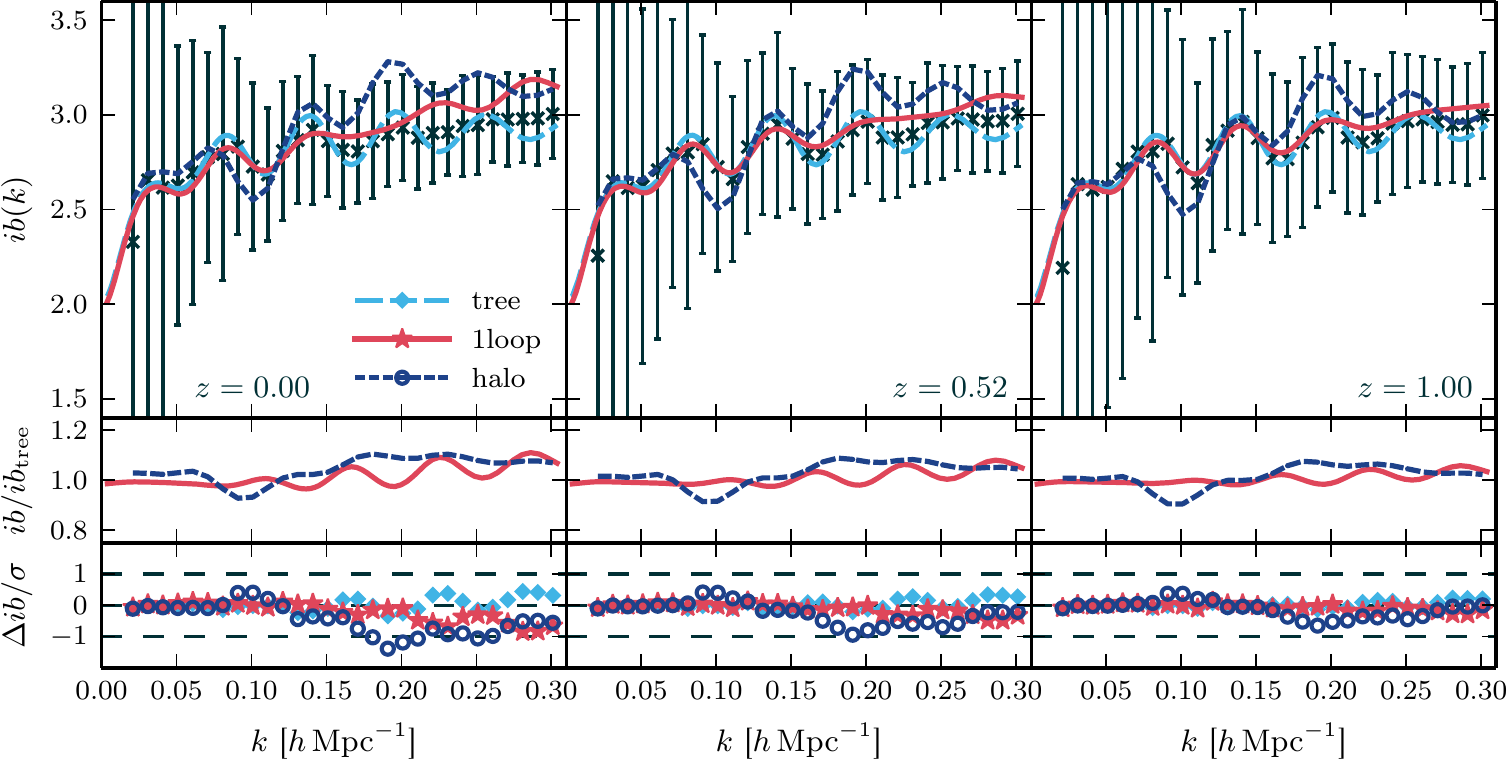}
\caption{Same configuration as Fig.~\ref{fig:bspec},
    showing values for the normalized integrated bispectrum.
    Error bars show the $1\sigma$ interval.}
\label{fig:ibspec}
\end{figure}

\begin{figure}
\centering
\includegraphics{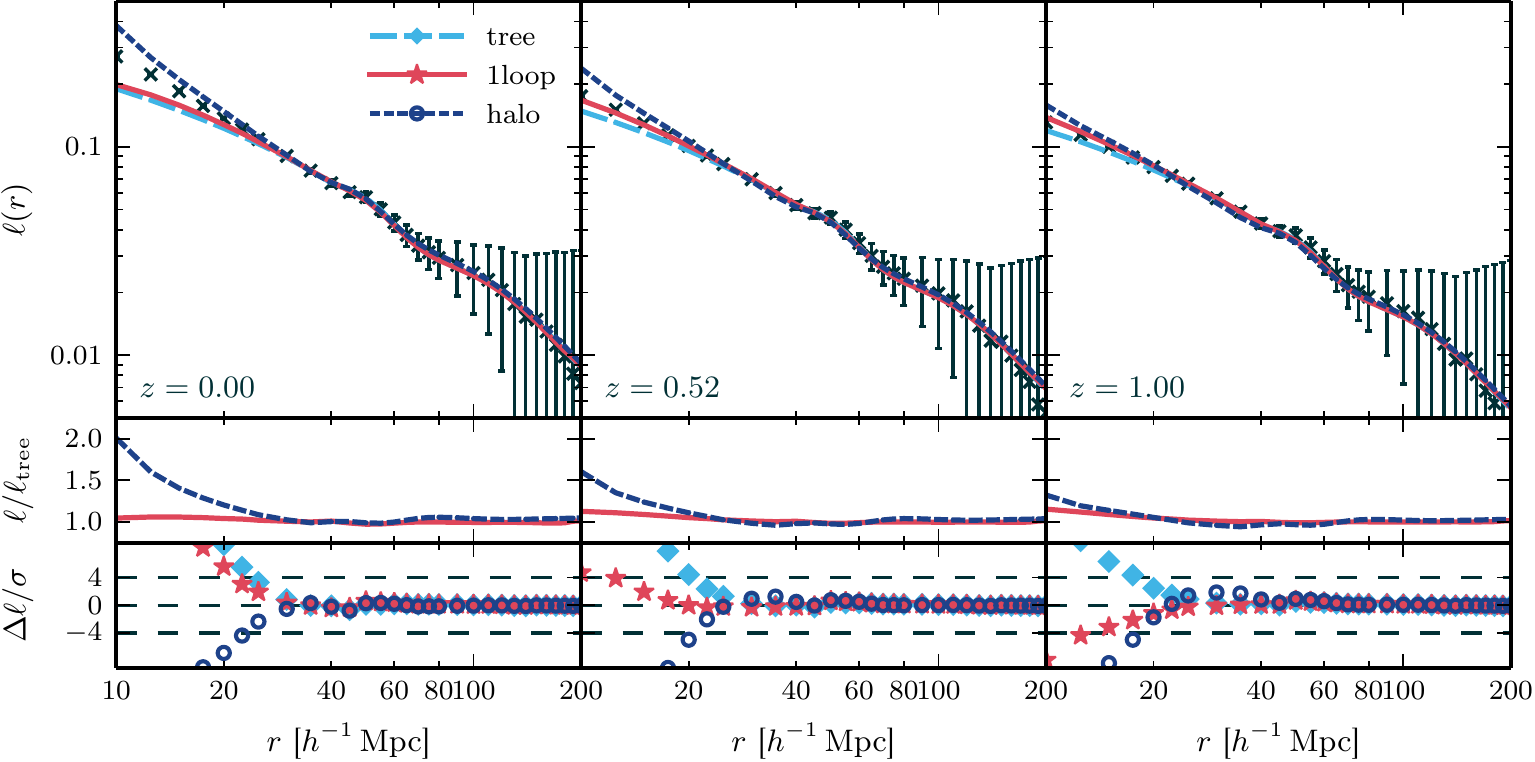}
\caption{Same configuration as Fig.~\ref{fig:bspec},
    showing values for theline correlation function at scale $r$.
    Error bars show the $1\sigma$ interval.}
\label{fig:linecorr}
\end{figure}

In Figs.~\ref{fig:bspec}--\ref{fig:linecorr} we show measurements of
each proxy for all three redshifts, averaged over the
$200$ different realizations.
We do not explicitly display our power spectrum measurements,
which have been well-studied by previous
authors
\citep[e.g.][]{Makino:1991rp, Lokas:1995xf, Scoccimarro:1996se,
  Scoccimarro:1997st, Scoccimarro:2000gm, Smith:2002dz, Seljak:2000gq,
  Peacock:2000qk, Scoccimarro:2001cj, Mead:2015yca}.
In each figure, the top row
contrasts our {\Nbody} measurements with the
tree-level, one-loop and halo model predictions.
The middle row displays
the one-loop and halo model predictions
relative to the tree-level prediction,
and the bottom row shows the difference between the {\Nbody}
measurements and the theoretical prediction in units of the
standard deviation of the {\Nbody} estimate.

\para{Fourier bispectrum}
We find that both of the SPT predictions are more accurate at large scales and high
redshifts. The halo model prediction is a better match at low redshift.
The differences between each theoretical estimate and the typical values
measured from simulation
are broadly consistent with previous analyses;
see~\citet{Scoccimarro:1997st,Scoccimarro:2000gm,Schmittfull:2012hq,Lazanu:2015rta}.

\para{Modal bispectrum}
In Fig.~\ref{fig:bspec-modes} we plot the
Fourier bispectrum reconstructed from~\eqref{eq:Bmodal}
using
our measurements of the $\beta_n^Q$ coefficients.
This is easier to interpret than the $\beta$-values themselves.
The scatter between predicted and measured values (most clearly visible in the
bottom row) is similar to the scatter for the directly-measured
Fourier bispectrum (Fig.~\ref{fig:bspec}), and indicates that differences
between the reconstructed and directly-measured values
are small.
We give a more detailed analysis of the accuracy of the modal bispectrum
in Section~\ref{sec:reconstructions}.

\para{Integrated bispectrum}
We give values for the normalized integrated bispectrum in Fig.~\ref{fig:ibspec}.
Except for a few $k$-bins
the error bars are too large to show any preference for a particular theoretical
model.
In contrast to Figs.~\ref{fig:bspec}--\ref{fig:bspec-modes},
the bottom row shows that tree-level SPT is a good match to the measured $\ib$
at all three redshifts. Conversely, the halo model prediction is a better match
at high redshift.
Our theoretical predictions are consistent with those reported by~\citet{Chiang:2014oga},
but our measured values have larger error bars because we work with a
smaller simulation volume.

\para{Line correlation function}
Finally, we present our measurements of the line correlation function in
Fig.~\ref{fig:linecorr}.
The one-loop and halo-model predictions appearing here are new, and have not
previously been studied.
The most striking feature is the discrepancy between the halo model and SPT-based
predictions in the smallest $r$-bins.
This is consistent with the analyses of~\cite{Wolstenhulme:2014cla}
and~\citet{Eggemeier:2016asq},
which both found differences between the tree-level prediction
and values measured from simulation
on scales with $r \lesssim 30\,h^{-1}\,\Mpc$.
The agreement is good
for larger $r$.

\para{Theory error}
The bottom panels of Figs.~\ref{fig:bspec}--\ref{fig:linecorr} show that our
theoretical predictions are accurate within a restricted range of scales.
Outside this range
it becomes progressively more difficult to model the observables.
This mis-modelling should be regarded as an additional source of systematic
error---a \emph{theory error}---when forecasting constraints, or analysing data,
using any of these theoretical models.
In particle phenomenology
such theory errors are routinely estimated when performing fits to data,
but their use in cosmology is less common.
In this paper we construct Fisher forecasts for parameter error bars
using both SPT-based models and the halo model.
Comparison of these error bars
enables us to estimate the impact of theoretical uncertainties
on future constraints that incorporate three-point statistics
(see Section~\ref{sec:theory-dep}).

An alternative prescription for estimating theory errors
was used by~\citet{Baldauf:2016sjb} and~\citet{Welling:2016dng}.
In their approach the theoretical uncertainty in one-loop SPT is estimated
from the next-order term in the loop expansion.
We find that this prescription gives noticeably larger estimates than the
difference between one-loop SPT and the values we measure from simulations.
Therefore, although~\cite{Baldauf:2012hs} and~\citet{Welling:2016dng}
concluded that (for example)
constraints on some types of primordial non-Gaussianity would be weakened significantly
after accounting for theory errors, our
numerical comparison suggests that
the attainable error may degrade by less
than their analysis would suggest.

\subsubsection{Accuracy of modal reconstruction}
\label{sec:reconstructions}

\begin{figure}
  \centering
  \includegraphics{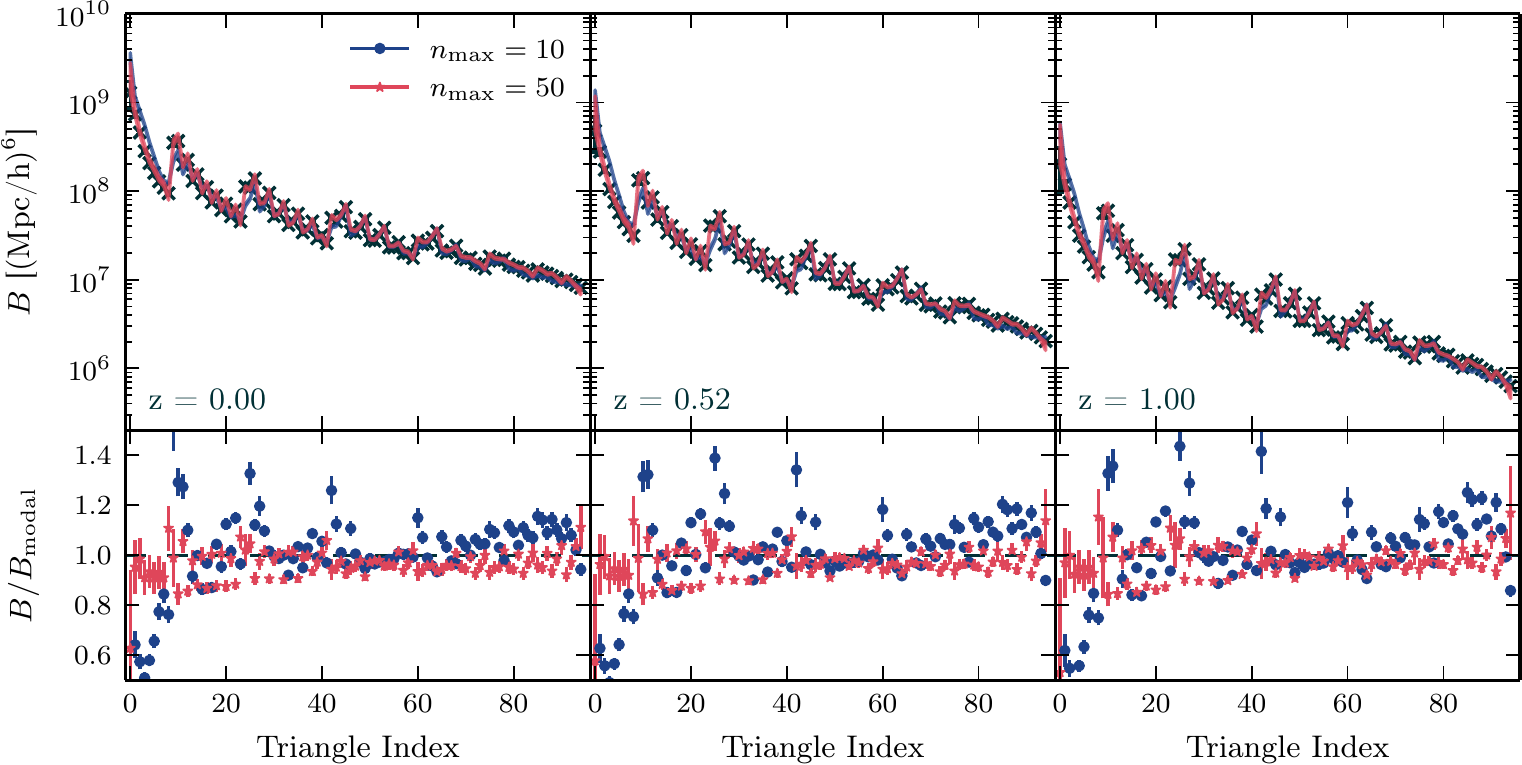}
  \caption{Modal bispectra reconstructed using $10$ modes (blue) and $50$ modes (red) at redshifts $z=0$, $0.52$ and $1$. The lower panels show the ratio of the measured normal bispectrum and modal bispectrum.}
  \label{fig:bspecrecon}
\end{figure}

Comparison of Figs.~\ref{fig:bspec} and~\ref{fig:bspec-modes}
demonstrates that the Fourier bispectrum reconstructed from our
measurements of the $\beta^Q_n$ accurately reproduces the
correct amplitude and shape dependence.
This information is embedded in the modal coefficients.
For example, the zeroth basis mode $R_0 \propto Q_0$ is a constant
and therefore $\beta^R_0 \propto \beta^Q_0$ captures information about
the mean amplitude of the Fourier bispectrum over all configurations---or,
equivalently, the skewness of $\delta$.
The next few modes are slowly varying functions of configuration.
Taken together, these low-order modes
carry the principal amplitude information
and for reasonably smooth bispectra we expect they exhibit the strongest
dependence on background cosmological parameters.
The higher modes capture more subtle detail.
As with any basis decomposition, their inclusion increases the
accuracy of the reconstruction.

To see this in detail, consider a reconstruction using only $\nmax = 10$
modes.
In Fig.~\ref{fig:bspecrecon} we plot the Fourier bispectrum reconstructed
in this way (blue line) compared to the
reconstruction using $\nmax = 50$ described above (red line).
Black crosses mark the measured data points.
In the lower panel we plot the ratio between these measured values and the
reconstructions.
The accuracy is good whether we use $\nmax = 10$ or $\nmax = 50$, but the scatter is
smaller for $\nmax = 50$.
We conclude that, in this case, the first 10 modes
are sufficient to capture the main behaviour of the Fourier bispectrum,
but extra modes are helpful if we wish to reproduce the precise configuration
dependence to within $\lesssim 10\%$ accuracy.

\subsection{Derivatives with respect to cosmological parameters}
\label{sec:derivs}

\begin{figure}
  \centering
  \includegraphics[scale=0.98]{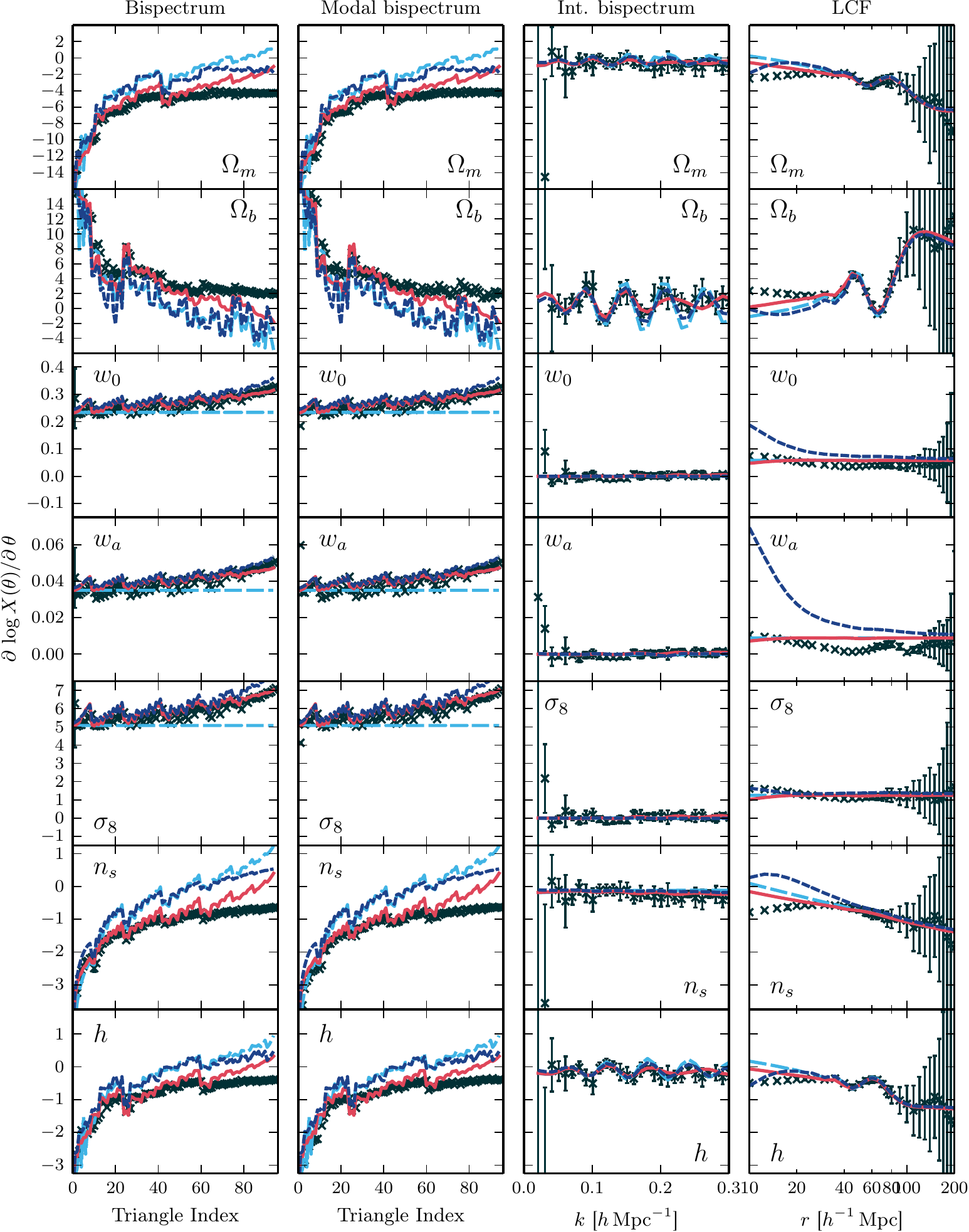}
  \caption{Derivatives of the Fourier bispectrum and its proxies with respect
  to the parameters at $z=0$. The four columns show (from left to right) the derivatives of: $B$, $\Bmodal$ (reconstructed from $\beta$), $\ib$ and $\LCF$. As in Figs.~\ref{fig:bspec}--\ref{fig:linecorr}, measured values are in black, while light blue dashed, red solid and dark blue short-dashed lines are the tree-level, 1-loop and halo model predictions, respectively.}
  \label{fig:derivatives}
\end{figure}

In the remainder of this paper our aim is
to obtain Fisher forecasts of error bars for a parameter set
$\theta_\alpha$,
where the index $\alpha$ labels one of the cosmological
parameters of Table~\ref{tab:parameters}.
For this purpose the role of a
theoretical model is to
predict the derivatives of observables with respect to each
parameter,
and
the accuracy of the forecast depends on the reliability of these predictions.
In this section we study how well our three theoretical models
reproduce the derivatives estimated from our simulation suite.
We compute the derivative of some estimator
$\hat{X}$ at wavenumber $k$ with respect to a parameter $\theta_\alpha$
by the rule
\begin{equation}
    {\frac{\D{\hat{X}}(k\,|\,\B{\theta})}{\D{\theta_{\alpha}}}}
    =
    \hat{\bar{X}}(k\,|\,\B{\theta})
    {\frac{\D{\ln{\hat{X}}}(k\,|\,\B{\theta})}{\D{\theta_{\alpha}}}}
    ,
\end{equation}
where $\hat{\bar{X}}(k|\B{\theta})$ is the average over the                                                               
$200$ fiducial simulations of set (1) (described in
Section~\ref{ssec:simulations})
for $X \in
\{P, B, \beta, \ib, \LCF \}$,
and
the logarithmic derivative
with respect to $\theta_\alpha$
is computed
using
\begin{equation}
    \label{eq:means+derivs.logderiv}
    {\frac{\D{\ln{\hat{X}}}(k \mid \B{\theta})}{\D{\theta_{\alpha}}}}
    =
    \frac{1}{4}
    \sum_{i=1}^4
    \frac{\hat{X}^{(i)}(k \mid \B{\theta} + \Delta\theta_{\alpha})
            - \hat{X}^{(i)}(k \mid \B{\theta} - \Delta\theta_{\alpha})}
        {2\Delta\theta_{\alpha} \hat{X}^{(i)}(k \mid \B{\theta})}
    .
\end{equation}
The sum is over the four realizations used in simulation set (2),
and the derivative is constructed using the $+\Delta\theta_\alpha$ and
$-\Delta\theta_\alpha$
offset simulations described in Section~\ref{ssec:simulations}.
The advantage of
the logarithmic derivative is that
both realizations in the numerator
on the right-hand side of~\eqref{eq:means+derivs.logderiv}
share initial
conditions with their fiducial partner
in the denominator.
Therefore, division by the fiducial estimate
$\hat{X}^{(i)}(k \mid \B{\theta})$
minimizes dependence on the specific realization.%
    \footnote{This strategy is less
    successful for the line correlation function.
    In this case the fiducial value could be very close to zero
    on some scales.
    In turn, this produces large errors in the logarithmic derivative.
    Therefore, for the line correlation function,
    we estimate the linear derivative
    $\D{\LCF} / \D{\theta_\alpha}$ instead.}

In Fig.~\ref{fig:derivatives} we plot the derivatives of each observable with respect
to the cosmological parameters at $z=0.52$.
Our forecasts use three redshift bins, but their behaviour is similar
to the $z=0.52$ bin and the statements made below can be taken to apply
at all three redshifts.
We do not include the power spectrum, for which the derivatives
appeared in~\citet{Smith:2012uz}.

\para{Modal bispectrum}
To simplify comparison of the modal bispectrum with the Fourier bispectrum,
Fig.~\ref{fig:derivatives} plots derivatives of the reconstructed
bispectrum rather than derivatives of $\beta_n^Q$ or $\beta_n^R$.
Comparison of the first two columns shows that the cosmology-dependence
is accurately captured using $\nmax = 50$, either for theoretical predictions
or the measured values.

There is a significant spread in performance of the theoretical models,
with tree-level SPT and the halo model generally offering the poorest match.
For the derivatives with respect to $\Omega_m$, $\Omega_b$, $n_s$ and $h$
these models give similar predictions.
The probable reason is that, in the standard halo model,
the halo mass function and halo profile
are fixed to the fiducial cosmology.
Only the input power spectrum is taken to vary with the cosmological
parameters, and since it matches the tree-level SPT prediction its
derivatives will be equal.
Therefore the halo-model derivatives will differ from those of tree-level
SPT only via a (possibly scale-dependent) prefactor.
More complex halo models with cosmology-dependent
halo parametrizations
have been studied (see, eg., \citet{Mead:2016zqy} for
an application to dark energy models).
However, determining which variation of the halo model captures the
cosmological parameter dependence of the bispectrum most
accurately is outside the scope of this paper.
We simply note that, if the halo model is to be used for
analysis or forecasting of the Fourier bispectrum,
its implementation should be chosen with care
because its performance depends on these details.

\para{Integrated bispectrum}
The derivatives of the integrated bispectrum are shown in the
third column of Fig.~\ref{fig:derivatives}.
The errors bars on the measured values
are again too large to show a clear
preference for any model---and they are generally so
large that the measurement is not
significantly different from zero.
These results are consistent
with those reported by~\citet{Chiang:2015pwa}
for a range of values of $\Omega_m$, $\sigma_8$
and $n_s$.
We conclude that the integrated bispectrum is
rather insensitive to the background cosmology
and is therefore a comparatively poor tool
to constrain it.
While this means we must expect a Fisher forecast to
predict weaker error bars for the parameters
of Table~\ref{tab:parameters}, this insensitivity
could be an advantage if the intention is to
use the integrated bispectrum as a probe of
other physics.
For example, in addition to the background cosmology
we may wish to use the large-scale structure bispectrum
to constrain the possibility of \emph{primordial}
three-point configurations produced by inflation on
squeezed configurations.
Insensitivity to the background cosmology would
reduce the likelihood of degeneracies in these
measurements.

\para{Line correlation function}
The last column of Fig.~\ref{fig:derivatives}
shows the derivatives of the line correlation function.
As for the typical values discussed above,
the values predicted by our theoretical
models
are significantly discrepant with the measured
values in the smallest $r$ bins.
Also, the derivative with respect to the
dark energy parameter $w_a$
is particularly discrepant
for the halo model.
One possible explanation is the
construction of the halo model
as described above,
with its fixed halo mass function and halo profile.
Alternatively, it is possible that the
halo model power spectrum and bispectrum
that we use
are subtly inconsistent in a way that
produces inaccuracies in the line correlation
function on small scales.

\subsection{Non-Gaussian covariance}
\label{sec:ngcovariance}

\begin{figure}
  \centering
  \includegraphics[width=\textwidth]{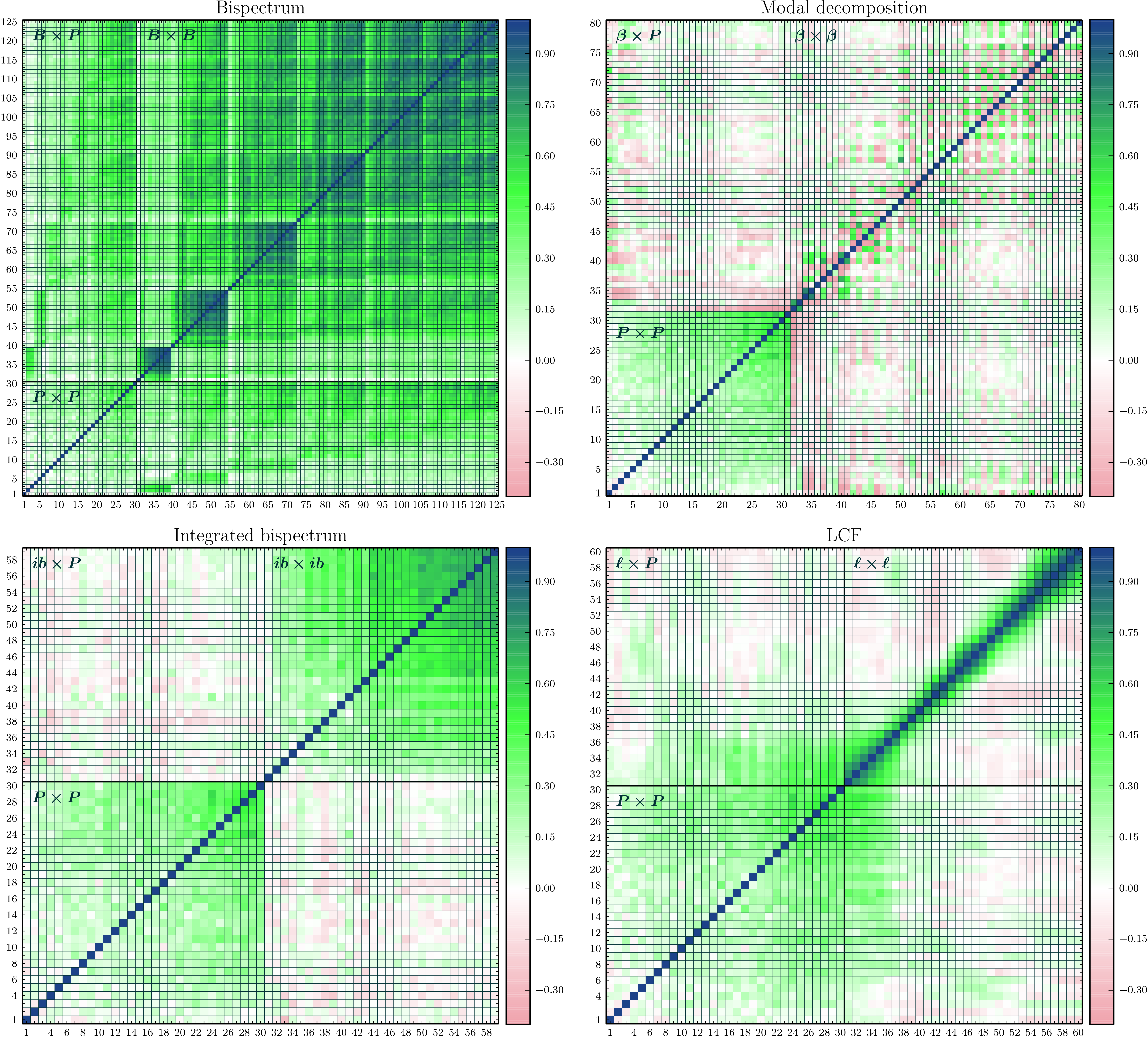}
  \caption{Correlation matrices for (clockwise from top left) $P+B$, $P+\beta$, $P+\LCF$ and $P+\ib$ and at redshift $z=0.0$. In each panel, the lower-left quadrant
    contains the power spectrum auto-correlation ($P \times P$), while the upper-right quadrant contains the auto-correlation of the corresponding 3-point correlation
    measure. The upper-left and lower-right quadrants contain the cross-covariance.
  }
  \label{fig:correlation}
\end{figure}

The analytic, Gaussian covariance of each proxy is most
accurate at high redshifts and on large scales, where
the matter fluctuations are more nearly Gaussian
and therefore more accurately described by the power spectrum
alone.
At low redshifts and on small scales, however, the
Gaussian approximation fails due to non-linear evolution
of matter fluctuations.
This evolution generates additional contributions
to the covariance through higher-order $n$-point correlations.

The simplest and most robust approach to obtain accurate non-Gaussian
covariances has been to analyse large suites of {\Nbody} simulations.
This method was used by
\citet{Takahashi:2009bq}, \citet{Takahashi:2009ty},
\cite{Blot:2015cvj}, and~\citet{Klypin:2017iwu} to study the
non-Gaussian covariance of the power spectrum.
Other authors have performed analogous studies for the
bispectrum~\citep{Sefusatti:2006pa,Chan:2016ehg},
the real-space partner of the integrated bispectrum~\citep{Chiang:2015eza},
and the line correlation function~\citep{Eggemeier:2016asq}.
In this section, we present our measurements of the non-Gaussian covariance
for each proxy, estimated from our suite of simulations.
We also discuss the cross-covariance between pairs of proxies. 

In Sections~\ref{sec:covariance} and~\ref{sec:paramEstim}
we quantify the impact of these complex non-diagonal
covariances
on estimates of signal-to-noise
and Fisher forecasts.

\para{Correlation matrices}
We plot
correlation matrices for the measurements $P+B$, $P+\beta$, $P+\ib$, and $P+\LCF$
in Fig.~\ref{fig:correlation}.
We show measurements only at $z=0$ where differences between the Gaussian and
non-Gaussian covariances are largest.

The correlation coefficient $\CorrMatrix_{ij}$ between two data bins $i$ and $j$
is defined to satisfy
$\CorrMatrix_{ij} \equiv \hat{\CovMatrix}_{ij}/\sqrt{\hat{\CovMatrix}_{ii}\,\hat{\CovMatrix}_{jj}}$, 
where $\hat{\CovMatrix}$ is the covariance matrix estimated from the simulation suite,
\begin{equation}
	\hat{\CovMatrix}_{ij}
	=
	\frac{1}{\Nreal}
	\sum_{n=1}^{\Nreal}
	\Big[
		\hat{S}_i^{(n)}
		- \hat{\bar{S}}_i
	\Big]
	\Big[
		\hat{S}_j^{(n)}
		- \hat{\bar{S}}_j
	\Big]
	,
	\label{eq:covariance-def}
\end{equation}
and $\Nreal = 200$ is the number of realizations.
To measure an auto-covariance the data vector
$S$ contains all measurements of a single proxy,
$S = ( X_{a,1}, \ldots, X_{a,n})$
or to measure a cross-covariance 
it contains all measurements from a pair,
$S = ( X_{a,1}, \ldots, X_{a,n_1}, X_{b,1}, \ldots, X_{b,n_2} )$,
where $X_a, X_b \in \{ P, B, \beta, \ib, \LCF \}$.
The correlation matrix measures the degree of
coupling between different measurements.
Its elements take values between $-1$
(where the bins are fully anti-correlated)
and $+1$
(where the bins are fully correlated).
A value of zero corresponds to independent measurements.
For comparison, the Gaussian covariance matrices
for $P$, $B$, $\beta$ and $\ib$ are diagonal,
whereas for $\LCF$ there are correlations
between neighbouring bins with similar $r$
because it is a real-space statistic
and therefore includes
contributions from many Fourier configurations.
In the Gaussian approximation
the cross-covariance between $P$ and any
bispectrum proxy is zero.

\para{Fourier bispectrum}
For $P+B$ (upper-left panel of Fig.~\ref{fig:correlation}) the correlation matrix
has an approximate block structure due to the ordering of the 95
triangle configurations that we measure.
The blocks correspond to groups of adjacent configurations
with shared values of $k_1$ or $k_2$.
While the power spectrum $P(k)$
shows mild correlations between different bins
at high $k$, the bispectrum
exhibits much stronger correlations.
There are also non-zero cross-correlations between
power spectrum and bispectrum bins.
The correlation between power spectrum and bispectrum
tends to be higher when $P(k)$
and $B(k_1, k_2, k_3)$ have wavenumber bins that overlap.
Similarly, the correlation between different bispectrum
bins is higher when
the configurations share at least one wavenumber.
However, even configurations that have no wavenumbers
in common can be strongly correlated,
with correlation coefficient as large as $\sim 0.8$,
due to non-linear growth.

\para{Modal bispectrum}
In the upper-right panel of Fig.~\ref{fig:correlation}
we present measurements of the correlation coefficients for $P+\beta^R$.
These have not previously been reported.
As explained in Section~\ref{sec:predict-modal}
these measurements apply to the $R$-basis,
for which the covariance matrix is \emph{constructed} to be
diagonal in the Gaussian approximation.
We find that only the first two modes are correlated with the majority
of $P(k)$ bins. 
This is reasonable because the lowest modes probe the most scale-independent
features of the phase bispectrum.
The remainder show low-to-moderate 
correlation or anti-correlation
due to non-linear effects.

\para{Integrated bispectrum and line correlation function}
Correlation measurements for the integrated bispectrum appear
in the lower-left panel of Fig.~\ref{fig:correlation}.
The $\ib(k)$ measurements show stronger
auto-correlations than $P(k)$
as $k$ increases,
while the $P \times \ib$ cross-correlation
is relatively featureless.
This indicates that the two data sets are nearly independent.
Similarly, we find that the $P \times \LCF$ cross-correlation
is nearly featureless except where the smallest $r$
bins and highest $k$ bins show significant correlation.
Relative to the Gaussian covariance matrix for $\LCF$,
the $r$ bins with
$r \lesssim 50 \, h^{-1} \, \Mpc$
are more strongly correlated due to non-linear growth.

\para{Cross-covariances}
Finally, we have computed the correlation matrices between the
bispectrum and its proxies.
These enable us to identify which bispectrum configurations
contribute most to individual bins of
$\beta^R$, $\ib$ or $\LCF$.
We find that the first two $\beta^R$ modes are strongly
correlated with the bispectrum over large range
of triangles, while the remainder
are generally more correlated with triangles on the
largest scales
(that is, lower triangle index).
This structure is similar to the $P+\beta^R$ correlation matrix.

We find that $B$ and $\ib$ are very weakly correlated,
which we attribute to $\ib$ being dominated by
more strongly squeezed triangles than any we include
in the 95 measured configurations of $B$.
Finally, the line correlation function is correlated with
a majority of bispectrum configurations when $r \lesssim 40 \, h^{-1} \, \Mpc$.
This indicates that the line correlation function
is sensitive to many different shapes of Fourier triangle.
We do not find particularly strong correlations
for $\LCF \times \ib$,
but $\LCF \times \beta^R$ shows
that the line correlation function at small $r$
is highly correlated with the first two
$\beta^R$ modes.
This is consistent with the observation that both
are sensitive to a wide range of Fourier configurations.

\section{Cumulative signal-to-noise of the bispectrum proxies}
\label{sec:covariance}

Before discussing the constraining power
of each proxy we first compute the available
signal-to-noise.
This is an intermediate step that characterizes
the significance with which measurements of
each proxy can be extracted from a data set.
Negligible signal-to-noise would normally
imply poor prospects for parameter constraints.
For example,
\citet{Chan:2016ehg}
and~\cite{Kayo:2012nm}
studied the signal-to-noise
as a proxy for the information content of the
Fourier bispectrum in the context of
large-scale structure and weak lensing,
respectively.

\para{Numerical procedure}
The cumulative signal-to-noise $\mathcal{S} / \mathcal{N}$
up to a maximum wavenumber $\kmax$ is defined by
\begin{equation}
	\left( \SignalToNoise \right)^2
	\equiv
	\sum_{k_i, k_j \leq \kmax} S_i \CovMatrix^{-1}_{ij} S_j ,
	\label{eq:signal-noise-def}
\end{equation}
where $S$ is the vector of typical values for either a single proxy
or a combination of proxies,
defined below equation~\eqref{eq:covariance-def}.
In this and subsequent sections we drop
the use of a hat to denote an estimated value,
and an overbar to denote a mean.
The sum in~\eqref{eq:signal-noise-def}
runs over all bins containing wavenumbers that satisfy the
condition $k \leq \kmax$.
For the Fourier bispectrum a bin corresponds to a triplet
of wavenumbers $(k_1, k_2, k_3)$, all of which are required to be
smaller than $\kmax$.

We use the non-Gaussian covariance matrix measured from simulations,
described in Section~\ref{sec:ngcovariance},
which we denote by $\CovMatrix_*$.
Its inverse $\CovMatrix^{-1}_*$ is not an unbiased estimator of
$\CovMatrix^{-1}$.
A simple prescription to approximately correct for this bias
is to rescale $\CovMatrix^{-1}_*$
by an Anderson--Hartlap factor~\citep{Anderson2003,Hartlap2007},
which yields
\begin{equation}
	\CovMatrix^{-1}
	\approx
	\frac{\Nreal-\Nbin-2}{\Nreal-1}\,\CovMatrix^{-1}_*
	,
\end{equation}
where $\Nreal$ is the number of realizations used to estimate the covariance matrix
and $\Nbin$ is its dimensionality.%
	\footnote{Although the Anderson--Hartlap prescription is
	simple to apply, it has been pointed out by~\citet{Sellentin2016} that this
	rescaling simply broadens the Gaussian likelihood of the data.
	These authors argued that the distribution of the data is more accurately
	modelled by a $t$-distribution.}
Care should be taken when computing the numerical inverse $\CovMatrix^{-1}_*$,
especially for combinations of measurements with signals of widely disparate magnitude.
To avoid issues associated with ill-conditioning
we first compute the correlation matrix
$\CorrMatrix_{*,ij} = \CovMatrix_{*,ij}/\sqrt{\CovMatrix_{*,ii}\,\CovMatrix_{*,jj}}$,
whose entries
lie between $-1$ and $+1$.
We determine the inverse $\CorrMatrix^{-1}_{ij}$ using a singular
value decomposition
and check that all singular values are above the noise.
Finally, we compute the inverse covariance using
\begin{equation}
	\CovMatrix^{-1}_{*,ij}
	=
	\frac{\CorrMatrix^{-1}_{*,ij}}{\sqrt{\CovMatrix_{*,ii} \CovMatrix_{*,jj}}}
	.
\end{equation}

\begin{figure}
\centering
\includegraphics{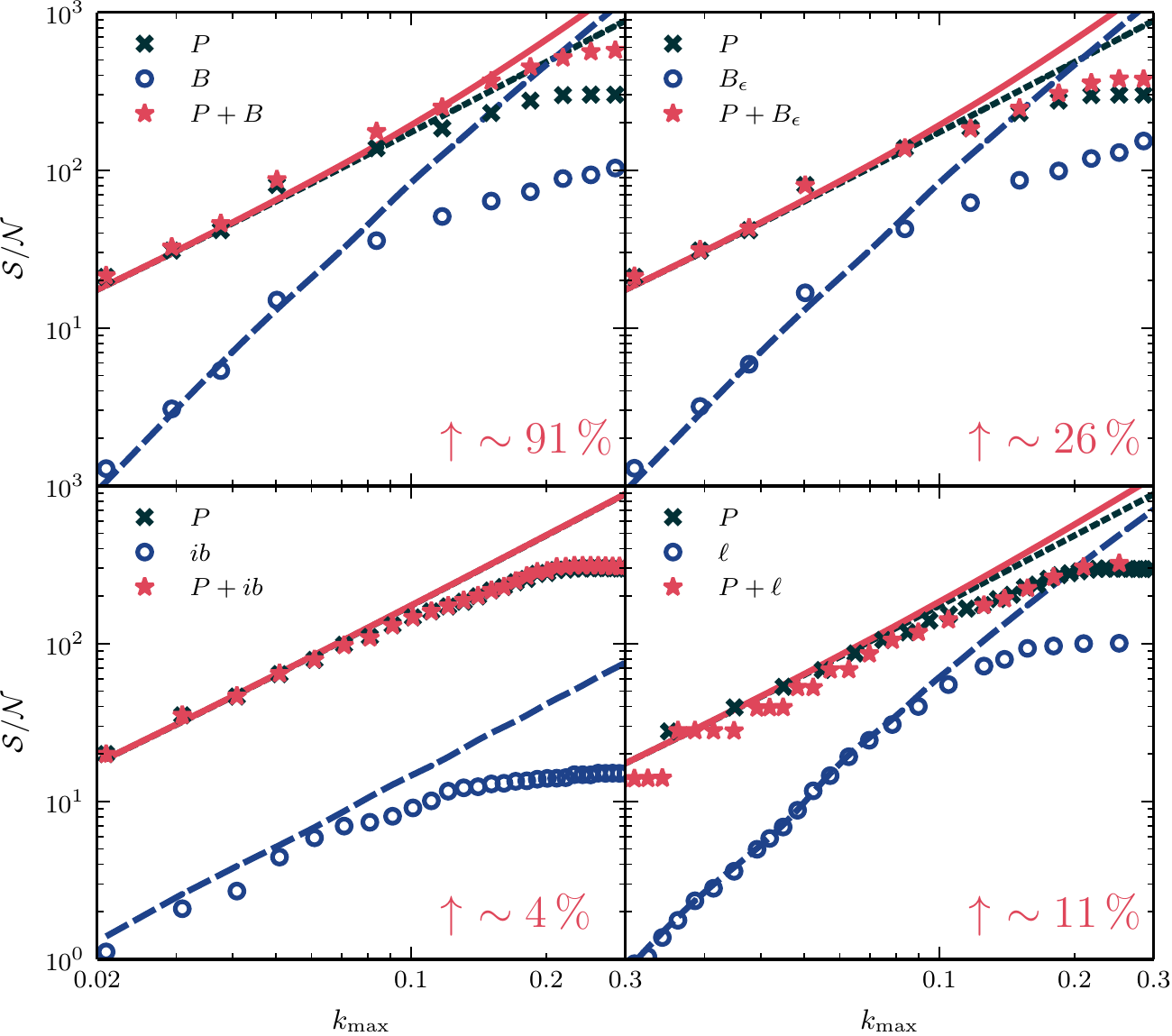}
\caption{Cumulative signal-to-noise at redshift $z=0$ as a function of the maximal mode $\kmax$ for the measure $X$---equal to
    the Fourier bispectrum, phase bispectrum, line
    correlation function or integrated bispectrum (clockwise, starting from the top left panel).
    In each panel, blue circles refer to the measured signal-to-noise for
    $X$,
    while black crosses represent the
    signal-to-noise for the power spectrum.
    We plot the signal-to-noise for the combination $P+X$, including cross-covariance,
    as red stars.
    The blue, black and red lines give
    the theoretical prediction using the Gaussian approximation and
    tree-level SPT.
    The percentage quoted in the bottom right corner gives the increase
    in signal-to-noise relative to the power spectrum alone
    at $\kmax = 0.3 \, h \, \Mpc^{-1}$.}
\label{fig:s2n}
\end{figure}

\para{Results}
In
Fig.~\ref{fig:s2n} we plot the resulting signal-to-noise measurements
for the Fourier bispectrum, integrated bispectrum, line correlation function
and the quantity $\Bphase$ defined in~\eqref{eq:phaseB}
and used in the construction of the line correlation function and the modal
bispectrum.
(The signal-to-noise from $\Bphase$ and the reconstructed
modal bispectrum give almost identical results.)
We estimate $\Bphase$ using the prescription
\begin{equation}
	B_{\epsilon}(k_1,\,k_2,\,k_3) = \frac{B(k_1, k_2, k_3)}{\sqrt{P(k_1) P(k_2) P(k_3)}} .
\end{equation}
Each panel of Fig.~\ref{fig:s2n} shows the cumulative signal-to-noise of the Fourier
bispectrum or a proxy (blue circles), together with the power spectrum (black crosses)
and their combination including the cross-covariance matrix (red stars).
The first four data points in the $B$ and $\Bphase$ panels use a bin size
$\Delta k = 2\kf$ in order to probe the low-$k$ regime.
The remainder derive from the measurements
presented in Section~\ref{sec:comparison} and use $\Delta k = 8 \kf$.
Our
measurements of the integrated bispectrum and line correlation function
carry forward the binning procedure used in~Section~\ref{sec:comparison}.
The step-like structure that occurs for $P+\ell$ is due to a mismatch
of scales between the power spectrum and the bins of the line
correlation function.
In each panel, for comparative purposes,
we plot lines of matching colour to show the
signal-to-noise computed using a Gaussian approximation to the
covariance matrix
and tree-level SPT to evaluate any correlation
measures it contains.

\para{Discussion}
First, we note that the Gaussian approximation overpredicts the signal-to-noise
for each proxy $X$ and its combination $P+X$ with the power spectrum.
This is consistent with the results reported by~\citet{Chan:2016ehg}.
The overprediction occurs because bins become coupled
by non-linear evolution, and therefore do not provide
independent information
as the Gaussian approximation assumes.
The effect can be quite severe:
while the power spectrum signal-to-noise
at $\kmax = 0.3 \, h \, \Mpc^{-1}$ is overpredicted
by a factor of three, the
impact on the Fourier bispectrum and its proxies
is much larger.
In these cases the overprediction ranges from a factor
of $\sim 5$ or $8$ for $\ib$ and $\LCF$
up to more than an order of magnitude for the Fourier
bispectrum.
At smaller $\kmax$ the overprediction is less, becoming
significant for $\kmax \gtrsim 0.1 \, h \, \Mpc^{-1}$.

The Fourier bispectrum, phase bispectrum, and line correlation function
\emph{individually} contribute $\sim 30\%$ of the signal-to-noise of $P(k)$
at $\kmax = 0.3 \, h \, \Mpc^{-1}$, while the integrated bispectrum achieves only
$5\%$ of the $P(k)$ signal-to-noise. For the Fourier bispectrum, this result
is consistent with~\citet{Chan:2016ehg}.

However, for estimating parameter constraints from the joint combination of $P$
and $B$, or one of its proxies, the individual signal-to-noise contributed by one
of these measurements is less important than whether it contains information
that is not already present in the power spectrum.
This is determined by the signal-to-noise of the combination
$P+X$ compared to $P$ alone.
The different proxies show significant
variation
in the improvement from use of $P+X$,
which
we indicate as a percentage
in the bottom-right corner of each panel.
Although $B$, $\Bphase$ and $\LCF$
individually carry roughly the same signal-to-noise,
the uplift in $P+X$ varies from $\sim 91\%$
to $\sim 11\%$.
Note that the signal-to-noise of $P+B$ receives a large
improvement from the cross-covariance, which was ignored in~\citet{Chan:2016ehg}.

The discrepancy in uplift between $B$ and $\Bphase$ is striking.
If this discrepancy were to carry over to parameter constraints
it would imply that the Fourier bispectrum carries
\emph{significantly} more constraining power than $\Bphase$,
even though both statistics are equivalent in the
approximation of Gaussian covariance.
If true, this would be very surprising.
We return to this question in Section~\ref{sec:signoiseProxy}
after we have obtained forecast parameter
uncertainties for $B$ and its proxies,
which enable us to
precisely quantify the constraining power of
each statistic.

\section{Parameter uncertainty forecasts}
\label{sec:paramEstim}

In this section we collect our major results, which are
Fisher forecasts of the error bars achievable on the parameter
set $\theta_\alpha = ( \Omega_m, \Omega_b, w_0, w_a, \sigma_8, n_s, h )$
of Table~\ref{tab:parameters},
based on a fiducial flat {\LambdaCDM} cosmology.
We perform these forecasts with and without inclusion of the
bias parameters $(b_1, b_2)$.

In Section~\ref{sec:forecasting-method} we summarize our implementation
of the Fisher forecasting method, and in Section~\ref{sec:information_content}
we present and compare the forecasts from each proxy.
By comparing forecasts with and without non-Gaussian covariances,
and using different theoretical models to describe the dark matter density,
we are able to characterize their influence on the final parameter
constraints.
These discussions appear in Sections~\ref{sec:ng-cov} and~\ref{sec:theory-dep}, respectively.
Finally, we return to the discussion of Section~\ref{sec:covariance}
and examine to what extent the signal-to-noise provides a reliable
metric by which to estimate improvements in
parameter constraints (Section~\ref{sec:signoiseProxy}).

\subsection{Forecasting method}
\label{sec:forecasting-method}

The Fisher formalism can be used to forecast the precision with
which cosmological parameters could be measured in a future survey.
Consider a data vector $\bx$ containing measurements of any combination
of statistical quantities.
The likelihood function $\Likelihood(\B{\theta} \mid \bx)$ is defined to be the
probability of the data given the parameters $\B{\theta}$, so
$\Likelihood(\B{\theta} \mid \bx) = P(\bx \mid \B{\theta})$.
Then the Fisher matrix $\FisherMatrix_{\alpha\beta}$ satisfies
\begin{equation}
    \FisherMatrix_{\alpha\beta}
    \equiv
    -
    \left\langle
        \frac{\partial^2 \ln \Likelihood(\B{\theta} \mid \bx)}
            {\partial\theta_{\alpha}\,\partial\theta_{\beta}}
    \right\rangle
    .
\end{equation}
The expected $1\sigma$
error on each parameter $\theta_\alpha$,
marginalized over all other parameters,
can be obtained from the diagonal elements of the inverse Fisher matrix
using $\sigma^2(\theta_\alpha) = (\FisherMatrix^{-1})_{\alpha\alpha}$.
To simplify the computation of $\FisherMatrix_{\alpha\beta}$
we make the assumption that the likelihood function is a multivariate
Gaussian,
\begin{equation}
    \label{eq:paramEstim.likelihood} 
    \Likelihood
    =
    \frac{1}{\sqrt{(2\pi)^n|\CovMatrix|}}
    \exp
    \left[
        -\frac{1}{2}(\bx-\B{\mu})^\transpose \CovMatrix^{-1} (\bx-\B{\mu})
    \right]
    ,
\end{equation}
where $\transpose$ denotes a matrix transpose
and $|\CovMatrix| = \det \CovMatrix$ is the determinant of $\CovMatrix$.
We have written the mean of the data vector as $\B{\mu} = \langle\bx\rangle$,
and its covariance matrix is
$\CovMatrix_{ij} = \langle x_i\,x_j\rangle - \mu_i\,\mu_j$.
With these assumptions it can be shown that~\citep{Tegmark:1996bz},
\begin{equation}
	\label{eq:paramEstim.fisherG}
    \FisherMatrix_{\alpha\beta}
    =
    \frac{1}{2}
    \Tr
    \left[
        \CovMatrix^{-1}
        \frac{\partial\CovMatrix}{\partial\theta_{\alpha}}
        \CovMatrix^{-1}
        \frac{\partial\CovMatrix}{\partial\theta_{\beta}}
    \right]
    +
    \frac{\partial \B{\mu}^\transpose}{\partial \theta_{\alpha}}
    \CovMatrix^{-1}
    \frac{\partial \B{\mu}}{\partial \theta_{\beta}}
    .
\end{equation}
The first term measures variation of the covariance matrix
with respect to the parameters, which is often
a smaller effect than the variation of the means
represented by the second term.
In the approximation that this first term
may be neglected
the Fisher matrix can be computed in terms of the inverse
covariance matrix for the fiducial model.
Our procedure to obtain this matrix
from the simulation suite
has already been described in Sections~\ref{sec:comparison}
and~\ref{sec:covariance}.

\para{Survey configuration}
The Fisher formalism depends explicitly on details of the survey
under discussion, both through the specification of the
data vector $\bx$---such as how many redshift bins
are used
and
which Fourier configurations are included---and the properties of the covariance
matrix $\CovMatrix$.
In the following we adopt the parameters
of an idealized survey of large-scale structure
consisting of three independent redshift slices
at $z=0$, $z=0.52$ and $z=1$.
Each slice has volume $V = 3.375 \,h^{-3}\,\Gpc^3$
and a mode cutoff at
$\kmax = 0.3 \, h \, \Mpc^{-1}$.
The total Fisher matrix can be written as a sum of
the Fisher matrix in each slice,
\begin{equation}
    \FisherMatrix^{\text{LSS}}_{\alpha\beta}
    =
    \FisherMatrix_{\alpha\beta}(z=0)
    + \FisherMatrix_{\alpha\beta}(z=0.52)
    + \FisherMatrix_{\alpha\beta}(z=1)
    .
\end{equation}
We assume that,
in each redshift bin,
the number density of galaxies is
sufficiently high
that the effect of shot noise is small.
We do not include redshift-space distortions or the effect of
complex survey geometry.
In general, all of these effects will be significant
for a realistic survey and cannot be neglected.
However, in this paper our intention is
to address the question of
whether the proxies described in Section~\ref{sec:estimators}
can be competitive with measurements of the Fourier bispectrum
\emph{in principle}.
Survey-specific effects will generally reduce the number of
configurations that can be measured, or increase the noise
on those for which measurements are possible.
This will typically weaken the performance of the proxies,
meaning that their neglect
gives us an estimate of the best-case scenario.
While we do not anticipate that astrophysical or
observational systematics will affect any one proxy
more than the others, this is an interesting question
to explore in future.

Each of the constraints we present includes a prior from the
cosmic microwave background power spectrum. We implement this
prior by adding a fourth Fisher matrix,
\begin{equation}
    \FisherMatrix^{\text{tot}}
    =
    \FisherMatrix^{\text{LSS}}
    + \FisherMatrix^{\text{CMB}}
    .
\end{equation}
Details of the computation of $\FisherMatrix^{\text{CMB}}$ for our choice
of fiducial parameters were given by~\citet{Smith:2012uz}.

\subsection{Constraining power of the bispectrum and its proxies}
\label{sec:information_content}

In this section we present our forecasts.
To minimize modelling errors
we construct the Fisher matrix for each proxy using
quantities measured from simulation,
except for derivatives with respect to the bias parameters
which cannot be obtained in this way.
For the Fourier bispectrum
we compute these derivatives analytically
by differentiating
the one-loop power spectrum~\eqref{eq:models.Pg-1loop}
and the tree-level bispectrum~\eqref{eq:models.Bg-tree}.
Once the derivatives have been obtained
we replace occurrences of the dark matter power spectrum and
bispectrum with their measured values.
Our prescription for the proxies is similar,
using the one-loop power spectrum
to estimate derivatives of $P(k)$
and tree-level formulae together with the
formulae of Section~\ref{sec:modelling}
to estimate derivatives of the proxy.

\begin{figure}
\centering
\includegraphics{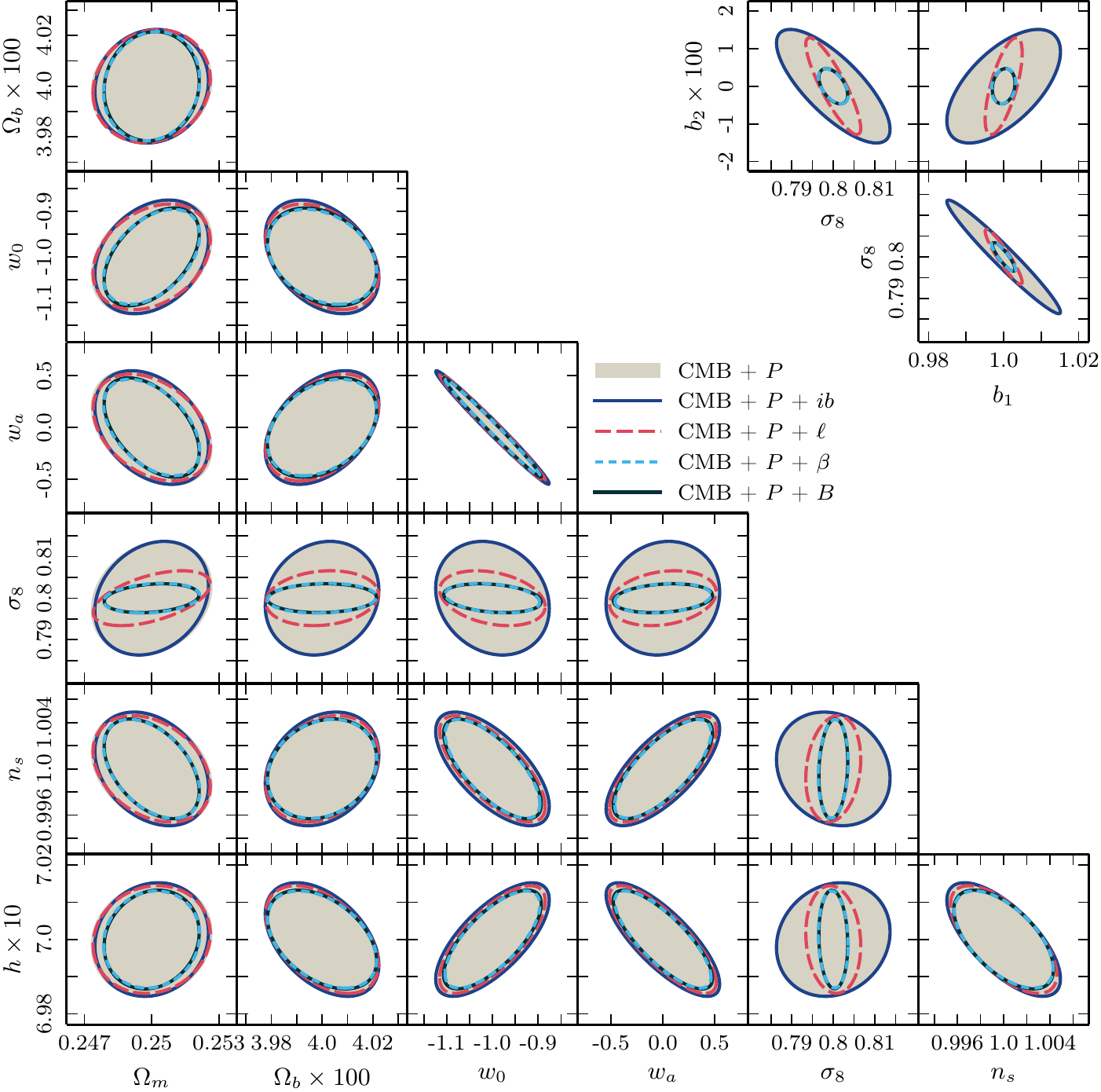}
\caption{Comparison of marginalized $1\sigma$ likelihood contours forecast
    from a combination of the power spectrum and one of the following 3-point
    measures: integrated bispectrum (dark blue, solid), line correlation function
    (red, long-dashed), modal decomposition (light blue, short-dashed) and
    Fourier bispectrum (black, solid). The grey shaded regions show the error
    ellipses for the power spectrum alone. All forecasts include priors from a
    Planck-like CMB experiment and use a cut-off scale $\kmax = 0.3 \,h\,\Mpc^{-1}$.
    The covariance matrices and parameter derivatives for the Fisher forecasts shown
    here are all derived from our simulation results in Section~\ref{sec:comparison}.}
    \label{fig:fisher_forecast}
\end{figure}

We plot the forecast $1\sigma$ confidence contours in Fig.~\ref{fig:fisher_forecast}.
Each panel shows predicted joint constraints for
a pair of parameters
after marginalizing over all the others.
The grey shaded region marks the constraint predicted from
measurements of the power spectrum
only, except for inclusion of the CMB prior that we apply to all estimates.
The solid dark-blue line marks the constraint predicted from $P+\ib$;
the long-dashed red line marks the constraint predicted from $P+\LCF$;
the short-dashed light-blue line marks the constraint predicted from $P+\beta$;
and the solid black line marks the constraint predicted from $P+B$.
We summarize the marginalized $1\sigma$ error bars in Table~\ref{tab:fisher}.
The value in parentheses following each uncertainty
indicates the percentage improvement compared to use of $P(k)$ alone.

\para{Improvement from three-point correlation data}
First consider the joint constraints from $P+B$
(solid black lines in Fig.~\ref{fig:fisher_forecast}).
These demonstrate that substantial improvements can be achieved compared
to measurement of the power spectrum only. This is especially evident for
$\sigma_8$ and the two bias parameters, for which the improvement is
roughly $70\%$--$80\%$; compare the second column of Table~\ref{tab:fisher}.
This is perhaps unsurprising: the bispectrum constrains a different combination
of $\sigma_8$ and $b_1$ than the power spectrum, and therefore assists in breaking
their degeneracy~\citep{Fry1994,Matarrese:1997sk}.
Nevertheless, other parameters that do not participate in this degeneracy
also experience improvements in the range $13\%$--$22\%$, with the exception of
$\Omega_b$.
This is already very well-measured by the CMB prior, and large-scale structure
measurements can add little
new information.
These conclusions are similar to those reported by~\citet{Sefusatti:2006pa},
who suggested that inclusion
of Fourier bispectrum measurements
could reduce uncertainties on $\Omega_m$ and $\sigma_8$
by a factor in the range $1.5$ to $2$.

Next, the forecast for the integrated bispectrum (solid dark-blue lines)
shows that it offers negligible improvement,
of order $\sim 2\%$,
in comparison to $P$ alone.
This is consistent with the very small dependence on cosmological parameters
discussed in Section~\ref{sec:derivs}, and the low signal-to-noise
obtained in Section~\ref{sec:covariance}.
On the other hand, the line correlation function offers comparable constraints
to the Fourier bispectrum for $\sigma_8$ and $b_1$, which receive
improvements of $53\%$ and $68\%$, respectively.
\citet{Eggemeier:2016asq}
demonstrated that this occurs because the line correlation function
is nearly independent of $b_1$ and therefore probes a different direction
in parameter space than $P$ or $B$.
Also, inclusion of $\LCF$ measurements
increases sensitivity to the dark energy parameters $w_0$ and $w_a$ by
$\sim 9\%$.
These improvements are only marginally degraded
compared to those from $P+B$, which are of order $15\%$.

Finally, Fig.~\ref{fig:fisher_forecast} demonstrates that the modal bispectrum with
$\nmax = 50$
(short-dashed light-blue lines)
is predicted to yield error bars nearly equivalent to the Fourier bispectrum
with $95$ triangles.
Note especially that there is no sign of the significant difference in
constraining power between $B$ and $\Bphase$---%
which is the quantity implicitly measured by $\beta$ with our choice of basis---%
that was suggested by
our analysis of signal-to-noise in Section~\ref{sec:covariance}.
We return to this apparent discrepancy in Section~\ref{sec:signoiseProxy}
below.
Just as important,
the differences between the cases
$\nmax = 10$ and $\nmax = 50$ are mostly negligible.
Therefore, even with as few as $\nmax = 10$ modes, the modal
decomposition retains nearly the full constraining power of the
bispectrum.
However, it should be remembered that Fig.~\ref{fig:bspecrecon}
suggests the Fourier bispectrum reconstructed with
so few modes will introduce
more significant scatter.
In a realistic analysis, these reconstruction errors could manifest themselves
as a bias on the best-fit cosmological parameters.
Unfortunately we cannot account for this bias in our Fisher analysis, but
it deserves further investigation.

\begin{table*}
  \centering
  \caption{Marginalized $1\sigma$ parameter uncertainties for the power spectrum and its
    combination with a 3-point correlation measure,
    including CMB priors. All quoted values are derived from the measured
    covariance matrices and parameter derivatives with $\kmax = 0.3\,h\,\Mpc^{-1}$.
    The percentages in parentheses refer to the improvement over the
    $P$-only results.}
  \begin{tabularx}{\textwidth}{cllXlXlXlXlX}
    \toprule
    & \multicolumn{1}{>{\hsize=1\hsize\centering}X}{\multirow{2}{*}{$P$}} & \multicolumn{2}{>{\hsize=2\hsize\centering}X}{\multirow{2}{*}{$P+B$}} & \multicolumn{2}{>{\hsize=2\hsize\centering}X}{$P+\beta$} & \multicolumn{2}{>{\hsize=2\hsize\centering}X}{$P+\beta$} &
      \multicolumn{2}{>{\hsize=2\hsize\centering}X}{\multirow{2}{*}{$P+\LCF$}} & \multicolumn{2}{>{\hsize=2\hsize\centering}X}{\multirow{2}{*}{$P+\ib$}} \\
        & & & & \multicolumn{2}{>{\hsize=2\hsize\centering}X}{$\nmax = 50$} & \multicolumn{2}{>{\hsize=2\hsize\centering}X}{$\nmax = 10$} & & &
    \\
    \cmidrule(r){2-2}
    \cmidrule(lr){3-4}
    \cmidrule(lr){5-6}
    \cmidrule(lr){7-8}
    \cmidrule(lr){9-10}
    \cmidrule(l){11-12}
        $\Omega_m$ & $0.00179$ & $0.00140$ & $(22\%)$ & $0.00141$ & $(21\%)$ & $0.00144$ & $(19\%)$ & $0.00172$ & $(4\%)$ & $0.00167$ & $(7\%)$ \\
        $\Omega_b$ & $0.00015$ & $0.00014$ & $(5\%)$ & $0.00014$ & $(5\%)$ & $0.00014$ & $(4\%)$ & $0.00015$ & $(2\%)$ & $0.00015$ & $(1\%)$ \\
        $w_0$ & $0.084$ & $0.070$ & $(16\%)$ & $0.068$ & $(19\%)$ & $0.069$ & $(17\%)$ & $0.076$ & $(9\%)$ & $0.082$ & $(2\%)$ \\
        $w_a$ & $0.370$ & $0.315$ & $(15\%)$ & $0.306$ & $(17\%)$ & $0.310$ & $(16\%)$ & $0.338$ & $(9\%)$ & $0.360$ & $(3\%)$ \\
        $\sigma_8$ & $0.0092$ & $0.0023$ & $(75\%)$ & $0.0024$ & $(74\%)$ & $0.0025$ & $(73\%)$ & $0.0043$ & $(53\%)$ & $0.0090$ & $(2\%)$ \\
        $n_s$ & $0.00327$ & $0.00284$ & $(13\%)$ & $0.00281$ & $(14\%)$ & $0.00284$ & $(13\%)$ & $0.00303$ & $(7\%)$ & $0.00323$ & $(1\%)$ \\
        $h$ & $0.00103$ & $0.00087$ & $(15\%)$ & $0.00086$ & $(16\%)$ & $0.00087$ & $(15\%)$ & $0.00095$ & $(7\%)$ & $0.00101$ & $(2\%)$ \\
        $b_1$ & $0.0103$ & $0.0020$ & $(81\%)$ & $0.0021$ & $(79\%)$ & $0.0022$ & $(79\%)$ & $0.0032$ & $(68\%)$ & $0.0100$ & $(3\%)$ \\
        $b_2$ & $0.0100$ & $0.0031$ & $(69\%)$ & $0.0031$ & $(69\%)$ & $0.0031$ & $(69\%)$ & $0.0085$ & $(15\%)$ & $0.0100$ & $(1\%)$
    \\
    \bottomrule
  \end{tabularx}
  \label{tab:fisher}
\end{table*}

\para{Combination with other observables}
The strong degeneracy between $\sigma_8$ and $b_1$ can be broken by
other means.
For example, it is possible to use
weak lensing measurements that probe the matter power spectrum directly.
Given that inclusion of 3-point correlation data yields the largest improvements
for $\sigma_8$ and the bias, it is worthwhile considering what improvements
should be expected were the bias to be fixed by other cosmological observations.

In a scenario of this kind the power spectrum constraints
would not be weakened by marginalization over the bias parameters,
and
therefore inclusion of 3-point correlation data would no longer yield such
a dramatic improvement for $\sigma_8$.
However, we still find encouraging improvements for many parameters.
For example, inclusion of either Fourier or modal bispectrum measurements
would decrease uncertainty on $\sigma_8$ by $\sim 25\%$ and all other parameters
except $\Omega_b$ by $10\%$--$15\%$.
Inclusion of $\LCF$ measurements would decrease uncertainty on $\sigma_8$ by
$20\%$, on the dark energy parameters by $\sim 10\%$, and for all other parameters
by $\lesssim 5\%$.
We conclude that, even in the extreme case that $b_1$ and $b_2$ can somehow be
determined exactly, inclusion of 3-point correlation data still provides
valuable additional information.

These Fisher forecasts should be interpreted with some care.
As explained above,
we do not include a number of astrophysical and observational effects that
complicate the analysis of realistic galaxy survey data.
These include redshift uncertainties, redshift-space distortions, irregular
survey geometries and shot noise.
In particular, for the forecasts presented here the effective shot noise is
set by the number density
$\bar{n} = 0.125 \, h^3 \, \Mpc^{-3}$
of particles in our simulation suite.
This is substantially larger than the galaxy number densities that will
be achieved by upcoming surveys.
We return to this issue in Section~\ref{sec:shotnoise}, where we discuss
how our predictions would be modified by
a more realistic number density.

\subsection{Effect of non-Gaussian covariance and cross-covariance}
\label{sec:ng-cov}

The non-Gaussian covariance measured in simulations differs from the
Gaussian approximation in two ways: (1) it includes
additional contributions to the variance of each bin from higher-order correlations,
and (2) it adds or enhances coupling between different bins of a single proxy,
and between bins of different proxies.
These non-Gaussian corrections generally lead to weaker parameter constraints
when compared to forecasts constructed using the Gaussian approximation,
because this assumes that every bin contributes independent information.
In this section we compare the relative impact of non-Gaussian
covariance for the different proxies
by contrasting Fisher forecasts made with and without its inclusion.
We give results for the combinations
$P+\ib$, $P+\LCF$, $P+\beta$ and $P+B$ and each choice
of theoretical model---tree-level SPT, 1-loop SPT, or the halo model.

\begin{figure}
\centering
\includegraphics[width=\textwidth]{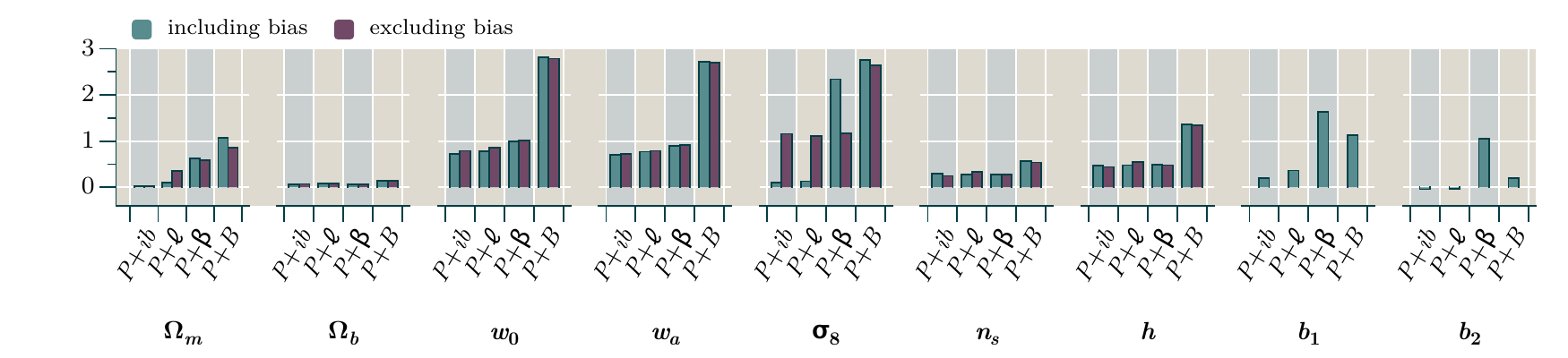}
\caption{Increase in parameter uncertainties from non-Gaussian covariances, measured using
    $\sigma_{NG}/\sigma_{G}-1$, where $\sigma_{NG}$ ($\sigma_{G}$) is the predicted
    error bar
    using the non-Gaussian (Gaussian-like) covariance from simulations.
    Predictions that include (do not include) a marginalization over the
    bias parameters are in blue-green (purple).}
\label{fig:fisher-ngcov}
\end{figure}

\para{Increase in uncertainty from non-Gaussian contributions}
Fig.~\ref{fig:fisher-ngcov} shows the relative increase
$\sigma_{NG}/\sigma_G -1$
in predicted uncertainty
for each parameter when non-Gaussian contributions are included.
To estimate $\sigma_G$
we use the expressions for Gaussian covariance given in
Section~\ref{sec:modelling}
with each quantity replaced by its value measured from our simulations.
For example, to construct the Gaussian covariance for $\ib$
we use equation~\eqref{eq:ib_covg} with $\sigma_L^2$ replaced by
its measured value.
We could equally well have constructed similar estimates using
one of the theoretical models to calculate such
values, but the result is not very different.
The discussion in this section would continue to apply
if we were to reproduce Fig.~\ref{fig:fisher-ngcov}
using estimates generated by any of these prescriptions.

The increase in uncertainty induced by inclusion of non-Gaussian effects
depends on the measure of 3-point correlations used to generate
constraints, the method used to estimate
the Gaussian covariance matrix, and the parameter in question.
In general we find that the Gaussian approximation
underpredicts the uncertainty for the Fourier bispectrum
more strongly than for its proxies.
Note also that---although $P+\beta$ and $P+B$ yield nearly identical
constraints when the non-Gaussian covariance is used,
as described in Section~\ref{sec:information_content}---the
importance
of the non-Gaussian covariance for these combinations is not the same.
Since the quantity $\Bphase$ measured by $\beta$ is not the same as
$B$, neglecting cross-covariance with $P$ (as the Gaussian
covariance does) will leave out different information
for $P+\beta$ compared to $P+B$.

Inclusion of non-Gaussian covariance impacts  uncertainties for $w_0$, $w_a$ and $\sigma_8$
more significantly than the other parameters.
This non-uniformity means that it is not obvious how
inclusion of non-Gaussian covariance might impact constraints
from 3-point correlations
on further parameters not considered here. 
For instance, a number of authors have used Gaussian
covariances to forecast future constraints
on a primordial bispectrum generated by inflation;
see \citet{Scoccimarro:2003wn}, \citet{Sefusatti:2007ih},
\citet{Sefusatti:2011gt}, \citet{Baldauf:2016sjb}, \citet{Welling:2016dng}
and \citet{Tellarini:2016sgp}. It is not yet clear how these forecasts will change when
more realistic non-Gaussian covariances are used.

\begin{figure}
\centering
\includegraphics[width=\textwidth]{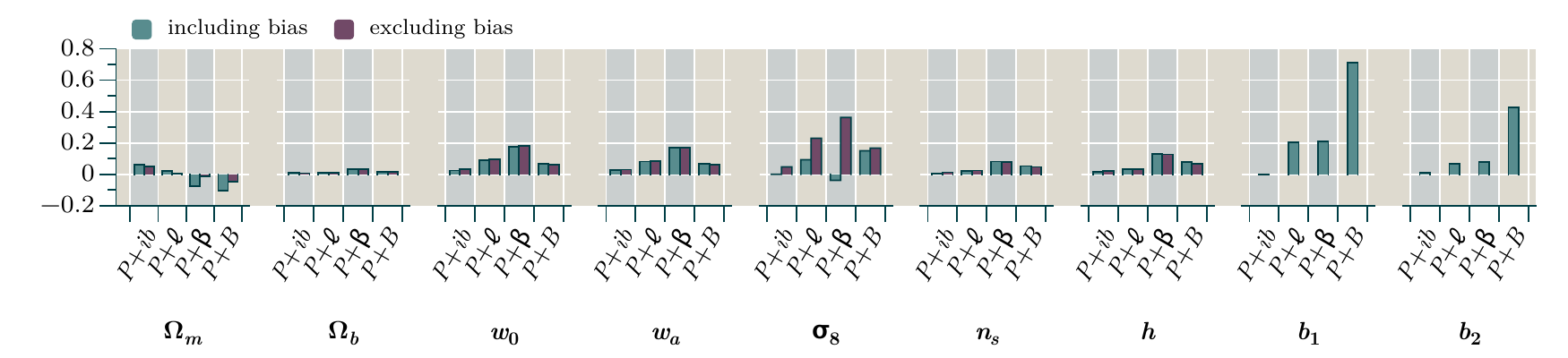}
\caption{Improvement in parameter uncertainties
    from the inclusion of cross-covariance, measured using
    $\sigma_{NG-\text{no}-CC}/\sigma_{NG-\text{with}-CC}-1$,
    where $\sigma_{NG-\text{with}-CC}$ is the error bar predicted using
    non-Gaussian covariance measured from simulations and
    $\sigma_{NG-\text{no}-CC}$ is the error bar predicted
    from the same covariance matrix,
    except with cross-covariances between $P$ and each 3-point statistic set to zero.
    Predictions that include (do not include) a marginalization over the bias
    parameters are in blue-green (purple).}
\label{fig:fisher-ngcov-cc}
\end{figure}

\para{Inclusion of cross-covariance}
In Fig.~\ref{fig:fisher-ngcov-cc} we summarize the influence of cross-covariance
between $P$ and the 3-point measures by comparing constraints using the full
non-Gaussian covariance to constraints where the cross-covariance has been set to zero. 
We find that inclusion of cross-covariances \emph{reduces} the predicted
uncertainties for nearly all parameters and choices of combination $P+X$,
whether or not we marginalize over galaxy bias. 
In the few cases where inclusion of cross-covariance did not reduce the uncertainties
(e.g. constraints on $\Omega_m$ from $P+B$ and $P+\beta$), the predicted error bar
is weakened by less than 12\% of the error bar without cross-covariance.
Overall, we find that ignoring cross-covariances can overestimate uncertainties
by up to $\sim 40\%$ when we do not marginalize over the bias, and by $40-70\%$ for
the special case of bispectrum constraints on the bias parameters themselves.

This reduction of uncertainties due to inclusion of cross-covariances
may be surprising.
While we have not explicitly identified the source of the improved
constraining power, this is not a new feature of Fisher forecasts
using non-Gaussian covariances.
For example, a number of authors
using cross-correlations between cluster counts, weak lensing power
spectra
and the weak lensing bispectrum
have found that parameter constraints can improve when cross-covariances
between strongly-coupled measurements are
included~\citep{Takada:2007fq,Sato:2013mq,Kayo:2012nm}.
But it is also possible that our improvements are partly due to the
galaxy biasing model we have chosen.
A simulation of halos, rather than dark matter alone, could be used to verify
the effect when simultaneously constraining both cosmological parameters and galaxy bias.

The conclusion of this discussion is that an accurate estimate for the
covariance matrix, including non-Gaussian contributions and off-diagonal terms,
is important if we wish to obtain reliable constraints.
Unfortunately, this is especially true for the Fourier bispectrum for which
the Gaussian approximation most significantly underestimates
the true parameter uncertainties.
This implies that surveys aiming to generate constraints from inclusion of $B$ measurements
cannot evade the computational difficulties
associated with estimating their covariance matrix.

To mitigate these difficulties we could consider use of $P+\beta$ rather than
$P+B$. As we have seen in Section~\ref{sec:information_content},
these combinations yield nearly equivalent constraints using $95$ Fourier
configurations and $50$ modal coefficients respectively,
and therefore the modal decomposition makes the information content of the
bispectrum more accessible by reducing the size of the covariance matrix
needed to obtain it.
We consider the efficiency with which each proxy can compress
the information carried by $B$ in Section~\ref{sec:sufficient-statistic}.

\subsection{Theory-dependence of the forecasts}
\label{sec:theory-dep}

\begin{figure}
\centering
\includegraphics[width=0.8\textwidth]{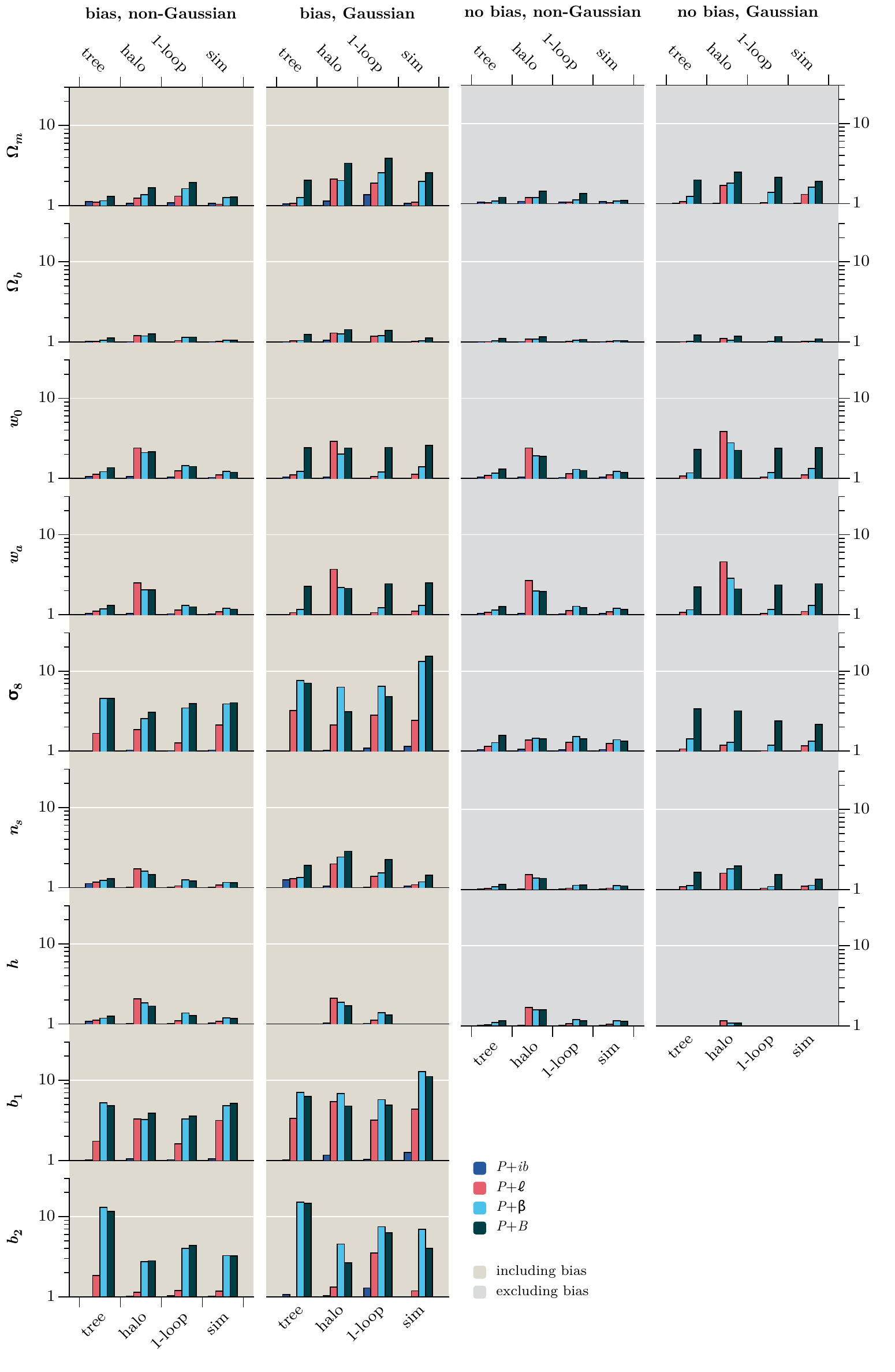}
\caption{Improvements in parameter uncertainties
    from the addition of 3-point statistics are shown as bars with
    height $\sigma(P)/\sigma(P+X)$, the ratio of parameter errors from $P$ only and
    the combination $P+X$. The labels at the top of each column indicate
    whether bias parameters are included or excluded,
    and whether Gaussian or non-Gaussian covariances are used.
    Each group of four bars corresponds to a different choice
    of theoretical model, and the colour of each bar indicates
    the $P+X$ combination. We note that, since the tree-level
    power spectrum does not depend on $b_2$, for the two
    tree-level bar groups in the last row, the bar heights measure
    $\sigmamax/\sigma(P+X)$, where $\sigmamax$ is the maximum error on
    $b_2$ among the four $\sigma(P+X)$ values.
}
\label{fig:fisher-biggrid}
\end{figure}

In Section~\ref{sec:information_content}
we have presented our Fisher forecasts based on simulated data,
and in Section~\ref{sec:ng-cov}
we have discussed the influence of non-Gaussian covariance
and cross-covariances.
These results enable us to assess the information content
carried by the Fourier bispectrum and its proxies, but the
question of how easily these statistics can be deployed
remains open. In particular, we would like to know
whether the use of simulated data is essential, or whether any of the
models described in Section~\ref{sec:modelling}
are sufficient.
In this section we study the dependence of our forecasts
on the choice of theoretical model used to
estimate the derivatives
$\partial \B{\mu}/\partial \theta_\alpha$
in equation~\eqref{eq:paramEstim.fisherG}.

\para{Match to forecast from simulations}
First, we consider whether there is a model that provides
a clear best-match to the forecast using simulated data.
Fig.~\ref{fig:fisher-biggrid} compares the forecasts for each parameter
using different prescriptions for the covariance matrix
and for different choices of
theoretical model, with marginalization over the bias
included or excluded.
The bar heights represent the reduction in
the predicted uncertainty
provided by a given combination, relative to the base
model of power spectrum data only combined with a CMB prior.
The results of Section~\ref{sec:information_content}
are labelled `sim'.
Unfortunately, for each combination $P+X$ there is no single choice
of theoretical model yielding forecasts
that provide the best match to the `sim' outcome
for all parameters---with or without marginalization over bias.

For example, consider the combination $P+B$ in the first column
of Fig.~\ref{fig:fisher-biggrid}. This summarizes forecasts
generated by
including
non-Gaussian covariance and marginalization over the bias.
For $\sigma_8$ it is 1-loop SPT that gives the best match to the
`sim' result, but for the linear bias parameter $b_1$
the best match comes from tree-level SPT.

Alternatively, one could ask whether any one model provides
uniformly conservative or uniformly optimistic forecasts.
If so, that model could be used to estimate upper or
lower limits on the uncertainty for any chosen parameter.
But Fig.~\ref{fig:fisher-biggrid}
demonstrates that there are no models with such properties.
For example, focusing again on the first column, there is no
single choice of theoretical model for $P+B$ that
forecasts the largest or smallest improvement
for all parameters.

\begin{figure}
\centering
\includegraphics[width=0.55\textwidth]{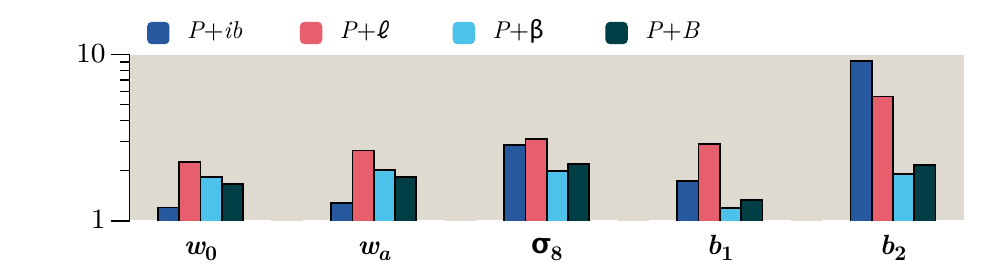}
\caption{Sensitivity factors, defined as the ratio between the largest and
    smallest forecast parameter uncertainty among the three theoretical models,
    for each $P+X$ combination.
    The forecasts compared here include bias parameters and use
    non-Gaussian covariances.
}
\label{fig:fisher-theory-sens}
\end{figure}

\begin{figure}
\centering
\includegraphics[width=\textwidth]{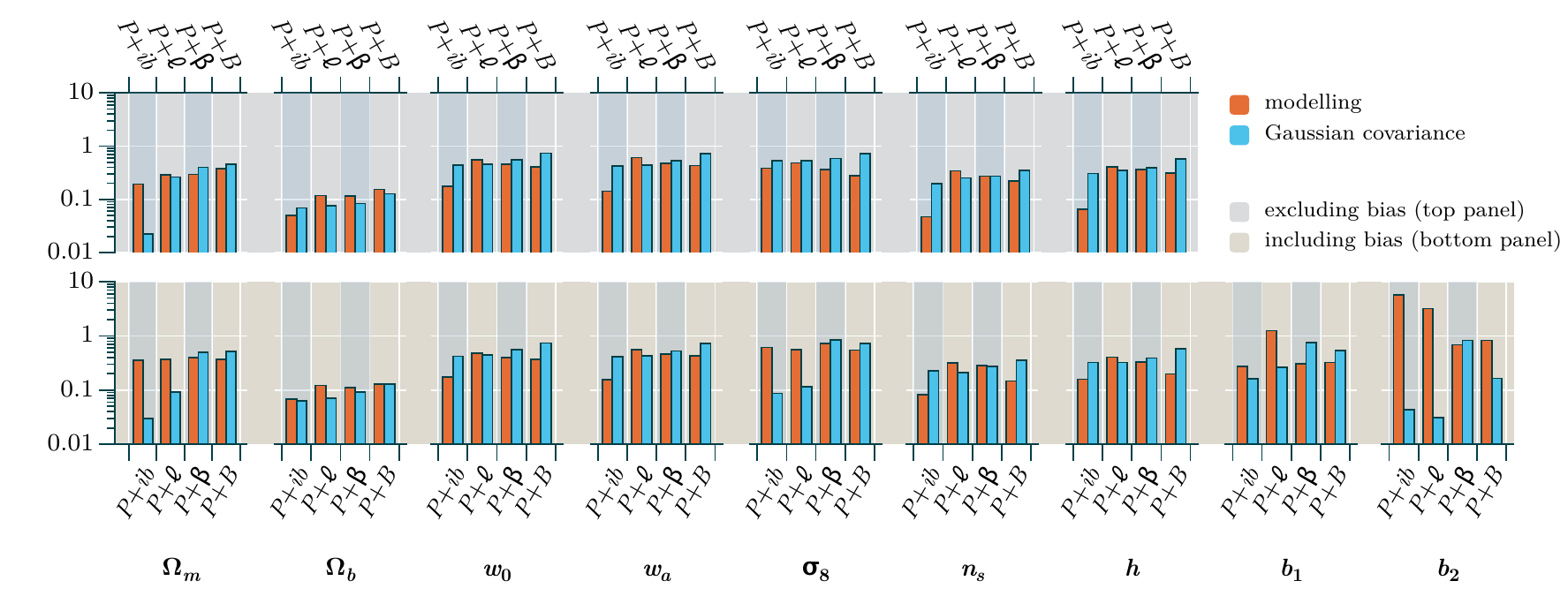}
\caption{Fractional difference in predicted uncertainties induced by theoretical modelling
    of derivatives (orange) or by using a Gaussian approximation to the covariance (blue).
    (See text for details of how the fractional differences are defined.)}
\label{fig:fisher-ngcov-vs-theory}
\end{figure}

\para{Sensitivity to theory error}
Next, we study the variation in forecasts for the Fourier bispectrum and its proxies
when we change the model used to compute $\partial \B{\mu} / \partial \theta_\alpha$.
To understand the sophistication required to obtain
accurate models we will
need to understand
which of these statistics (if any) are especially sensitive or immune
to theoretical mis-modelling.
We measure this dependence by a \emph{sensitivity factor}, which we define
to be the ratio between the largest and smallest forecast uncertainties
taken over the models of Section~\ref{sec:modelling}.
A sensitivity factor close to unity indicates that a forecast uncertainty
depends only weakly on the choice of theoretical model,
while a large value indicates that the model has a strong influence
on the final outcome.

We plot these sensitivity factors in Fig.~\ref{fig:fisher-theory-sens},
computed with inclusion of all bias parameters and using
non-Gaussian covariances.
Therefore the sensitivity factor solely reflects the variation in
uncertainty produced by different choices for theoretical model.
We conclude that there is no single measure of 3-point correlations that
consistently yields the largest or smallest sensitivity to variations in
modelling.
Therefore, there is apparently no single combination $P+X$ that should be
preferred to minimize the effect of theory errors on
inferred parameter constraints.

\para{Ranking by constraining power}
Neither of these criteria provide a rationale to prefer a
choice of theoretical model.
Nevertheless, we do find some general trends.
Irrespective of theoretical model, we find the largest
reductions in parameter uncertainties when the bias
is constrained simultaneously with the
cosmological parameters.
Also, the Fourier bispectrum and modal bispectrum
consistently offer the most significant improvements
compared to $P$-only measurements,
with very similar predicted uncertainties.
The line correlation function achieves
moderate improvement compared to $P$-only,
while the integrated bispectrum
has very weak constraining power---at least for the
parameter set we consider.
We conclude that $P+B$ or $P+\beta$ should be preferred
for constraints on {\LambdaCDM} parameters, with $P+\beta$
offering similar information at reduced computation cost
as discussed at the end of Section~\ref{sec:ng-cov}.

\para{Relative importance of modelling and non-Gaussian covariance}
Finally, we consider the relative importance of non-Gaussian covariance
and theoretical modelling for obtaining quantitatively accurate forecasts.
In Fig.~\ref{fig:fisher-ngcov-vs-theory} we show the fractional difference
in Fisher forecasts induced by variation of theoretical model
(orange bars) and use of the Gaussian approximation (blue bars).
To quantify the significance of theoretical modelling we plot
$\max(|\sigma_{NG,i}/\sigma_{NG}(\text{sim}) -1|)$,
where $i \in \{ \text{tree}, \text{1-loop}, \text{halo} \}$.
Therefore larger orange bars reflect more significant
deviation from the simulated forecast due
to theoretical uncertainty.
Meanwhile we quantify the role of the covariance matrix
by plotting  $|\sigma_G(\text{sim})/\sigma_{NG}(\text{sim})-1|$,
so increasing blue bars show that the Gaussian approximation
generates more significant errors in the forecast.

Fig.~\ref{fig:fisher-ngcov-vs-theory} shows that
the impact of theoretical uncertainty for
$P+\beta$ and $P+B$ is generally less significant
than neglect of non-Gaussian covariance,
whether or not we marginalize over the bias.
In contrast, for $P+\LCF$ the effect of modelling
nearly always dominates because of the difficulties
with the halo model discussed in Section~\ref{sec:derivs}.
For $P+\ib$ the non-Gaussian covariance plays an
important role if the bias parameters are not
included, but theoretical modelling
dominates when they are.

On balance, these results indicate that our forecasts
are slightly less sensitive to theory error than to the
approximation of Gaussian covariance.
This could be because the inverse covariance weighting
suppresses contributions from the non-linear regime where the theoretical
predictions are most discrepant.
But the difference is not large:
the average variation in our predicted uncertainties
from $P+B$ and $P+\beta$ 
due to theory modelling is $36\%$, whereas the
variation due to Gaussian covariances is $49\%$.
Therefore, we conclude that both issues must be
addressed in order to obtain quantitatively accurate
results.

\subsection{Signal-to-noise as a proxy for the information content}
\label{sec:signoiseProxy}
It is now necessary to address the question of why
the large discrepancy in uplift between
the signal-to-noise of
$B$ and $\Bphase$ (equivalently $\beta$) observed in Section~\ref{sec:covariance}
did not translate into significant differences
in the forecast for
parameter uncertainties in Section~\ref{sec:information_content}.

Consider a vector of values $S$ combining measures $P$
and $X$ of the 2- and 3-point
correlation data, respectively,
as defined below equation~\eqref{eq:covariance-def}.
For a given parameter $\theta$
the reduction in uncertainty compared to
measurements from $P$ alone can be estimated in the Fisher framework
by
\begin{equation}\label{eq:fisherratio}
    \frac{\FisherMatrix_\theta(S)}{\FisherMatrix_\theta(P)}
    = 
    \sum_{i,j}
        \frac{\partial S_i}{\partial \theta}
        \CovMatrix_{ij}^{-1}
        \frac{\partial S_j}{\partial \theta}
    \Big/
    \sum_{i,j}
        \frac{\partial P_i}{\partial \theta}
        (\CovMatrix^\text{P})_{ij}^{-1}
        \frac{\partial P_j}{\partial \theta}
    .
\end{equation}
To avoid ambiguity we use the notation $\CovMatrix^\text{P}$ to
denote the covariance matrix of the power spectrum \emph{only}.
Meanwhile, the increase in signal-to-noise in the same scenario is given by
\begin{equation}\label{eq:signoiseratio}
    \frac{(\mathcal{S}/\mathcal{N})^2_{S}}{(\mathcal{S}/\mathcal{N})^2_P}
    = 
    \sum_{i,j}
        S_i
        \CovMatrix_{ij}^{-1}
        S_j
    \Big/
    \sum_{i,j}
        P_i
        (\CovMatrix^\text{P})_{ij}^{-1}
        P_j
    .
\end{equation}
The uplift in signal-to-noise is often taken as an approximation to the
reduction in parameter uncertainty, which avoids
the need to compute
$\partial S_i / \partial \theta$.
As we have seen in Section~\ref{sec:derivs}, these derivatives can be rather fragile
and are susceptible to significant errors caused by theory mis-modelling.
Unfortunately, when applied to $S = P+B$ and $S = P+\beta$
our analysis demonstrates that the ratios
$\FisherMatrix_\theta(P+B) / \FisherMatrix_\theta(P)$
and
$\FisherMatrix_\theta(P+\Bphase) / \FisherMatrix_\theta(P)$
are nearly equal, whereas
the same ratios constructed using $\mathcal{S}/\mathcal{N}$ are very discrepant.
Therefore we must conclude that
improvements in signal-to-noise cannot always be
interpreted as a predictor of the improvement
in Fisher information.

\para{Invariance of the Fisher matrix}
First consider the Fisher matrix. Suppose we perform a redefinition
so that $S_i \rightarrow S'_i = S'_i(S_j)$,
where $S'_i$ may be an arbitrary nonlinear function
of the original measurements. For example, the transformation from
$B$ to $\Bphase$ is of this type. The derivative $\partial S_i / \partial \theta_\alpha$
transforms `contravariantly' on its index $i$, in the sense
$\partial S'_i / \partial \theta_\alpha =
\sum_m (\partial S'_i / \partial S_m) (\partial S_m / \partial \theta_\alpha)$.
Meanwhile, the covariance matrix becomes
\begin{equation}
    \CovMatrix^S_{ij} \rightarrow \CovMatrix^{S'}_{ij}
    =
    \langle (S'_i - \bar{S}'_i) (S'_j - \bar{S}'_j) \rangle
    =
    \sum_{m,n}
    \frac{\partial S'_i}{\partial S_m}
    \frac{\partial S'_j}{\partial S_n}
    \CovMatrix^S_{mn}
    +
    \cdots ,
    \label{eq:covmatrix-tensor}
\end{equation}
where `$\cdots$' denotes terms involving
higher order correlations that we have not
written explicitly.
Provided these are small compared to the
$\CovMatrix^S_{mn}$ term, equation~\eqref{eq:covmatrix-tensor}
shows that the covariance matrix also transforms
`contravariantly', and therefore that its inverse
transforms `covariantly'. Subject to these
approximations we conclude that the Fisher matrix
should be roughly \emph{invariant}.
This agrees with our observation that $\FisherMatrix_\theta(P+B)$
and $\FisherMatrix_\theta(P+\Bphase)$ are nearly equal,
demonstrated numerically in Table~\ref{tab:fisher}.

Now consider the signal-to-noise.
Since $S_i$ has neither a co-
or contravariant transformation law, the
combination $\sum_{i,j} S_i \CovMatrix^{-1}_{ij} S_j$
appearing in the signal-to-noise will typically \emph{not}
be invariant. Therefore different choices $S_i$ and $S'_i$ may yield
inequivalent results for $\mathcal{S}/\mathcal{N}$.
For example, we have verified that using $P + \ln B$ predicts
a significant increase in the signal-to-noise compared to 
$P+B$, whereas their Fisher matrices continue to agree. 
In Table~\ref{tab:unmargin_improve}
we summarize the improvement in unmarginalized constraints from the addition of $B$ or $\Bphase$.
This demonstrates that empirically the increase in
signal-to-noise from $\Bphase$ provides a more accurate estimate of the Fisher information than $B$. This property holds for both proxies of $\Bphase$, namely the modal bispectrum, and the line correlation function. 

\para{Gaussian limit}
This outcome
is not inconsistent with the result that $B$ and $\Bphase$
show an equivalent uplift in signal-to-noise in the Gaussian approximation.
In this case the covariance matrix for $\Bphase$ is
$\CovMatrix_{ij}^{B_\epsilon}=\BispectrumDegeneracy \Kronecker_{i j}$,
where the constant $\BispectrumDegeneracy$ takes the values $1$, $2$ or $6$
for scalene, isosceles and equilateral configurations, respectively,
as described in Section~\ref{sec:modelling}.
In the same approximation the covariance matrix for the
Fourier bispectrum is
$\CovMatrix_{ij}^{ B}=\BispectrumDegeneracy P(k_{i_1})P(k_{i_2})P(k_{i_3}) \Kronecker_{i j}$.
Therefore we conclude that the signal-to-noise
for $B$ and $\Bphase$ is identically equal as
\begin{equation}
    B_i (\CovMatrix^B)_{ij}^{-1} B_j
    =
    B_{\epsilon i} (\CovMatrix^{\Bphase})_{ij}^{-1} B_{\epsilon j}
    =
    \frac{1}{\BispectrumDegeneracy}
    \frac{B_i^2 \Kronecker_{ij}}{P(k_{i_1})P(k_{i_2})P(k_{i_3})}
    .
\end{equation}
In the Gaussian approximation the power spectrum is
an independent source of information, which explains the agreement.
However, once off-diagonal contributions in the covariance matrix
are included,
$B$ and $P$ are no longer independent and non-linear combinations
may give very different results for the signal-to-noise.

\para{Comparison with Chan \& Blot}
Our signal-to-noise for $P+B$ differs from that reported by~\citet{Chan:2016ehg}
because we include cross-covariance (Section~\ref{sec:covariance}).
Since empirically the signal-to-noise of $P+\Bphase$ 
gives a more accurate estimate of the information gain from 3-point correlation
data, the $\sim 26\%$ expected improvement from the 3-point information in 
$\Bphase$ is in
good agreement with the $\sim 30\%$ improvement suggested by~\citet{Chan:2016ehg}.
However, the details of these calculations are rather different.
The unmarginalized constraints in Table~\ref{tab:unmargin_improve} and most of
the marginalized constraints in Table~\ref{tab:fisher} support
this conclusion.
For $\sigma_8$, $b_1$ and $b_2$, for which the effect
in Table~\ref{tab:fisher}
is substantially larger than
$\sim 30\%$, we ascribe the improvement
to degeneracies of $P$ that are broken by 3-point correlation data.

\begin{table*}
  \centering
  \caption{Percent improvement of unmarginalized constraints using $P+B$ compared to
    $P$ only at $z=0$.}
  \begin{tabular}{ccccccccc}
      \toprule
        $\Omega_M$ &     $\Omega_B $    &   $ w_0  $   &   $ w_a $  & $ \sigma_8 $   &    $ n_s $     &    $ h$    &     $b_1$    &     $b_2$ \\
        \midrule
		$12.9\%$ &     $19.4\%$ &     $26.0\%$ &     $27.0\%$ &     $26.4\%$ &     $15.1\%$ &     $15.6\%$ &     $42.4\%$ &     $43.4\%$ \\
	\bottomrule
  \end{tabular}
  \label{tab:unmargin_improve}
\end{table*}

\section{Discussion}\label{sec:discussion}

\subsection{Compression and efficiency of the Fourier bispectrum proxies}
\label{sec:sufficient-statistic}

In an ideal survey aiming to measure the Fourier bispectrum we should
clearly choose a bin width $\Delta k$ that is sufficiently small to
reproduce all small-scale features of interest.
However,
because the number of Fourier configurations in a volume with mode cut-off $\kmax$
scales as $\sim (\kmax / \Delta k)^3$ this task will quickly become
computationally expensive. And, as we have emphasized several times,
a more serious problem is that we must estimate and invert the
covariance matrix for all these measurements.
This requires us to perform at least as many
{\Nbody} simulations as the number of configurations that we retain.

In this section we consider how well this large number of Fourier
configurations can be compressed by the proxies described in Section~\ref{sec:estimators}.
Suppose that available resources limit the number of simulations
that can be performed
in such a way that we can estimate an accurate covariance matrix
for $\sim 30$ bins of
the Fourier bispectrum or one of its proxies, in combination with another $30$
measurements of the power spectrum $P(k)$.
Among the measures of 3-point correlations that we consider, is there
a preferred choice that provides optimal constraints on our set of
cosmological parameters?
If so, this measure would provide the most successful compression of
the full Fourier bispectrum into a manageable number of measurements.

\begin{figure}
  \centering
  \includegraphics{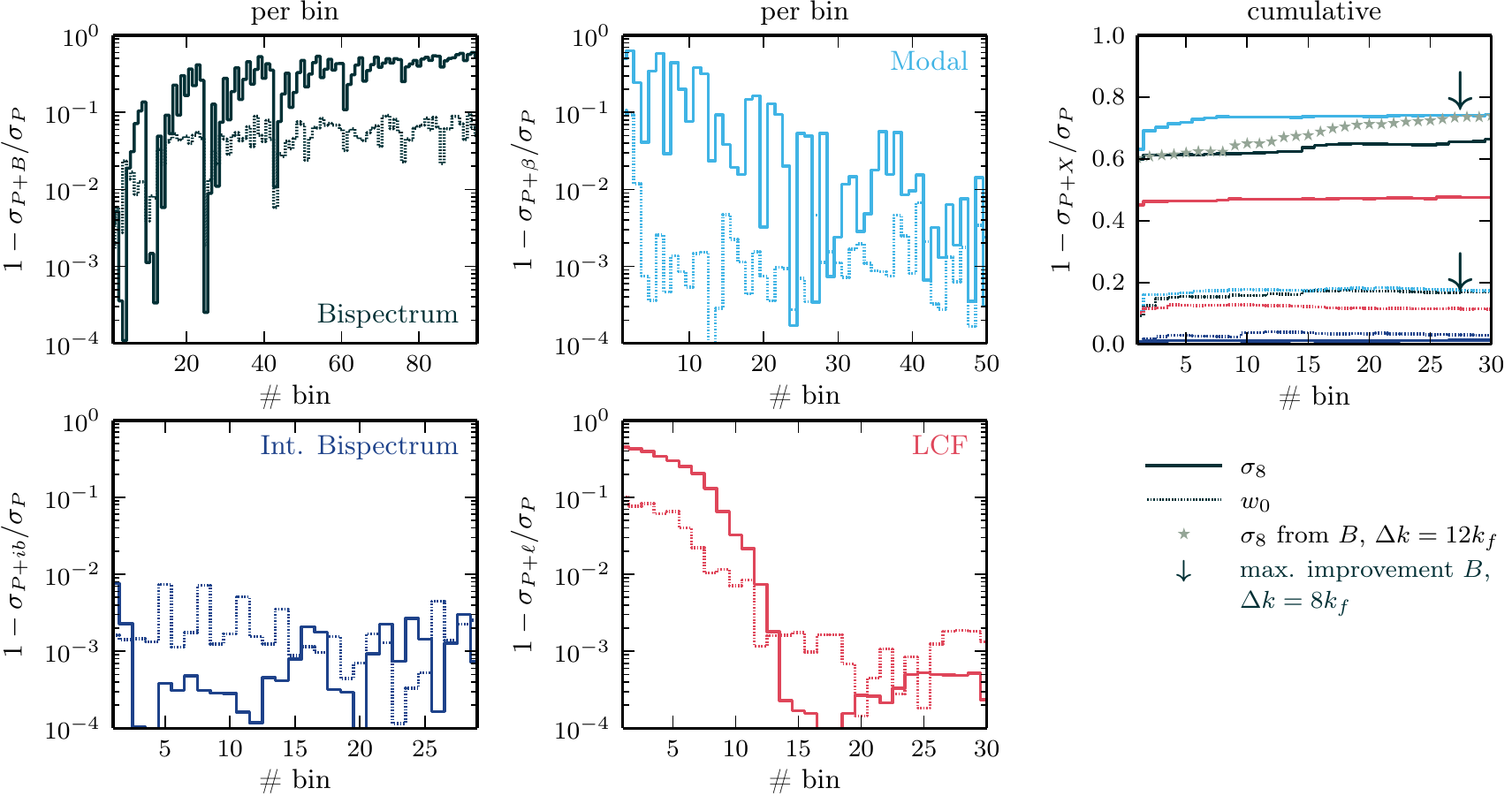}
  \caption{\semibold{First two columns}: decrease in forecast
  parameter uncertainty (improvement) from combining the power
  spectrum with a single bin
  of a 3-point correlation measure,
  compared to the power spectrum alone. The Fisher matrix was
  computed from the non-Gaussian covariance matrix and the
  measured parameter derivatives $\partial\B{\mu} / \partial \theta_\alpha$.
  Solid (dotted) lines show $\sigma_8$ ($w_0$) with all other parameters
  (including bias) marginalized.
  \semibold{Third column}: cumulative improvement
  from adding the $30$ best bins. Arrows indicate the maximal
  improvement obtained from the Fourier
  bispectrum with $\Delta k = 8 \kf$, while stars show the uncertainty for
  $\sigma_8$ using Fourier
  bispectrum measurements with the larger bin width $\Delta k = 12 \kf$. }
  \label{fig:data_compression}
\end{figure}

\para{Compression by reduction to $\leq 30$ bins}
To this end we combine the power spectrum bins with
a single additional configuration from
the Fourier bispectrum or one of its proxies, and compute the corresponding
Fisher matrix (as in Section~\ref{sec:forecasting-method})
using values for $\partial \B{\mu} / \partial \theta_\alpha$ estimated
from our simulation suite.
The four left panels of Fig.~\ref{fig:data_compression}
show the reduction in predicted uncertainty---%
defined as the shrinkage of the error bar, $1-\sigma_{P+X} / \sigma_P$---%
for the representative parameters $\sigma_8$ (solid lines) and $w_0$ (dotted lines)
for each of the possible bins.
Using these reductions as a measure of the information stored in each bin
we conclude that most of the information carried by the Fourier
bispectrum $B$ is contained in small-scale triangles
(towards larger triangle index).
A similar conclusion applies for the line correlation function,
for which significant reductions occur only for the first $\sim 12$ bins,
corresponding to the range of scales $10\,h^{-1}\,\Mpc$ -- $50\,h^{-1}\,\Mpc$.
This is reasonable, because the line correlation is constructed
to give a negligible signal on large scales.
Finally, while the modal decomposition exhibits some variability,
smaller mode numbers typically provide larger gains.
The integrated bispectrum shows consistently weak improvements
over all bins.

Second, for each combination $P+X$ we identify a set of $30$ bins for $X$ that provide
the largest improvements.
Adding them cumulatively to the power spectrum, starting from the bin carrying most information,
we obtain the plot on the right-hand side of Fig.~\ref{fig:data_compression}.
Both the line correlation function and the modal bispectrum converge rapidly to the maximal
improvement available from the entire set of bins that we measure (this is
$30$ bins for $\ell$ and $50$ modes for $\beta$---see Table~\ref{tab:binning}).
For example, the line correlation is already within $2\%$ of the maximum after we have added
$\sim 2$ bins, while only $\sim 5$ modes of $\beta$ are required to arrive at a similar
value for the modal bispectrum.
In comparison the Fourier bispectrum converges much more slowly to the maximum
provided by the $95$ bins that we measure. This is especially evident for $\sigma_8$,
for which the improvement from the Fourier bispectrum has not yet converged to its
maximum value after the $30^{\mathrm{th}}$ bin.
(For guidance, we mark this maximum value with black arrows on the plot.)
However, it should be noted that our procedure to select the set of $30$ bins is not optimal because it
does not account for covariances between them. By 
analysing
random subsets of the $95$ possible bispectrum
bins we find that faster convergence is possible,
giving up to $\sim 90\%$ of the maximum reduction after $30$ bins.

\para{Compression by broadening bins}
Rather than reducing the number of configurations by restriction to a subset,
we might alternatively increase the width of each bin.
The same volume of data would then be compressed into fewer measurements.
To compare the performance of this strategy we repeat the analysis
described above for the Fourier bispectrum with a broader
bin width $\Delta k = 12 \kf$, which gives $34$ rather than $95$ Fourier
configurations with $\kmax = 0.3 \, h \, \Mpc^{-1}$.
We plot the corresponding cumulative reduction in uncertainty
for $\sigma_8$ as star-shaped symbols in the right-hand panel of Fig.~\ref{fig:data_compression}.
After $30$ bins the improvement is similar to that obtained from the modal
bispectrum, with the same caution about rate of convergence due to correlation between
bins.
Therefore---rather surprisingly---in this case we find no clear preference for the bin
width $\Delta k = 8\kf$ or $\Delta k = 12 \kf$, except that $\Delta k = 8\kf$ is more
computationally expensive, and it is more difficult to find an optimal subset of configurations.
However, it is not clear whether this conclusion would survive in a more realistic
analysis, where the signal can be noisy and demands finer binning. To explore these
issues in detail would require a more comprehensive analysis.

\para{Results}
This analysis agrees with the conclusions of Sections~\ref{sec:ng-cov} and~\ref{sec:theory-dep},
and supports the modal bispectrum as a good choice of proxy for 3-point correlation data.
In addition to the advantages discussed in previous sections, it requires the fewest bins
and loses almost no information.

These results could be modified in cases where it is possible to compute a
covariance matrix for $\gg 30$ configurations of the Fourier bispectrum,
as done (for example) by~\citet{Gil-Marin:2016wya}.
However, the mock catalogues used to produce such covariance matrices are often
generated using perturbation theory and therefore are likely to be
inaccurate on small scales.
We expect that it is a better strategy to use fewer bins and obtain high-quality
measurements of the covariance matrix from catalogues generated using
full {\Nbody} simulations.
The significant benefit of the modal decomposition is that it
facilitates construction of the smallest set of bins that still carry a
majority of the information.

Finally, although the line correlation function provides weaker improvements than either
the Fourier bispectrum or modal bispectrum, it has the advantage that
it clearly separates the scales carrying useful information from those that do not---%
all bins with $r \gtrsim 50 \, h^{-1} \, \Mpc$ have negligible impact.
It is also possible that the performance of the line correlation function could be
improved by relaxing the condition of strict collinearity,
which would increase the range of Fourier configurations it is able to aggregate.

\subsection{Shot Noise}
\label{sec:shotnoise}

\begin{figure}
\centering
\includegraphics[width=0.8\textwidth]{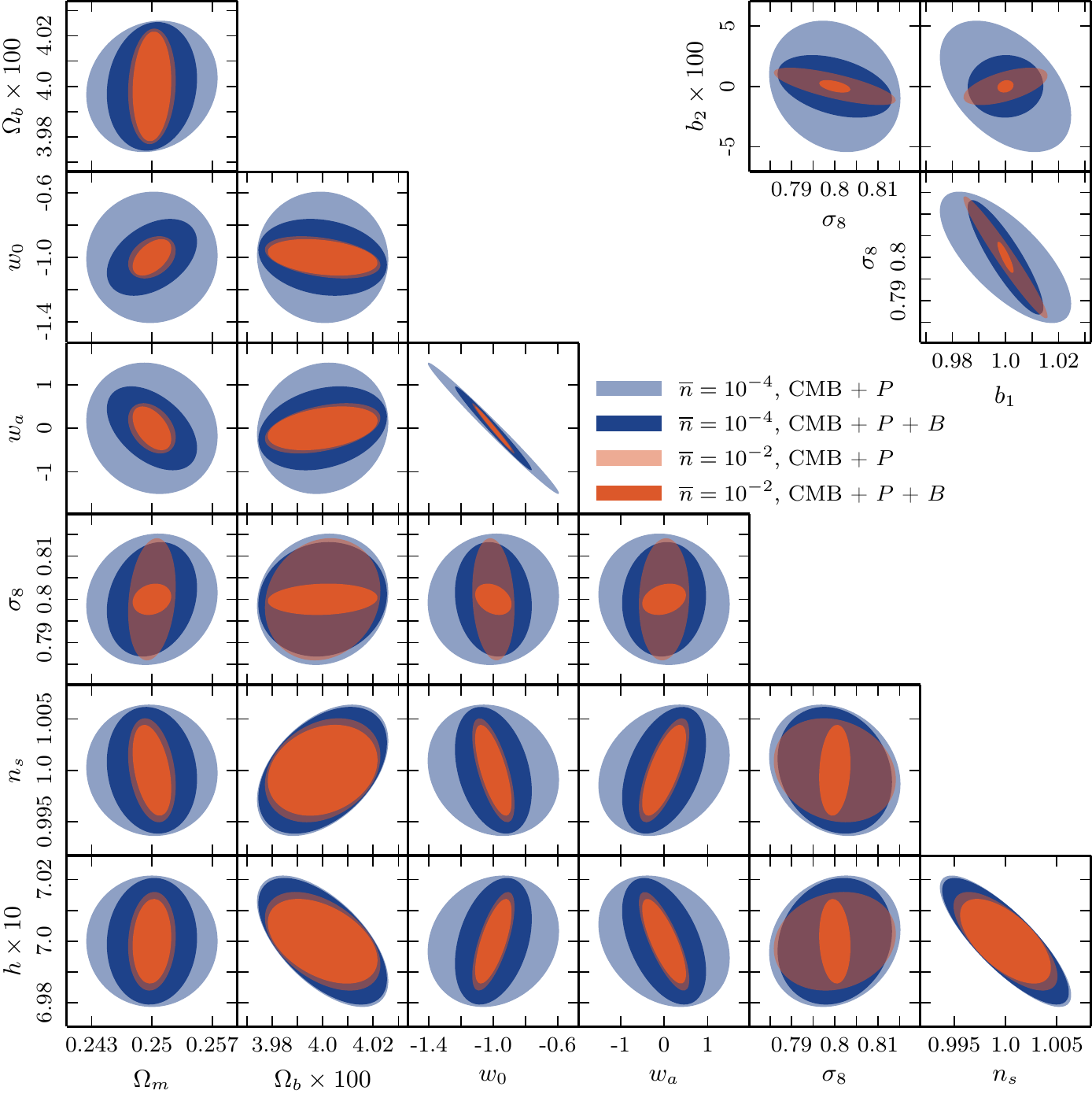}
\caption{Comparison of the Fisher forecasts with shot noise corresponding to $\bar{n}_1 =10^{-2}\,h^3\,\Mpc^{-3}$ (orange) and $\bar{n}_2 = 10^{-4}\,h^3\,\Mpc^{-3}$ (blue). The pale ellipses correspond to uncertainties using the power spectrum only, while the dark
    ellipses show the predicted uncertainty when 3-point correlation
    information is included.
}
\label{fig:fisher-shotnoise}
\end{figure}

Galaxies are discrete, point-like tracers of the underlying matter fluctuations,
and therefore samples of their abundance are affected by shot noise.
This noise is expected to impact
higher-order statistics more significantly than the power
spectrum~\citep{Sefusatti:2004xz,Chan:2016ehg}.
Up to this point our analysis has implicitly used the low effective shot noise provided by
our simulations, and therefore there
is some concern that our forecasts will degrade with larger, more realistic noise.
In this section we perform an approximate analysis of this degradation
and quantify its effect on our predicted parameter uncertainties.

Assuming Poisson statistics,
we may correct for
shot-noise contributions to the observed discrete
power spectrum $\hat{P}^{\text{disc}}$ and bispectrum $\hat{B}^{\text{disc}}$
by subtraction~\citep{Peebles1980,Matarrese:1997sk},
\begin{subequations}
\begin{align}
  \hat{P}(k) & = \hat{P}^{\text{disc}}(k) - \frac{1}{\bar{n}} , \label{eq:Pshot}\\
  \hat{B}(k_1,k_2,k_3) & = \hat{B}^{\text{disc}}(k_1,k_2,k_3) - \frac{1}{\bar{n}}\Big[\hat{P}(k_1)+\hat{P}(k_2)+\hat{P}(k_3)\Big] - \frac{1}{\bar{n}^2}
  .
  \label{eq:Bshot}
\end{align}
\end{subequations}
Here, $\bar{n}$ is the average number density of the discrete tracers. We use the upper and lower limits $\bar{n}_1=10^{-2} \, h^3 \, \Mpc^{-3}$
and $\bar{n}_2 = 10^{-4} \, h^{3}\,\Mpc^{-3}$
to represent optimistic and pessimistic levels of shot noise for upcoming galaxy surveys.
To measure $\hat{P}^{\text{disc}}$ and $\hat{B}^{\text{disc}}$ we downsample the number of particles in our
simulation suite by selecting random subsets matching the desired averaged density $\bar{n}$,
and use this to compute corrected estimators
$\hat{P}$ and $\hat{B}$ from
equations~\eqref{eq:Pshot}~and~\eqref{eq:Bshot}.
Although this downsampling procedure will not introduce exactly Poisson shot noise,
we have checked that it is nearly Poisson by verifying that the corrected quantities
agree with measurements made using the full set of particles to within a few percent.
Strictly speaking, the covariance matrix of $\hat{P}$ and $\hat{B}$ obtained in this way
is the matter covariance with Poisson shot noise, but for our fiducial biasing model
we may interpret it as the covariance of the galaxy power spectrum and
bispectrum with Poisson shot noise.
We use this covariance, leaving the parameter derivatives unchanged from
Section~\ref{sec:information_content},
to compute the Fisher matrices.

We plot forecasts using the fiducial number densities $\bar{n}_1$ and $\bar{n}_2$
in Fig.~\ref{fig:fisher-shotnoise},
with orange ellipses corresponding to the lower noise level
(higher number density)
and blue ellipses corresponding to the higher noise level
(lower number density).
The orange ellipses show good agreement with the forecasts
for the idealized scenario of Section~\ref{sec:information_content},
indicating that relatively little degradation occurs.
However, it is unlikely that such high number densities will
be attained in the near future.
By contrast
the blue ellipses represent a conservative view of what should be possible.

If shot noise degrades the signal from 3-point correlations more strongly
than for 2-point correlations then the fractional improvement from its
inclusion should be smaller for low $\bar{n}$.
In terms of Fig.~\ref{fig:fisher-shotnoise}
this means that the difference between the light and dark
blue ellipses should be smaller
than the difference between the light and dark orange ellipses.
This effect is visible for some parameters, such as $\sigma_8$.
However, in the case of
$\Omega_m$, $w_0$ and $w_a$ the fractional improvement from inclusion of 3-point
correlation data is \emph{larger} at lower $\bar{n}$.
The effect for $w_0$ and $w_a$ is particularly striking.
Using all particles in our simulations, the addition of $B$ data decreased
measurement uncertainties
by $16\%$ and $15\%$, respectively (see Table~\ref{tab:fisher}).
With $\bar{n} = 10^{-4} \, h^3 \, \Mpc^{-3}$
we find improvements of
$41\%$ and $36\%$.
We interpret this to mean that recovery of cosmological information
in the presence of shot noise
depends significantly on cross-covariances between measurements.
These cross-covariances themselves depend on the shot noise and
can partially subtract its effect.

\section{Conclusions}\label{sec:conclusions}
As large scale structure surveys grow in size and sophistication, the
rapidly-approaching cosmic variance limit on 2-point statistics
encourages us to look to higher-order correlations,
such as the 3-point function, as a new source of information.
Previously, \citet{Sefusatti:2006pa}
suggested that considerable additional constraining power
could be achieved by combining the power spectrum and bispectrum.
On the other hand, the signal-to-noise analysis given by~\citet{Chan:2016ehg}
pointed to no more than modest improvements.
Our results show that there is a significant benefit from
inclusion of three-point correlation data,
but its benefits must be balanced against the challenges
it brings.

In this paper, we focus on two particular challenges:
(1) The number of measurable configurations of the Fourier
bispectrum is generally very large unless one
coarse-grains the data.
We have investigated whether the \emph{modal bispectrum},
\emph{line correlation function}
and \emph{integrated bispectrum} can act as `proxies' for
the Fourier bispectrum, compressing its information into
fewer configurations without unacceptable information loss.
(2) Bispectrum observations are difficult to model to the same accuracy as the power spectrum.
Errors in clustering predictions from theoretical models, in addition to
assumptions about covariances and noise properties, generally propagate into
inaccurate error bars or a bias on inferred parameters.
We have quantified how our forecasts are influenced by both
the
assumption of Gaussian covariance and theoretical errors.

To do so
we have measured the power spectrum, Fourier bispectrum and each of its proxies
from a suite of 200 dark matter {\Nbody} simulations at redshifts $z=0$, $z=0.52$ and $z=1$
to obtain fully non-Gaussian covariances and cross-covariances.
We measure the dependence of each measurement on the cosmological parameters
$\{\Omega_m,\Omega_b, w_0, w_a, \sigma_8, n_s, h\}$
using additional simulations displaced from our fiducial model.
We assume an local Lagrangian biasing scheme that includes two bias parameters, $\{b_1,b_2\}$.
Using all these components, in combination with theoretical predictions for each proxy
from tree-level and 1-loop SPT and the halo model, we have conducted a signal-to-noise
analysis and implemented the Fisher forecasting method for an idealized survey scenario.
Our main results on the constraining power and future viability
of each measure of 3-point correlations are:

\para{Comparison of 3-point correlation measures}
Section~\ref{sec:information_content} presented our main results.
Our forecasts show that
inclusion of
the Fourier bispectrum offers significant improvements over the power spectrum alone,
with $\BigO(10\%-30\%)$ improvement on cosmological parameter constraints, and up to
$\BigO(80\%)$ improvement when it is used to break degeneracies
with the bias parameters. 
The \textit{modal bispectrum} offers an attractive alternative,
achieving equivalent constraints with as few as 10 modes.
However, up to 50 modes may be necessary to reconstruct the Fourier
bispectrum to within $\lesssim 10\%$
accuracy on individual triangle configurations.
The \textit{line correlation function} appears to be slightly less optimal,
although a future extension
to sample more Fourier configurations by relaxing the requirement of strict collinearity
may improve its performance.
The \textit{integrated bispectrum} offers little constraining power for our
set of cosmological parameters.
It is sensitive to highly squeezed triangles, whereas the gravitational bispectrum peaks
on equilateral triangles.
This property of $\ib$ is a disadvantage for our purposes, but may be an advantage if one
is interested in
studying squeezed-mode primordial non-Gaussianity with minimal degeneracies.

\para{Data compression}
In Section~\ref{sec:sufficient-statistic}, we explored how the total constraining
power of each
measure is distributed over the total number of data bins.
While the Fourier bispectrum and modal bispectrum
give nearly equivalent parameter constraints when
$\sim 30$ bins are used, the modal method converges to its full constraining power
with a smaller
subset of bins.
We conclude that the modal bispectrum provides more efficient access to the information
carried by 3-point correlations.

We note that more realistic survey
scenarios---for example, accounting for noisy data---may require finer binning.
Increasing the binning resolution of the Fourier bispectrum by a factor of $n$ in
each $k$-dimension
corresponds to a factor $\BigO(n^3)$ increase in configurations.
The number of simulations required to
accurately capture their covariance would increase similarly.
If the number of modal coefficients required to capture fine features of the bispectrum
does not grow so dramatically, it is possible that
the modal bispectrum could accumulate an even larger advantage
compared to the Fourier bispectrum.

\para{Signal-to-noise ratio as a measure of information content}
In Sections~\ref{sec:covariance}, \ref{sec:information_content}
and~\ref{sec:signoiseProxy}
we argue that
use of the signal-to-noise ratio to predict the
constraining power of 3-point correlation
data can be misleading.
We show that
the bispectrum and phase bispectrum---which is probed by the modal
bispectrum---give significantly
different signal-to-noise ratios, but still yield nearly identical forecasts.
As we describe in Section~\ref{sec:signoiseProxy},
for the scenarios considered in this paper,
the improvement shown by these forecasts is empirically
better predicted by the
signal-to-noise ratio of the
phase bispectrum $\Bphase$ than the Fourier bispectrum $B$.
The $\sim \BigO(30\%)$ uplift in signal-to-noise
from the phase bispectrum translates to the same improvement
in cosmological parameter constraints, except for those where
degeneracies play a significant role. 
As we explain in Section~\ref{sec:signoiseProxy}, while this improvement is 
numerically consistent with~\citet{Chan:2016ehg}, our procedure is rather 
different.
For a general parameter set and a given measure of the 3-point correlations, the 
signal-to-noise will not typically give an
accurate estimate of its constraining power.

\para{Impact of non-Gaussian covariances}
Accounting for non-Gaussian covariance is essential for optimally constraining
cosmological parameters. In Section~\ref{sec:ng-cov} we showed that the Fourier
bispectrum estimator is particularly sensitive to the covariance:
our predicted uncertainties may be nearly a factor of 4 too small if the
Gaussian approximation is used.
At the same time, we find that the non-Gaussian cross-covariance between the power
spectrum and the Fourier bispectrum or its proxies
generally results in parameter errors that are $\BigO(10\%)$ \emph{smaller} than
if cross-covariances are ignored.

\para{Impact of theoretical modelling uncertainties}
Our results in Section~\ref{sec:theory-dep} indicate that the impact of theory errors on
our predicted uncertainties
is smaller than the impact of assuming Gaussian covariance, although both approximations
change the forecasts by $\sim 30\%$ to $50\%$ on average.
In this paper we measure the effect of theoretical uncertainty by comparing forecasts using
SPT and the halo model to forecasts derived purely from {\Nbody} measurements.
Our approach differs from that of~\citet{Baldauf:2016sjb} and~\citet{Welling:2016dng},
who incorporated estimates of the theory error into their Fisher forecasts
by taking the error in each data bin to be the sum of statistical and theoretical errors.

\para{Impact of shot noise}
To assess the impact of shot noise, in Section~\ref{sec:shotnoise} we down-sample our
simulation suite to averaged number densities of $\bar{n}=10^{-2}\,h^3\,\Mpc^{-3}$
and $10^{-4}\,h^3\,\Mpc^{-3}$, and compute forecasts using non-Gaussian covariance
matrices that include low and high levels of Poisson shot noise.
Contrary to na\"{\i}ve expectations, we find that the addition of 3-point
correlation information can become \emph{more} significant at high levels of shot noise
owing to the non-trivial dependence of the cross-covariance on $\bar{n}$.
This appears most significant for the dark energy parameters $w_0$ and $w_a$, and suggests
that 3-point correlation information may be crucial to distinguish between dark energy models.
More generally, our result implies that 3-point correlation
measurements may yield significant additional constraining power even when shot noise levels
are high.

\medskip\noindent
To make robust inferences with 3-point correlation information, future surveys will require
refinement of the methods we have considered here.
For example, while we have demonstrated that the modal decomposition provides efficient
data compression of the matter bispectrum in an idealized survey, it will be important to
verify that this remains true when halo distributions, redshift-space distortions and the
complex noise properties of realistic surveys are introduced.
We have emphasized the importance of including non-Gaussian covariances and
theory uncertainties in our forecasts.
Realistic analyses will likely require more efficient ways to obtain covariances,
and a consistent approach to inclusion of theory errors in software pipelines.
Achieving each of these aims will be important milestones ahead of upcoming
surveys of
large-scale structure. 

\section*{Acknowledgements}\label{sec:acknow}

The work reported in this paper has been supported by the
European Research Council under the European Union's Seventh Framework Programme
(FP/2007--2013) and ERC Grant Agreement No. 308082 (JB, DR, DS).
This work was supported by the Science and Technology Facilities
Council [grant numbers ST/L000652/1, ST/P000525/1] (DS, RES).
AE acknowledges support from the UK Science and Technology
Facilities Council via Research Training Grant
[grant number ST/M503836/1], and thanks Roman Scoccimarro and the
Physics Department of New York University for hospitality
during the final phases of this project.
DR acknowledges useful conversations on the normalization of the
modal decomposition with Hemant Shukla.
JB would like to thank Benjamin Joachimi for useful discussions.

\para{Data availability statement}
To assist those wishing to replicate or extend our results, we have
made available measurements of
the power spectrum, bispectrum, integrated bispectrum,
line correlation function, and modal bispectrum
coefficients
that have been extracted from our {\Nbody}
simulation suite.
\\
\begin{tabular}{ll}
    \semibold{License} & \href{https://creativecommons.org/licenses/by/4.0/}{Creative Commons Attribution 4.0 International} \;
    \includegraphics[scale=0.12]{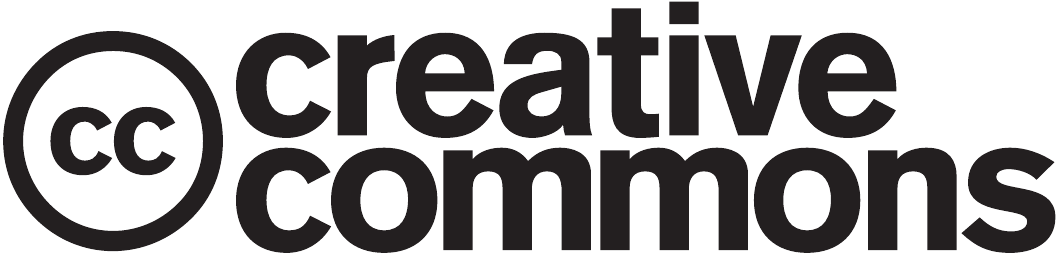}
    \includegraphics[scale=0.15]{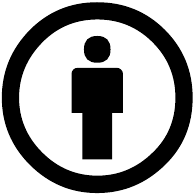} \\
    \semibold{Author} & $\copyright$ University of Sussex 2017. Contributed by Donough Regan \& Alexander Eggemeier  \\
    \semibold{Attribution} & Please cite \texttt{zenodo.org} DOI and this paper \\
    \semibold{Download} & \href{https://zenodo.org/record/438187}{\raisebox{-1mm}{\includegraphics[scale=0.65]{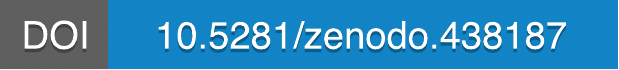}}}
\end{tabular}\\

\label{acknow}
\bibliography{paper}

\appendix

\section{Construction of the modal decomposition}
\label{app:modal}

\subsection{Construction of the $Q$-basis}
\label{app:poly}

The goal of the modal decomposition is to write the estimated bispectrum
in the form
\begin{equation}
	\label{eq:weighting}
	w(k_1,k_2,k_3) \hat{B}(k_1,k_2,k_3)
	=
	\sum_{n =0}^{\nmax-1}
	\hat{\beta}^Q_n Q_n(k_1,k_2,k_3)
	,
\end{equation}
where $w(k_1,k_2,k_3)$ is the arbitrary weighting function~\eqref{eq:weight-function},
and the $Q_n$ represent basis modes with
coefficients $\beta_n^Q$. The $Q_n$ then contain all the
information about the bispectrum.
They should span the possible functions
on wavenumbers $k_i$
that satisfy the triangle
condition, $\sum_i k_i \geq 2 \max\{k_1,k_2,k_3\}$
(denoted by $\TriangleRegion$ in the main text)
but are otherwise arbitrary.
For our concrete numerical results we
choose a basis built out of one-dimensional
polynomials $q_p(x)$ which are orthonormal within $\TriangleRegion$~\citep{Fergusson:2009nv}.
More precisely, in a unit box, we define the integral
$\mathcal{T}[f]=\int_{\TriangleRegion} f(x) \, \D{x}\, \D{y}\, \D{z}$, where $x,y,z$
satisfy the triangle condition within the box $x,y,z\in
[0,1]$. Evaluating the $y$ and $z$ integrals, one finds that
$\mathcal{T}[f]= 0.5 \int_0^1 f(x) \, x(4-3 x) \, \D{x}$. This allows one
to define an inner product, $\langle f , g\rangle \equiv \mathcal{T}[f g]$
(which is not equal to the inner product~\eqref{eq:inner-product-simple})
and set up a generating function for the one-dimensional polynomials,
$q_n$, using $w_n=\mathcal{T}[x^n]$, in the form of a secular determinant
\begin{equation}
	q_n(x)
	=
	\frac{1}{\mathcal{N}}
	\begin{vmatrix}
		1/2 & 7/24 &\dots & w_n \\ 
		7/24 & 1/5&\dots & w_{n+1} \\ 
		\dots & \dots &\dots & \dots\\
		w_{n-1}& w_{n}& . & w_{2n-1}\\
		1& x& \dots & x^n\\
	\end{vmatrix}
\end{equation}
where $\mathcal{N}$ is chosen such that $\langle q_n , q_m\rangle =
\Kronecker_{n m}$. The basis functions $Q_n(x,y,z)$ are defined as
symmetric combinations of combinations of these 1-dimensional polynomials, in the
form
\begin{equation}
	\label{eq:basis_expan}
	Q_n(x,y,z)
	=
	\frac{1}{6}
	\Big[
		q_r(x) q_s(y) q_t(z) +q_r(x) q_t(y) q_s(z)+\dots+q_t(x) q_s(y) q_r(z)
	\Big]
	\equiv
	q_{\{ r}(x) q_s(y) q_{t\}}(z)
	,
\end{equation}
with $n$ representing the triple of indices $\{ r,s,t \}$.
After choosing an ordering of these triples we can
exchange $n$ for a simpler integer label.
For a particular realization with wavenumbers
in the range $\kmin$ and $\kmax$ we use the notation
$Q_n(k_1,k_2,k_3)$ to represent $Q_n(x_1,x_2,x_3)$, where $x_i = (k_i
- \kmin)/(\kmax - \kmin)\in [0,1]$.

\subsection{Calculation of the modal coefficients using the voxel method}\label{app:voxel}

In Section~\ref{sec:modalEst} we explained how
equation~\eqref{eq:wBhat_Qn} reduces estimation of the modal coefficients from
simulation or data to a single 3-dimensional integral over a product of
three Fourier transforms $\Mfactor{n}(\bx)$.
If the bispectrum is given analytically, however,
we may instead use the simpler equation~\eqref{eq:inner-product-simple}
and compute the inner product using a sum of volumes of all
`voxels' within a cubic grid with linear spacing along each
axis $(k_1, k_2, k_3)$.

To calculate the volume of each voxel we relabel the coordinates
as $(x, y, z)$, rescaled so that $0 \leq x, y, z \leq 1$.
We associate each of the 8 possible vertices of the voxel
with a value
$p_1,\dots, p_8$, given
by the product of $Q_m$ and $wB$ (or $Q_m$ and $Q_n$ in the case
of $\llangle Q_m | Q_n \rrangle$) at that vertex.
Finally, we define
an interpolation function
$f$
by writing
\begin{equation}
	f(x,y,z)=a_1+a_2 x + a_3 y + a_4 z + a_5 x y + a_6 x z + a_7 y z + a_8 x y z .
\end{equation}
The coefficients $a_i$ may be obtained analytically in terms of the $p_i$.
We assign the volume of the voxel to be zero if fewer than four of its
vertices satisfy the triangle condition, while
if all $8$ vertices satisfying the triangle condition
its volume is
\begin{equation}
	\int_{0 \leq x,y,z \leq 1} f(x,y,z) \, \D{x} \, \D{y} \, \D{z}
	=
	\frac{1}{8} \sum_{i=1}^8 p_i
	,
\end{equation}
as expected.
For intermediate cases we write the volume in the form
\begin{equation}
	\int_{\mathcal{C}} f(x,y,z) \, \D{x} \, \D{y} \, \D{z}
	,
\end{equation}
where $\mathcal{C}$ indicates that only those points satisfying the
triangle condition and forming a closed volume within the voxel should
be included. In the case of $4$ points there are 3 possible volumes
given by
\begin{equation}
  \mathcal{C}_a^{(4)}=\{x,y,z\mid x+1 \leq y+z\}\,,\quad\mathcal{C}_b^{(4)}
  =\{x,y,z \mid y+1 \leq x+z\}\,,\quad\mathcal{C}_c^{(4)}=\{x,y,z \mid z+1 \leq x+y\}\,.
\end{equation}
For $5$ points the only possibility is that $x+y+z \geq 2 \max\{x,y,z\}$, while for $6$ and $7$ points there are again $3$
possibilities, given respectively by,
\begin{equation}
\begin{split}
  \mathcal{C}_a^{(6)} & =\{x,y,z \mid x \leq y+z , y\leq x+z\} ,
  &
  \mathcal{C}_b^{(6)} & =\{x,y,z \mid x \leq y+z , z\leq x+y\} ,
  &
  \mathcal{C}_c^{(6)} & =\{x,y,z \mid y \leq x+z ,z \leq x+y \} ,
  \\
  \mathcal{C}_a^{(7)} & =\{x,y,z \mid x \leq y+z\} ,
  &
  \mathcal{C}_b^{(7)} & =\{x,y,z \mid y \leq x+z\} ,
  &
  \mathcal{C}_c^{(7)} & =\{x,y,z \mid z \leq x+y\} .
\end{split}
\end{equation}
In each case the analytic form of the integral in terms of the vertex
values $p_i$ can be calculated easily. Computation of each integral using
this voxel method is highly accurate and efficient.

\end{document}